\documentclass[aps,amsmath,amssymb,prd,floatfix,preprint,superscriptaddress,nofootinbib,12pt]{JHEP3}
\usepackage{epsfig}
\usepackage{amsmath}
\usepackage{amssymb,amsfonts}
\usepackage{comment}
\usepackage{graphicx}
\usepackage{latexsym}
\usepackage{slashed}
\usepackage{xcolor}
\usepackage{subfigure}
\usepackage{subfig}

\let\normalcolor\relax

\relax
\renewcommand{\theequation}{\arabic{section}.\arabic{equation}}
\def\be{\begin{equation}}
\def\ee{\end{equation}}
\def\bs{\begin{subequations}}
\def\es{\end{subequations}}
\newcommand{\een}{\end{subequations}}
\newcommand{\ben}{\begin{subequations}}
\newcommand{\beq}{\begin{eqalignno}}
\newcommand{\eeq}{\end{eqalignno}}

\DeclareMathOperator\Tr{tr}

\DeclareMathOperator{\sgn}{sgn}

\def\l{\lambda}
\def\m{\mu}
\def\n{\nu}

\def\e{\epsilon}

\newcommand\fverb{\setbox\pippobox=\hbox\bgroup\verb}
\newcommand\fverbdo{\egroup\medskip\noindent%
                        \fbox{\unhbox\pippobox}\ }
\newcommand\fverbit{\egroup\item[\fbox{\unhbox\pippobox}]}
\newbox\pippobox

\def\hri#1#2{\href{http://arxiv.org/abs/#1}{[ArXiv:#1]#2}}
\def\hre#1#2{\href{http://arxiv.org/abs/#1/#2}{[ArXiv:#1/#2]}}

\def\beq{\begin{equation}}
\def\eeq{\end{equation}}

\def\4R{{{}^{(4)}R}}

\def\K5{{\kappa}}
\def\K52{{\kappa^2}}

\newcommand{\half}{\frac{1}{2}}

\def\bea{\begin{eqnarray}}
\def\eea{\end{eqnarray}}
\def\nn{\nonumber}
\newcommand{\ba}{\begin{aligned}}
\newcommand{\ea}{\end{aligned}}

\title{Liouville theory and Matrix models \\
A Wheeler DeWitt perspective}
\author{P. Betzios$^{\flat}$, O. Papadoulaki$^*$\\
~\\
$^\flat$ \href{http://hep.physics.uoc.gr}{Crete Center for Theoretical Physics}, Institute for Theoretical and Computational Physics,
Department of Physics,  P.O. Box 2208,\\
University of Crete, 70013, Heraklion, Greece
~\\
~\\
$^*$ \href{https://www.ictp.it/}{International Centre for Theoretical Physics} \\
Strada Costiera 11, Trieste 34151 Italy.
}

\preprint{CCTP-2020-4\\
ITCP-IPP-2020/4}

\abstract{We analyse the connections between the Wheeler DeWitt approach for two dimensional quantum gravity and holography, focusing mainly in the case of Liouville theory coupled to $c=1$ matter. Our motivation is to understand whether some form of averaging is essential for the boundary theory, if we wish to describe the bulk quantum gravity path integral of this two dimensional example. The analysis hence, is in a spirit similar to the recent studies of Jackiw-Teitelboim (JT)-gravity. Macroscopic loop operators define the asymptotic region on which the holographic boundary dual resides. Matrix quantum mechanics (MQM) and the associated double scaled fermionic field theory on the contrary, is providing an explicit ``unitary in superspace" description of the complete dynamics of such two dimensional universes with matter, including the effects of topology change. If we try to associate a Hilbert space to a single boundary dual, it seems that it cannot contain all the information present in the non-perturbative bulk quantum gravity path integral and MQM.}


\begin{document}

\section{Introduction}

The studies of SYK and its low energy (hydrodynamic) limit described by the one dimensional Schwarzian theory~\cite{KM,JE,Almheiri:2014cka} revealed a holographic connection with a bulk two dimensional Jackiw-Teitelboim (JT) gravity theory. In fact recent work~\cite{Saad:2019lba} elucidated that this connection continues to hold for bulk topologies other than the disk, and that the complete bulk genus expansion can be resummed using a particular limit of the (double scaled) Hermitean matrix model
\be\label{onematrix}
\mathcal{Z} = \int d H e^{- N V(H)} \, .
\ee
The argument supporting this connection, is that the usual double scaling limit of a single Hermitean matrix model can describe the $(2,p)$ minimal models coupled to gravity, and the physics of JT gravity can be reached as a $p \rightarrow \infty$ limit of these models\footnote{For a complementary description, and a more extended analysis of its relation to minimal models, see~\cite{Okuyama:2019xbv,Johnson:2019eik,Johnson:2020heh}.}. Actually it is quite reasonable to expect such a limiting connection between the JT gravity and Liouville theory. As an example the Liouville equation appears naturally after employing two steps, first the identification of the boundary mode Schwarzian action with the Kirillov coadjoint orbit action on $M = Diff/SL(2,R)$ that is then identified with a 2d bulk non-local Polyakov action together with appropriate boundary terms~\cite{Mandal:2017thl}. The classical solutions of this latter action are then in correspondence with those arising from Liouville theory, albeit in this case the conformal mode of the metric is a non-dynamical non-normalisable mode fixed by imposing certain appropriate Virasoro constraints. This is an indirect way to say that the only dynamical degrees of freedom left in this problem are those of the fluctuating boundary arising from large diffeomorphisms\footnote{Yet another connection of the Schwarzian action with Liouville quantum mechanics on the boundary of space was analysed in~\cite{Bagrets:2017pwq}.}. This then indicates that the various existing models of Liouville quantum gravity coupled to matter~\cite{Klebanov:1991qa,Ginsparg:1993is,Nakayama:2004vk}, are in fact \emph{richer} examples of two dimensional bulk quantum gravity theories. Since a lot is known for these models both at a perturbative and non-perturbative level, it is both conceptually interesting and feasible to elucidate the properties of their holographic boundary duals. That said, we should clarify that from this point of view the various matrix models do not play the role of their boundary duals, but should be instead thought of as providing directly a link\footnote{This link is exemplified by the passage to the appropriate second quantised fermionic field theory.} to a ``third quantised description" of the bulk universes splitting and joining in a third quantised Hilbert space~\cite{BabyUniverses}. This interpretation is even more transparent in the $c=1$ case for which there is a natural notion of ``time" in superspace in which universes can evolve. Simply put, the target space of the $c=1$ string plays the role of superspace in which these two dimensional geometries are embedded.

The matrix models provide a quite powerful description, since it is possible to use them in order to obtain the partition function or other observables of the boundary duals - from the matrix model point of view one needs to introduce appropriate loop operators that create macroscopic boundaries on the bulk geometry. Let us briefly discuss the case of the partition function. In this case for the precise identification, one should actually use a loop/marked boundary of fixed size $\ell$ that is related to the temperature $\beta$ of the holographic dual theory~\cite{Saad:2019lba}. The Laplace transform of this quantity then gives the expression for the density of states (dos) of the boundary dual. In the concrete example corresponding to the $(2,p)$ models, this was shown to reduce to the Schwarzian density of states in the limit $p \rightarrow \infty$~\cite{Saad:2019lba}. Since the Schwarzian theory captures only the IR hydrodynamic excitations of the complete SYK model, it is then natural to ponder whether and how one could connect various integrable deformations of the $(2,p)$  matrix models with corresponding corrections to the ``hydrodynamic" Schwarzian action.

In particular a matrix model with a general potential of the form $V(H) = \sum_k t_k H^k$ is still an integrable system, and it is known that its partition function corresponds to a $\tau$-function of the KP-Hierarchy~\cite{Klebanov:1991qa,Ginsparg:1993is}. Similar things can be said about two-matrix models (2MM), with which one can describe the more general $(q,p)$ minimal models~\cite{Seiberg:2003nm,Kazakov:2004du,Maldacena:2004sn,Nakayama:2004vk}. The partition function of such two matrix models takes the general form
\be\label{2MMtau}
\tau_N = \mathcal{Z}(N) = \int d M \, d \bar{M} e^{-N \left( \Tr ( M \bar{M}) + \sum_{k>0} ( t_k \Tr M^k + \bar{t}_k  \Tr \bar{M}^k) \right)} \, ,
\ee
and is a $\tau$-function of the Toda integrable Hierarchy with $t_k$'s , $\bar{t}_k$'s playing the role of Toda "times".  Very interesting past work on the integrable dynamics of interfaces (Hele-Shaw flow) has revealed a deep connection between the dynamics of curves on the plane and this matrix model~\cite{Conformalmaps}. In fact the Schwarzian universally appears in the dispersionless limit of the Toda hierarchy when $\hbar = 1/N \, \rightarrow 0$, and can be related with a $\tau$-function for analytic curves which in turn is related to \eqref{2MMtau}. We will briefly review some of these facts in appendix~\ref{conformalmaps}, since they are related tangentially to this work.

The main focus of the present paper will be the case of $c=1$ Liouville theory having a dual description in terms of Matrix quantum mechanics of $N$-ZZ $D0$ branes~\cite{McGreevy:2003kb}. We emphasize again that even though in this case there is a natural interpretation of the theory as a string theory embedded in a two dimensional target space, the Liouville theory being a worldsheet CFT, in the present paper we will take the 2-d Quantum Gravity point of view~\cite{Moore:1991ag,Ginsparg:1993is}, where the worldsheet of the string will be treated as the bulk spacetime. This is in analogy with our previous discussion and interpretation of Jackiw-Teitelboim gravity and the minimal models. In short we will henceforth interpret the combination of $c=1$ matter with Liouville theory as a bone fide quantum gravity theory for the bulk spacetime. Let us make clear again that we do not wish to reproduce the JT-gravity results for the various observables, the theory we analyse is a richer UV complete theory of two dimensional gravity with matter. In particular for the $c=1$ case at hand, at the semiclassical level the dynamical degrees of freedom are then the conformal mode of the two dimensional bulk metric - Liouville field $\phi(z, \bar{z})$ - together with that of a $c=1$ matter boson which we denote by $X(z, \bar{z})$. The Euclidean bulk space coordinates will then be denoted by $z, \bar{z}$. 

The possibility for connecting this bulk quantum gravity theory with holography is corroborated by the fact that once we introduce macroscopic boundaries for the Liouville CFT (corresponding to insertions of macroscopic loops of size $\ell$ on a worldsheet in the usual point of view), near such boundaries the bulk metric can take  asymptotically the form of a nearly-$AdS_2$ space. The adjective nearly here corresponds to the fact that we can allow for fluctuations of the loop's shape keeping its overall size fixed. This is quite important, since it allows for a holographic relation between the bulk quantum gravity theory with a quantum mechanical system on the loop boundary akin to the usual AdS/CFT correspondence\footnote{Such an interpretation could in principle dispense with the constraint of the \emph{bulk Liouville} theory being a CFT and we might now have the freedom of defining a richer class of two dimensional bulk theories with more general matter content if we do not insist on a string theory interpretation.}. It could also pave the way to understand the appropriate extension and interpretation of the correspondence in the case of geometries having multiple asymptotic boundaries (Euclidean wormholes). Even though several proposals already exist in the literature~\cite{Maldacena:2004rf,Betzios:2019rds,Saad:2019lba,Marolf:2020xie}, it is fair to say that no ultimate consensus on the appropriate holographic interpretation of such geometries has been reached (and if it is unique). We provide more details on what we have learned about this intricate problem in the conclusions. 

Another natural question from the present point of view, is the role of the original matrix quantum mechanics (MQM) of N-$D0$ branes ($ZZ$ branes) described by $N \times N$ Hermitean matrices $M_{i j}(x)$. As we describe in the main part of the paper, the dual variable to the loop length $\ell$ that measures the size of macroscopic boundaries of the bulk of space, is a collective variable of the matrix eigenvalues $\lambda_i(x)$ of $M_{ij}(x)$, while the coordinate $x$ is directly related to the matter boson in the bulk. Hence if we interpret the compact $\ell$ as the inverse temperature $\beta$ of the boundary theory, we are then forced to think of the collective matrix eigenvalue density $\rho(\lambda)$ as describing the energy spectrum of the dual theory. This is in accord with the duality between JT gravity and the Hermitean one matrix model of \eqref{onematrix}. From this point of view the $D0$ branes now live in ``superspace" and the second quantised fermionic field theory is actually a ``third quantised" description of the dynamics of bulk universes. This means that MQM and the associated fermionic field theory provide us with a specific non-perturbative completion of the $c=1$ bulk quantum gravity path integral\footnote{A more recent understanding of the non-perturbative completion/s of the model and the role of ZZ-instantons is given in~\cite{Balthazar:2019rnh}.}, as the one and two matrix models do in the simpler cases of JT-gravity and $(p,q)$ minimal models.

This discussion raises new interesting possibilities as well as questions. To start with, one can now try to understand at a full quantum mechanical level various asymptotically $AdS_2$ bulk geometries such as black holes (together with the presence of matter excitations) directly on what was previously interpreted as the worldsheet of strings\footnote{Similar considerations have previously been put forward by~\cite{deBoer:2017xdk,Vegh:2019any}.}. In fact one can go even further, using the free non-relativistic fermionic field theory. Based on the analysis of~\cite{BabyUniverses}, it is a very interesting and unexpected fact that this field theory is non-interacting but can still describe the processes of bulk topology change. This is made possible due to the fact that the field theory coordinate $\lambda$ is related with the conformal mode of the bulk geometry $\phi$ via a complicated non-local transform~\cite{Moore:1991ag}\footnote{It is an interesting problem whether some similar transform could encode compactly the process of topology change in higher dimensional examples, at least at a minisuperspace level.}. Therefore it seems that we have managed to find a (quite simple) \emph{non-disordered} quantum mechanical system that performs the full quantum gravity path integral and automatically sums over bulk topologies. Remarkably this system is dynamical and defined on superspace instead of being localised on a single boundary of the bulk space. More precisely, it is an integrable system on superspace where ``time" is related to the $c=1$ matter field. This point of view then inevitably leads to the following questions: Does the bulk theory actually contain states that can be identified with black holes? Can we then describe complicated processes such as those of forming black holes - what about unitarity if we can create baby universes? Is there any notion of chaos for the bulk theory even though the superspace field theory is an integrable system? Is there a quantum mechanical action defined directly on the boundary of the bulk space (or at multiple boundaries) that can encapsulate the same physics? Would this have to be an intrinsically disorder averaged system such as SYK or could it be a usual unitary quantum mechanical system with an associated Hilbert space? Our motivation hence is to try to understand and answer as many of these questions as possible.

\paragraph{Structure of the paper and results -}
Let us now summarise the skeleton of our paper as well as our findings. In section~\ref{Liouvilletheory} we review some facts about Liouville theory with boundaries, such as the various solutions to equations of motion and the minisuperspace wavefunctions that are related to such geometries. In appendix~\ref{Minimalmodels} we briefly describe the matrix models dual to the minimal models and their integrable deformations and describe the connections with conformal maps of curves on the plane and the Schwarzian. The main focus of our analysis is the $c=1$ case. We first review how the fermionic field theory can be used as a tool to extract various correlators in section~\ref{MQMandfermions}, and then move into computing the main interesting observables: First the boundary dual thermal partition function $Z_{dual}(\beta)$ both at genus zero and at the non-perturbative level in section~\ref{Partitionfunction} and then the \emph{dual density of states} $\rho_{dual}(E)$ in~\ref{DOSofdual}. 

At genus zero the partition function corresponds to the Liouville minisuperspace WdW wavefunction, while the non-perturbative result cannot be given such a straightforward interpretation, but is instead an integral of Whittaker functions that solve a corrected WdW equation encoding topology changing terms. We then observe that there is an exponential increase $\sim e^{2 \sqrt{\m} E}$ in the dos at low energies ($\mu$ plays the role of an infrared cutoff to the energy spectrum such that the dual theory is gapped), that transitions to the Wigner semicircle law $\sim \sqrt{E^2 - 4 \mu^2}$ at higher energies, combined with a persistent fast oscillatory behaviour of small amplitude. These oscillations might be an indication that whatever the boundary dual theory is, it has a chance of being a non-disorder averaged system. Similar non-perturbative effects also appear in~\cite{Saad:2019lba}, but in that case are related to a doubly-exponential non-perturbative contribution to various observables. We will comment on a possible interpretation of such non-perturbative effects from the bulk point of view in the conclusions~\ref{Conclusions}. 

We then continue with an analysis of the density of states two-point function $\langle \rho_{dual}(E) \rho_{dual}(E') \rangle$ and its fourier transform, the \emph{spectral form factor}  $SFF(t) = \langle Z(\beta + i t) Z(\beta - i t) \rangle$ in section~\ref{twoasregions}. The geometries contributing to these quantities are both connected and disconnected. The correlator of energy eigenvalues has a strong resemblance with the \emph{sine}-kernel, and  hence exhibits the universal short distance repulsion of matrix model ensembles such as the GUE. Nevertheless, its exact behaviour deviates from the sine-kernel slightly, indicative of non-universal physics. For the SFF, the disconnected part quickly decays to zero and at late times it is the connected geometries that play the most important role. These are complex continuations of Euclidean wormholes corresponding to loop-loop correlators $\langle \hat{\mathcal{W}}(\ell_1, q) \hat{\mathcal{W}}(\ell_2, - q) \rangle = M_2(\ell_1, \ell_2) $ from the point of view of the matrix model\footnote{From now on we denote with $q$ the momentum dual to the matter field zero mode $x$. The SFF is computed in the limit $q \rightarrow 0$, the generic loop correlator defines a more refined observable with Dirichlet ($x =$ fixed), or Neumann ($q = $fixed) boundary conditions for the matter field $X$.}. The relevant physics is analysed in section~\ref{SFFconnected} where it is found that at genus zero (and zero momentum - $q=0$), they lead to a constant piece in the SFF. The non-perturbative answer can only be expressed as a double integral with a highly oscillatory integrand. A numerical analysis of this integral exhibits the expected increasing behaviour that saturates in a plateau, but on top of this there exist persistent oscillations for which $\Delta SFF_c / SFF_c \rightarrow O(1)$ at late times\footnote{They are not as pronounced as in higher dimensional non-integrable examples, where they are extremely erratic and their size is of the order of the original signal.}. On the other hand for non-zero $q$ the behaviour is qualitatively different. The non-perturbative connected correlator exhibits an initially decreasing slope behaviour, that transitions into a smooth increasing one relaxing to a plateau at very late times. The main qualitative difference with the $q=0$ case is that the behaviour of the correlator is much smoother. A single boundary dual cannot capture the information contained in the connected correlator, since there is no indication for a factorisation of the complete non-perturbative result. The only possibilities left retaining unitarity, are that either the connected correlator describes a system of coupled boundary theories as proposed in~\cite{Betzios:2019rds}, or that it is inherently impossible for a single boundary dual to describe this information contained in the bulk quantum gravity path integral and MQM, and hence unitarity can be only restored on the complete ``third quantised Hilbert space".

In section~\ref{DSregime}, we analyse a possible cosmological interpretation of the wavefunctions, as WdW wavefunctions of two-dimensional universes. In order to do so we follow the analytic continuation procedure in the field space proposed and studied in~\cite{Hertog:2011ky,Maldacena:2019cbz,Cotler:2019nbi,Cotler:2019dcj}, that involves what one might call ``negative $AdS_2$, trumpet geometries". In our description this corresponds to using loops of imaginary parameter $z = i \ell$. The wavefunctions at genus zero are Hankel functions $\sim \frac{1}{z} H^{(1)}_{i q}(z)$. Nevertheless, it is known since the work of~\cite{DaCunha:2003fm}, that all the various types of Bessel functions can appear, by imposing different boundary conditions and physical restrictions on the solutions to the mini-superspace WdW equation. We provide a review of the two most commonly employed ones (no-boundary and tunneling proposals), in appendix~\ref{WdWcompendium}. The non-perturbative description seems to encode all these various possibilities in the form of different large parameter limits of the non-perturbative Whittaker wavefunctions \eqref{ExactwavefunctionsWdW}. In a geometrical language, this corresponds to complexifying the bulk geometries and choosing different contours in the complex field space. Obstructions and Stokes phenomena are then naturally expected to arise, for the genus zero asymptotic answers. 

We conclude with various comments and suggestions for future research.

\paragraph{A summary of relations -}

We summarize in a table the various relations that we understand between the various physical quantities from the bulk quantum gravity (Liouville) and matrix model point of view. More details can be found in the corresponding chapters. Empty slots correspond either to the fact that there is no corresponding quantity, or that we do not yet understand the appropriate relation. Notice that the bulk quantum gravity theory can also be interpreted as a string theory on a target space. The acronyms used are $DOS:$ for the density of states and $SFF:$ for the spectral form factor.

\vskip .3cm
\begin{tabular}{||c|c|c||}
\hline
\textbf{Quantum gravity} & \textbf{Matrix model} & \textbf{Boundary dual} \\ \hline\hline
Liouville potential $\mu\, e^{2\phi}$ & 
	Inverted oscillator potential &  - \\ \hline
 Cosmological constant $\mu$ &
	Chemical potential $-\mu$ & IR mass gap $\mu$ \\ \hline
$D0$ particle ($\phi$: D, $X$: N) &
	Matrix eigenvalue $\l_i$ & Energy eigenvalue $E_i$ \\ \hline  	

Boundary: $S_{ bdy}\,=\,\mu_B \oint\, e^{\phi}$ &
	 Loop operator: $\langle \Tr \log [\mathrm{z} - \l ] \rangle$ & Microcanonical $\langle \rho_{dual}(E) \rangle$ \\ \hline
Bdy. cosm. const. $\mu_B$ &
	Loop parameter $\mathrm{z}$ & Energy $E$ \\ \hline
fixed size bdy $\ell = e^{\phi_0}$  & Loop length $\ell$ &
	Inv. temperature $\beta$ \\ \hline	

WdW wavefunction $\Psi(\ell)$ & 
	Fixed size loop oper. $\langle M_1(\ell) \rangle$ & Partition func. $Z_{dual}(\beta)$ \\ \hline	

Third quantised vacuum & 
	Fermi sea of eigenvalues & - \\ \hline	

Closed surfaces &
	Fermionic density quanta & - \\ \hline
S-matrix of universes &
	S-matrix of density quanta & -  \\ \hline  	
Two boundaries: $\ell_{1,2}$ &
	Loop correlator $\langle M_2(\ell_1, \ell_2) \rangle$ &  SFF: $\ell_{1,2} = \beta \pm i t$  \\ \hline  	

Two boundaries: $\mu_B^{1,2}$ &
	Density corr. $\langle \rho(\lambda_1) \rho(\lambda_2) \rangle$ &  DOS. correlator  \\ \hline  	

\end{tabular}
\vskip .2cm

\section{Liouville theory}\label{Liouvilletheory}

We begin by briefly reviewing some general facts about Liouville theory, focusing mainly in the $c=1$ case. In this latter case, the Liouville action is to be completed with the action of one extra free bosonic matter field which we will label $X(z, \bar{z})$. In course we will also delineate the points of departure of the usual interpretation of the theory as a string theory embedded in the linear dilaton background. The Liouville action on a manifold with boundaries is~\cite{Fateev:2000ik,Teschner:2000md}  
\be\label{Liouvilleaction} 
S = \int_{\mathcal{M}} d^2 z \sqrt{g} \left(\frac{1}{4 \pi} g^{a b} \partial_a \phi \partial_b \phi + \frac{1}{4 \pi} Q R \phi + \mu e^{2 b \phi} \right) + \int_{\partial \mathcal{M}} d u g^{1/4} \left(\frac{Q K \phi}{2 \pi} + \mu_B e^{b \phi} \right)\, ,
\ee
with K the extrinsic curvature and the parameters $\mu, \mu_B$ the bulk-boundary cosmological constants. This interpretation for these parameters stems from the fact that the simplest bulk operator is the area $A = \int_{\mathcal{M}} d^2 z \sqrt{g} e^{2 b \phi} $ and the simplest boundary operator is the length of the boundary $\ell = \oint \, du \, g^{1/4}  e^{b \phi(u)}$, with $u$ parametrising the boundary coordinate of the surface. A natural set of operators are
\be
V_a \, = \, e^{2 a \phi(z, \bar{z})} \, , \qquad \Delta_a = a (Q - a) \, .
\ee
There exist special operators among them, for which $a = Q/2 + i P$, with $P$ real, that correspond to non-local operators that create macroscopic holes in the geometry. Their dimensions are $\Delta = Q^2/4 + P^2$ and correspond to Liouville primaries that are delta-function normalised. 
The various parameters of Liouville theory are related in the following way ($\mu_{KPZ}$ is the so called KPZ scaling parameter appearing in correlation functions)
\bea\label{liouvilleparameters}
c_L=1+6 Q^2\, , \quad \quad Q= b + b^{-1}\, , \nn \\
\quad \mu_B = \frac{\Gamma(1-b^2)}{\pi}\sqrt{\mu_{KPZ}} \cosh(\pi b \sigma)\, , \quad \quad \mu_{KPZ} = \pi \mu \frac{\Gamma(b^2)}{\Gamma(1-b^2)}\, .
\eea
If this action is completed together with a $c=1$ boson which we will denote by $X(z, \bar{z})$, then $c_{matter}=1 \Rightarrow b=1, \, Q=2$ and one finds a renormalization of $\mu_{KPZ}, \,  \mu_B$ such that
$\mu_B =  2 \sqrt{\mu} \cosh(\pi \sigma)$ becomes the correct relation between the bulk and boundary cosmological constants in terms of a dimensionless parameter $\sigma$.

To get into contact with the holographic picture, it is first important to discuss the various boundary conditions for the metric and matter fields $\phi$ and $X$ and then the properties of the relevant boundary states. The matter field being a free boson, can satisfy either Dirichlet or Neumann conditions in the usual fashion. It is easy to see from \eqref{Liouvilleaction}, that the analogous possibilities for the Liouville mode are ($n$ is the unit normal vector)
\be
\delta \phi |_{\partial M} = 0 \, ,  \quad \frac{ \partial  \phi}{\partial n} + Q K + 2 \pi \mu_B b e^{b \phi} \, |_{\partial M} = 0 \, .
\ee
The Dirichlet boundary conditions $\phi |_{bdy} = \phi_b$ are conformally invariant only asymptotically for $\phi = \pm \infty$. In the limit $\phi \rightarrow - \infty$ we have the weakly coupled region where the metric acquires an infinitesimal size and thus this is the regime in which we describe local disturbances of the bulk space. On the other hand for $\phi \rightarrow \infty$ (strongly coupled region of Liouville) distances blow up and we probe large scales of the bulk metric.

In addition, for a Holographic interpretation that is in line with the $AdS/CFT$ correspondence, there ought to be a possibility for the bulk geometry to asymptote to $AdS_2$. In fact this is precicely a solution of the Liouville theory equations of motion for which the asymptotic boundary is at $\phi \rightarrow \infty$
\be\label{Poincaredisk}
e^{2 b \phi}(z, \bar{z}) = \frac{Q}{ \pi \mu b} \frac{1}{(1 - z \bar{z})^2}
\ee
This solution is that of the constant negative curvature metric on the Poincare disk. The metric is invariant under the Moebius transformations of the group $PSL(2,R)$. Let us now discuss more general solutions. On a quotient of hyperbolic space $H_2/\Gamma$ with $\Gamma$ a discrete Fuchsian group, the general metric that solves the Liouville equations of motion is defined in terms of two arbitrary functions~\cite{Ginsparg:1993is}
\be\label{Generalmetric}
ds^2 = e^{2 b \phi}(z, \bar{z}) d z d \bar{z} = \frac{Q}{ \pi \mu b}  \frac{\partial A \bar{\partial}B}{\left(A(z) - B(\bar{z}) \right)^2} d z d \bar{z} \, .
\ee
There exist three types of monodromy properties of $A,B$ near non-trivial cycles of the manifold (three $SL(2,R)$ conjugacy classes). For the hyperbolic one the curve surrounds a handle and this class can be used to describe higher topologies. We also have the elliptic monodromy class, which corresponds to surfaces on which the curve surrounds local punctures and the parabolic class for which the curve surrounds macroscopic boundaries. The solution \eqref{Poincaredisk} is of the parabolic class. Solutions of the elliptic class can be useful to describe the presence of singularities in the bulk of space\footnote{In fact there is a competition and a transition between the various solutions, in an ensemble that depends on the ratio between the area of the surface $A$ and the length of the loop $\ell$~\cite{Moore:1991ag}.}. The boundary of \eqref{Generalmetric} is now at the locus $A(z) = B(\bar{z})$, instead of the previous $|z|=1$. Of course these more general metrics capture also the sub-case of Nearly-$AdS_2$ geometries~\cite{Mandal:2017thl} that correspond to slightly deforming the shape of the boundary. We notice that the important coordinate independent condition on the possible deformations that we allow\footnote{This should also hold away from the strict Schwarzian limit of the minimal models of appendix~\ref{Minimalmodels}.}, is that they keep the $SL(2,R)$ conjugacy class near the boundary to be of the parabolic type so that one still finds a macroscopic boundary on which the holographic dual can reside.

Another point of view for understanding such types of geometries, is to relate them to solutions to the bulk \emph{minisuperspace} WdW equation~\cite{Ginsparg:1993is}\footnote{We will henceforth work in the $c=1$ case ($b=1$) using the scaling parameter $\mu_{KPZ} = \mu$, in order not to clutter the notation.}
\be\label{MinisuperspaceWdW}
\left( - \frac{\partial^2}{\partial \phi_0^2} + 4 \mu e^{2 \phi_0} - q^2 \right) \Psi_q (\phi_0) = 0 \, .
\ee
In the expression above $q$ is the momentum conjugate to the zero mode $x$ of the matter field $X(z,\bar{z})$ and thus a real number. In case the surface has a boundary of size $\ell$, this can be expressed in terms of the zero mode $\phi_0$ as $\ell = e^{\phi_0}$ which is kept fixed. Notice that by fixing only the overall boundary size we can still allow the possibility of other non-trivial non-zero mode deformations to change its shape, eqn. \eqref{MinisuperspaceWdW} focuses only on the zero mode. The wavefunctions corresponding to loops with macroscopic sizes (at genus-zero) are given by
\be\label{MacroscopicWdW}
\Psi^{macro}_q (\ell) = \frac{1}{\pi} \sqrt{q \sinh \pi q} \, K_{i q} (2 \sqrt{\mu} \ell) \, ,
\ee
These are (delta-function) normalizable wavefunctions with the norm 
\be 
\int_0^\infty \frac{d \ell}{\ell} \Psi^{macro}_q (\ell) \Psi^{macro}_{q'} (\ell) = \delta(q-q') \, ,
\ee
which are exponentially damped for $\ell \rightarrow \infty$ and oscillate an infinite number of times for $\ell \rightarrow 0$. Different topologies of the bulk geometries are not described by these wavefunctions, but we will later see that the matrix model is able to resum the topologies automatically and express the wavefunctions in terms of integrals of Whittaker functions.

On the other hand, the microscopic states correspond to wavefunctions that diverge as $\ell \rightarrow 0$ and vanish as $\ell \rightarrow \infty$, given by an analytic continuation of the previous solutions
\be\label{MicroscopicWdW}
\Psi^{micro}_\omega (\ell) = \frac{1}{\pi} \sqrt{\omega \sinh \pi \omega} K_{\omega} (2 \sqrt{\mu} \ell) \, ,
\ee
where $\omega$ is a real number. These are non-normalisable and correspond to local punctures (short distance bulk singularities). Notice that we can assign two different interpretations for such wavefunctions. One is from the point of view of the bulk surface and in this case they correspond to solutions with elliptic monodromy around a puncture. The second is to consider an analytic continuation for the matter field $x \rightarrow i t $ with a subsequent interpretation of $\omega$ as a superspace frequency. In particular from the  point of view of third quantisation, these are good asymptotic wavefunctions to use, if we wish to describe the scattering of universes in superspace, and in particular they probe the region near what we would call a ``Big-Bang" or ``Big-Crunch" singularity.

Let us finally briefly discuss the boundary conditions for the matter field $X(z, \bar{z})$ and the various types of D-branes that appear in Liouville theory~\cite{McGreevy:2003kb,McGreevy:2003ep,Klebanov:2003km}\footnote{That are interpreted here as superspace D-branes or \emph{SD-branes}.}. As we noted these can be either Neumann or Dirichlet, which corresponds to the two possible choices of quantising matter fields on Euclidean $AdS_2$. Neumann boundary conditions for $X(z, \bar{z})$ correspond to ZZ boundary conditions and are relevant for describing $D0$ branes. Such branes are localised in the large $\phi \rightarrow \infty$ region. The FZZT-branes have Neumann conditions for $\phi$ (and either condition for $X$), stretch from $\phi \rightarrow - \infty$ and dissolve at the region of $\phi \sim - \log \mu_B$, so they can even penetrate the strongly coupled (large scale of the geometry) region depending on the value of $\mu_B$. This is for the fixed $\mu_B$ ensemble (that corresponds to unmarked boundary conditions according to~\cite{Saad:2019lba}). The relevant wavefunction (for $c=1$) is
\be
\Psi_\nu(\sigma) = \mu^{- i \nu} \frac{\left[\Gamma(1+ 2 i \nu)\right]^2 \cos (2 \pi \nu \sigma)}{2^{1/4}(- 2 i \pi \nu)}
\ee
If we instead perform a Laplace transform in order to keep the length of the boundary $\ell = e^{\phi_0}$ fixed, we are then describing surfaces with fluctuating boundaries of fixed size. Using the relation $\mu_B = 2 \sqrt{\mu} \cosh (\pi \sigma)$, the fixed $\ell$ wavefunction is~\cite{Martinec:2003ka} 
\be
\Psi_\nu(\ell) \sim K_{2 i \nu}(2 \sqrt{\mu} \ell) = \int_0^\infty d (\pi \sigma) e^{- 2 \sqrt{\mu} \ell \cosh (\pi \sigma) } \cos(2 \pi \nu \sigma) \, .
\ee
This is again a solution to the minisuperspace WdW equation \eqref{MinisuperspaceWdW} upon identifying $q = 2 \nu$. This was given as an argument~\cite{Fateev:2000ik} for the minisuperspace description being exact (up to overall normalisations of the wavefunctions). Of course this WdW equation is exact only for the genus zero result, whilst the non-perturbative result given in section~\ref{Partitionfunction} can be related to a corrected version of the WdW that incorporates a term related to changes in topology. 

We conclude this section briefly mentioning the last type of $D$-branes, the $D$-instantons~\cite{Giombi:2008sr,Balthazar:2019rnh} that correspond to Dirichlet conditions both in  $X(z, \bar{z})$ and $\phi(z, \bar{z})$. Their wavefunctions are labelled by two integers $m,n$ (as for the $D0$-branes). These instantons are important for properly defining the vacuum state of the theory (here this is the third quantised vacuum) and were also argued to be related to fragmented $AdS_2$ spaces but we will not analyse them further here.

\section{Matrix quantum mechanics and fermionic field theory}\label{MQMandfermions}

Let us now pass to the double scaled free fermionic field theory description of the model~\cite{Moore:1991sf,Klebanov:1991qa,Ginsparg:1993is} that allows for an exact computation of various observables. After diagonalising the variables of matrix quantum mechanics and passing to the double scaling limit, the dynamics can be equivalently described in terms of the second quantised non-relativistic fermionic field action
\be\label{secondfermifield}
S = \int d t \, d \lambda \, \hat \psi^\dagger(t,\lambda) \left( i \frac{\partial}{\partial t} + \frac{\partial^2}{\partial \lambda^2} + \frac{\lambda^2}{4} \right) \hat \psi(t, \lambda) \,.
\ee
The double scaled fermi fields are defined using the normalised even/odd parabolic cylinder functions (see appendix~\ref{Parabolic}), $\psi^s(\omega,\lambda) \, s=\pm$, as (summation over the signs s, is implicit)
\be
\hat \psi (t, \, \lambda) = \int d \omega \, e^{i \omega t} \, \hat{b}_{s}(\omega) \, \psi^s(\omega, \lambda)
\ee
where the fermi-sea vacuum $| \mu \rangle$  ($\mu$ is a chemical potential), is defined by 
\bea
\hat{b}_{s}(\omega) | \mu \rangle&=&0, \quad  \omega < \mu \nn \\
\hat{b}^\dagger_{s}(\omega) | \mu \rangle&=&0, \quad  \omega > \mu
\eea
and the continuum fermionic oscillators satisfy $\lbrace \hat{b}^\dagger_{s}(\omega), \, \hat{b}_{s'}(\omega')  \rbrace = \delta_{s, s'} \delta(\omega - \omega')$. We should mention at this point that there exist various choices of defining the vacuum. In the old works the two common vacua are the one related to the bosonic string that has only one side of the potential filled, and the 0B vacuum that has both sides of the potential filled~\cite{Takayanagi:2003sm,Douglas:2003up}. In the recent work~\cite{Balthazar:2019rnh} a new definition of the bosonic vacuum appeared in which there is no rightgoing flux from the left side of the fermi sea. These choices affect the non-perturbative physics of the model.  As we find in the next section, a fermi sea having support only on one side gives a non-perturbative WdW wavefunction that has a more natural interpretation as a partition function of a Euclidean theory on the boundary of $AdS_2$. On the other hand a fermi sea with both ``worlds" interconnected via non perturbative effects, seems to be better suited to describe after an analytic continuation, wavefunctions of geometries of a cosmological type.

So far we described the fermionic field theory in real time/energy $t \leftrightarrow \omega$. We can also pass to the Euclidean description via $t \rightarrow i x, \, \omega \rightarrow - i q$, but notice that this analytic continuation has a priori nothing to do with the bulk space/time notion of time. The natural interpretation of the matrix model time in our point of view is that of a time variable in superspace in which the universe is embedded. This means that in agreement with the discussion in the introduction, the natural interpretation of \eqref{secondfermifield}, is that of a third quantised action describing the evolution of two dimensional universes in superspace described by the coordinates $(t, \lambda )$. Notice also that this action can capture the process of topology change for the bulk geometry (since the observables that one can compute using it are known to incorporate a resummed bulk genus expansion). At the same time this action is simply quadratic in the superspace field $\psi(t, \lambda)$. This is quite interesting and unexpected since the third quantised superspace analysis in \cite{BabyUniverses}, indicated that one needs a non-linear modification of the WdW equation in order to describe topology changing processes. The reason of why such a non-linear modification is not essential for the fermionic field theory, is that the third quantised action is expressed in terms of $\lambda$, which is related via a complicated non-local transform to the metric conformal Liouville mode $\phi$. From this point of view the condition that the bulk theory is a CFT, fixes the possible geometries of superspace in which the bulk geometry can be embedded into.

Working now in Euclidean minisuperspace signature, we first define Matrix operators $\frac{1}{N}\Tr f(\hat M(x))$ and their fourier transform
\be
\hat{\mathcal{O}}(q)= \int d x e^{i q x} \frac{1}{N} \Tr f(\hat M(x)) \, .
\ee
A simple example is the macroscopic loop operator ($L$ is a discrete lattice variable)
\be\label{fixedsizeloopoperator}
\hat W(L, x) \, = \, \frac{1}{N} \, \Tr e^{L \hat M(x)} \, .
\ee
There is another description of the form~\cite{Martinec:2004td}
\be
\hat W(\mathrm{z}, x)\, = \, - \frac{1}{N} \, \Tr \log [\mathrm{z} - \hat M(x)] \, = \,  \frac{1}{N} \, \sum_{l=1}^\infty  \, \frac{1}{l} \, \Tr  \left[ \hat M(x)/\mathrm{z} \right]^l - \log \mathrm{z} \, . 
\ee
This description keeps fixed a chemical potential $\mathrm{z} = \mu_B$ dual to the loop size. This is the boundary cosmological constant in the Liouville side as described in section~\ref{Liouvilletheory}.

In terms of  the fermions, the most basic operator is the density operator
\be\label{fermiondensityoperator} 
\hat \rho(x, \lambda)=\hat \psi^\dagger (x, \lambda) \hat \psi (x, \lambda)\, .
\ee
In the double scaling continuum limit, we can employ the second quantised fermionic formalism to rewrite the Matrix operators in terms of the basic density operator. Since $\lambda$ is the coordinate parametrising the matrix eigenvalues, the general relation is as follows 
\be
\hat{\mathcal{O}}(q)= \int d x e^{i q x} \int d \lambda \, f(\lambda) \, \hat \rho(x, \lambda)\, .
\ee
In particular the macroscopic loop operators \eqref{fixedsizeloopoperator} with length $\ell$, have as a function $f(\lambda, \ell)= e^{-\ell \lambda}$. We can also describe local operators in terms of these, by shrinking the loop operators $\ell \rightarrow 0$ and dividing by appropriate powers of $\ell$. A technical complication is that the support of the density $\rho$, depending on the non-perturbative completion of the model, can be on either side of the quadratic maximum of the inverted oscillator potential $(-\infty, -2 \sqrt{\mu}]\cup[2 \sqrt{\mu}, \infty)$, so typically it is more convenient to consider the operators
\be\label{Loopoperatordfn}
\hat {\mathcal{W}}(z,x)= \int_{-\infty}^\infty d \lambda \, \hat \psi^\dagger (x, \lambda) e^{i z \lambda} \hat \psi(x, \lambda)	
\ee
and then Wick rotate $z= \pm i \ell$ for the various pieces of the correlator that have support in either side of the cut so that the corresponding integrals are convergent~\footnote{This procedure has been shown to give the correct results for the bosonic string theory~\cite{Moore:1991sf}. From the present point of view of the boundary dual, it will result in a positive definite spectral density and a well defined partition function.}. We will denote the expectation value of this loop operator as
\be
M_1(z,x) = \langle \hat{\psi}^\dagger e^{i z \hat{\lambda}} \hat{\psi} \rangle \, ,
\ee
and so forth for the higher point correlators $M_{n}(z_i, x_i)$. The details of the computation of such correlation functions are reviewed in appendices~\ref{Parabolic} and~\ref{fermioniccorrelators}. The results for the correlation functions of a compact boson $X$ can be obtained from those of the corresponding non-compact case, via the use of the formulae of appendix~\ref{finiteTcorrelators}. We will now directly proceed to analyse the one and two point functions of loop operators.

\section{One macroscopic loop}

\begin{figure}[tb!]
\begin{center}
\includegraphics[width=0.9\textwidth]{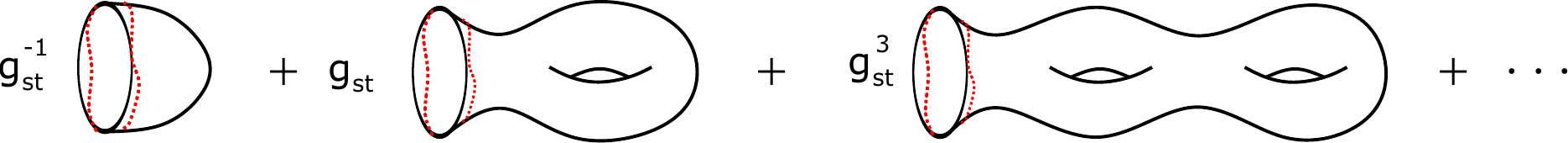}%
\caption{The perturbative expansion of the WdW wavefunction in powers of $g_{st} \sim 1/\mu$. The dashed lines indicate that we keep fixed only the overall size of the loop $\ell$.}
\label{fig:WdW}
\end{center}
\end{figure}

\subsection{The WdW equation and partition function}\label{Partitionfunction}

We first start analysing the result for one macroscopic loop. This is a one-point function from the point of view of the matrix model, but corresponds to a non-perturbative WdW wavefunction from the bulk quantum gravity path integral point of view\footnote{Here the non-perturbative choice is equivalent to having both sides of the potential filled, see appendix~\ref{fermioniccorrelators}, the one sided case is described in subsection \ref{OnesidedLL}.}, defined on a single macroscopic boundary of size $\ell$. The expression reads
\be\label{oneloopMQM}
\Psi_{WdW}(\ell, \mu) \, = \, M_1(z = i \ell,\mu) \, = \, \Re \left( i \int_0^\infty \frac{d \xi}{\xi} e^{i \xi } \frac{e^{i  \coth (\xi/2 \mu)  \frac{ z^2}{2}}}{\sinh (\xi/2 \m)} \right) \, .
\ee
We first notice that this expression does not depend on $q$, the superspace momentum dual to the matter field zero mode $x$, in contrast with all the higher point correlation functions. This means that this wavefunction will obey a more general equation compared to \eqref{MinisuperspaceWdW}, but for $q=0$. To find this equation, we can compute the $\mu$ derivative of this integral exactly in terms of Whittaker functions
\be
\frac{\partial M_1(z,\mu) }{\partial \mu} = - \Re \left((- i z^2)^{-\half} \Gamma \left(\half - i \mu \right)  W_{i \mu, 0}(i z^2)  \right) \, .
\ee
This has an interpretation as the one point function of the area (or cosmological) operator $ \langle \int e^{2 \phi} \rangle$. The first thing to observe is that the analytic continuation $z = i \ell$ merely affects the result by an overall phase. The expressions are also invariant under the $\mathbb{Z}_2$ reversal $z \leftrightarrow - z$. More importantly, the Whittaker functions that appear in these expressions obey the WdW equation 
\be\label{ExactwavefunctionsWdW}
\left( - \left[\ell \frac{\partial}{\partial \ell}\right]^2 + 4 \mu  \ell^2 + 4 \eta^2 - \ell^4 \right) \frac{W_{i \mu, \eta}(i \ell^2)}{\ell} = 0 \, ,
\ee
which is a generalisation of the minisuperspace Liouville result \eqref{MinisuperspaceWdW}. The last term in particular was argued~\cite{Moore:1991sf} to come from wormhole-like effects that involve the square of the cosmological constant operator $\sim (\int e^{2  \phi})^2$. This is also consistent with the fact that the genus zero WdW equation \eqref{MinisuperspaceWdW}, is missing precicely this term, while the exact result resummed the various topologies as shown in fig.~\ref{fig:WdW}. To get the genus zero result from the exact \eqref{oneloopMQM}, one should perform a $1/\mu$ expansion of the integrand and keep the first term to obtain the genus zero or disk partition function
\be\label{Genuszerowavefunction}
\Psi^{(0)}_{WdW}(\ell, \mu) \, = \, \Re \left( 2 i   \int_0^\infty \frac{d \xi}{\xi^2} e^{i \mu \xi } e^{ - i \frac{ \ell^2 }{\xi}} \right) \, = \,  {2 \sqrt{\mu}  \over \ell} K_{1}(2 \sqrt{\mu} \ell) \, .
\ee

\begin{figure}[!tb]
\begin{center}
\includegraphics[width=0.49\textwidth]{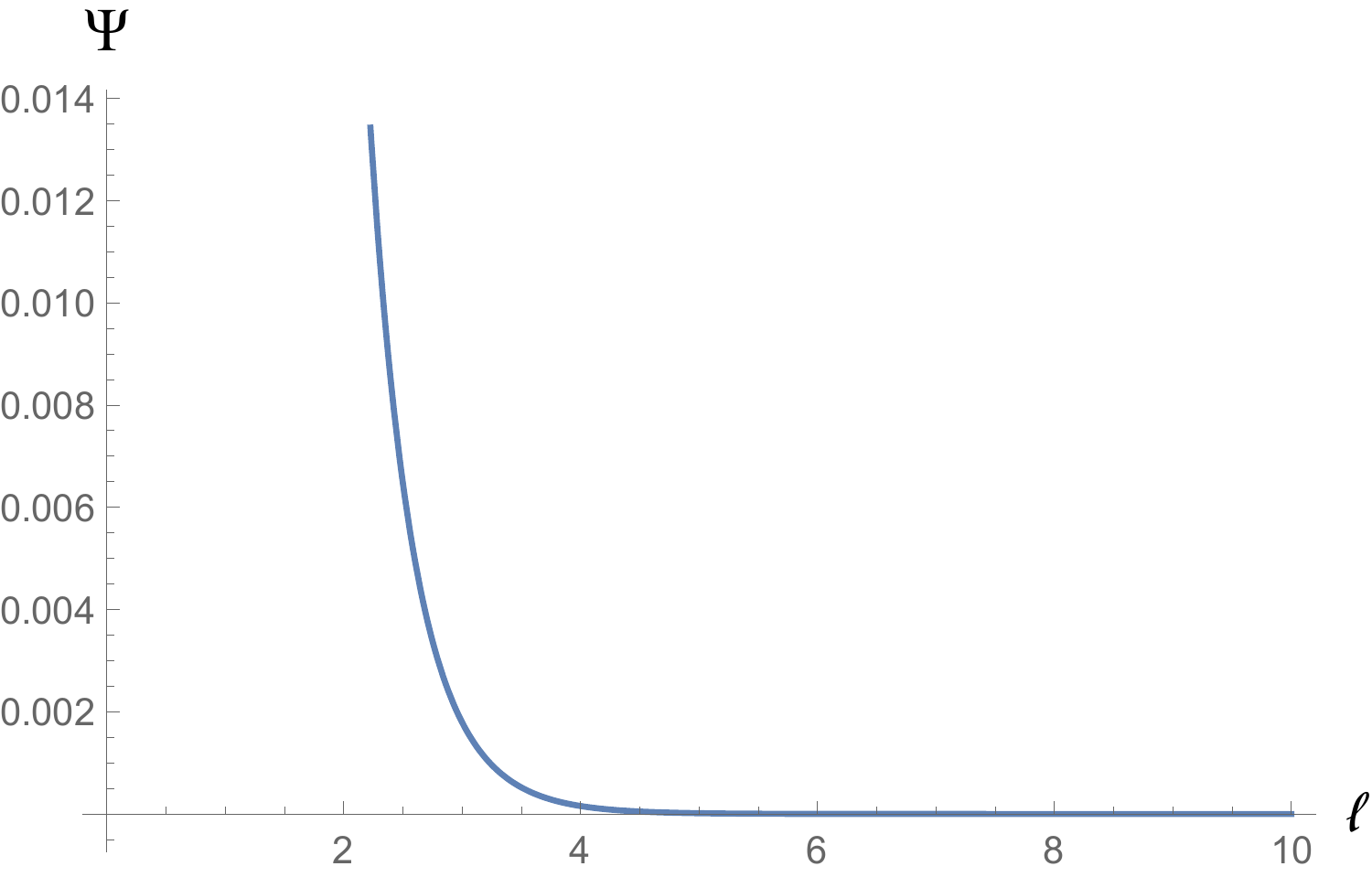}%
\hspace{2mm}\includegraphics[width=0.49\textwidth]{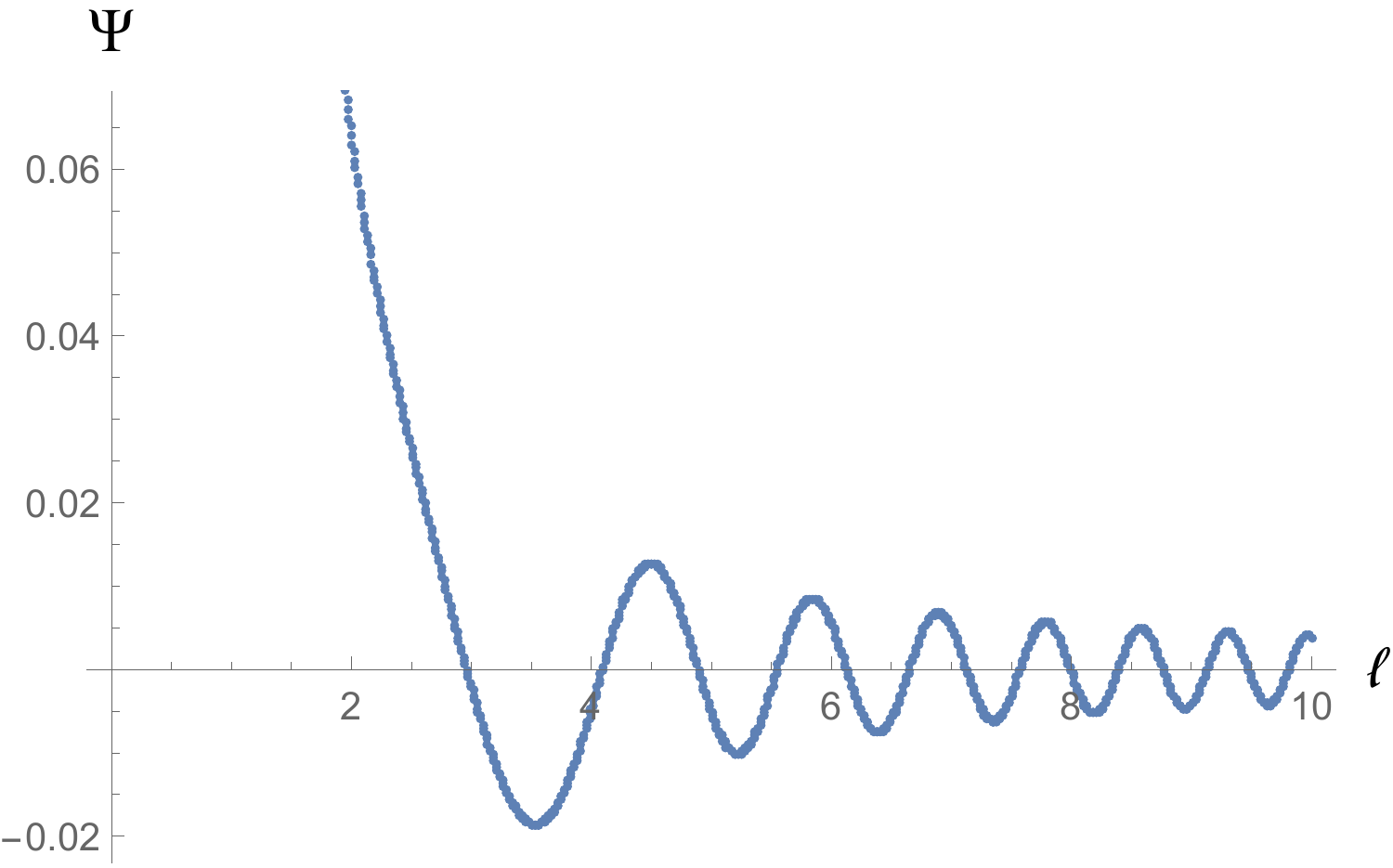}
\end{center}
\caption{Left: The genus-zero WdW wavefunction as a function of the size $\ell$. Right: The non-perturbative wavefunction (computed numerically) exhibiting a slowly decaying envelope with oscillations.}
\label{fig:WdWwavefunction}
\end{figure}

While the genus zero wavefunction has an exponential decaying behaviour for large $\ell$\footnote{An indication of the importance of non-perturbative effects, is the intriguing fact that the exponential decay holds in any finite order truncation of the genus expansion of the exact result.}, the exact wavefunction has an initially fast decaying behaviour that transitions to a slowly decaying envelope with superimposed oscillations, see fig.~\ref{fig:WdWwavefunction}. This can be also studied analytically by employing a steepest descent approximation (see appendix~\ref{SteepestDescent}) to the integral \eqref{oneloopMQM}. In the first quadrant of the complex $\xi$ plane the integrand vanishes exponentially fast at infinity and hence we can rotate the contour at will in the region $\Im \, \xi > 0$. The steepest descent contour goes actually along the direction of the positive imaginary axis, so we should be careful treating any poles or saddle points. The poles at $\xi = i 2 \pi n$ are combined with a very fast oscillatory behaviour from the exponent due to the factor $\coth \xi/2$\footnote{One can actually show that there is no contribution to the integral from these points by considering small semi-circles around them.}. There are two types of leading contributions to the integral as $\ell \rightarrow \infty$. The first is from the region near $\xi = 0$. The integral around this region is approximated by the genus zero result \eqref{Genuszerowavefunction} and decays exponentially. In order to study possible saddle points, it is best to exponentiate the denominator and find the saddle points of the function $S[u] = \log( \sin u/2 ) + \half \ell^2 \cot u/2 $ with $\xi = i u$\footnote{Taking all the terms into the action $S[u]$ and finding the exact saddles produces only small $\mu / \ell^2$ corrections to the leading result.}. There is a number of saddle points at $u^* =  \,  Arc \left[\sin (\ell^2) \right] \, + \,  2  n \pi$ or $u^* =  \, - Arc \left[\sin (\ell^2) \right] \, + \,  (2  n + 1) \pi$. The integral along the steepest descent axis is now expressed as
\be
I(\ell) = \Re \int_0^\infty \frac{du}{u} e^{- u \mu} e^{- S[u]}
\ee
To get the leading contribution we just need to make sure that our contour passes through the first saddle point $u^*(\ell)$. From that point we can choose any path we like in the first quadrant, since this only affects subleading contributions. An obvious issue is that this is a movable saddle point since it depends on $\ell$, which we wish to send to $\infty$. To remedy this according to the discussion of appendix \ref{SteepestDescent}, we define $u = u^* u'$ and perform the saddle point integral to find the leading contribution
\be
I(\ell) \sim \Re  \,\frac{2 \sqrt{2 \pi} \, e^{- \mu u^* \, - \, \half \ell^2 \cot (u^*/2)}}{(1- \ell^2)^{1/4} (u^*)^2} 
\ee
This expression is in an excellent agreement with the numerical result plotted in fig. \ref{fig:WdWwavefunction}. The amplitude decays as $1/\sqrt{\ell} \log^2 \ell$ and the phase oscillates as $e^{i \ell - 2 i \mu \log \ell}$ asymptotically for large $\ell$.

We can also compute exactly the wavefunction with the insertion of a local operator $V_q$~\cite{Ginsparg:1993is}, the result employing the exact wavefunctions for general $\eta$ \eqref{ExactwavefunctionsWdW}
\bea\label{Exactwavefunctionoperatorinsertion}
\langle  \mathcal{W}(\ell, -q) V_q \rangle \, &=& \, \Psi^{q}_{WdW}(\ell, \mu)  \, \nn \\
&=& {2 \Gamma(- |q|) \over \ell} \Im \left[ e^{\frac{3 \pi i}{4}(1+ |q|)} \int_0^{|q|} d t \Gamma \left(\half - i \mu + t \right) W_{i \mu - t + |q|/2, \, |q|/2}(i \ell^2) \right]  \nn \\
\eea
This reduces to the genus zero result
\be\label{Genuszerowavefunctionoperatorinsertion}
\Psi^{(0), q}_{WdW}(\ell, \mu)\, = \, 2 |q| \Gamma(-|q|) \mu^{|q|/2} K_q(2 \sqrt{\mu} \ell) \, ,
\ee
in accordance with the Liouville answer \eqref{MicroscopicWdW}, for the wavefunctions describing generic microscopic states (punctures of the surface).
Multiple insertions can be computed with the use of the more general formula for correlation functions \eqref{Anyloopcorrelator} by shrinking the size of all exept one of the loops and picking the appropriate terms.

We close this subsection with some remarks on the wavefunctions and the consequences of the identification $\Psi_{WdW}(\ell) = Z_{dual}(\beta = \ell)$. The non-perturbative wavefunction is of the Hartle-Hawking type (see appendix~\ref{WdWcompendium} and especially eqn. \eqref{HHwavefunction}). It is real by construction and exhibits a decaying behaviour at small $\ell$ indicative of being in the forbidden (quantum) region of mini-superspace\footnote{Notice that the spaces we study (ex. Poincare disk) have a \emph{Euclidean} signature and \emph{negative} cosmological constant. In a cosmological setting the resulting wavefunction is analysed in chapter~\ref{DSregime}.}. The genus zero large $\ell$ decaying behaviour is indicative of describing a space that reaches the Euclidean vacuum with no excitations when it expands to infinite size. This holds at any fixed genus truncation of the non-perturbative result. On the contrary, for large $\ell$ the non-perturbative fast oscillatory behaviour is usually indicative of a semi-classical space interpretation (see again the appendix~\ref{WdWcompendium} for some discussion on that). This is physically reasonable since large geometries are indeed expected to have a semi-classical description, while small geometries are highly quantum mechanical. Quite remarkably the same behaviour also appears in the case of a cosmological setting (upon continuing $\ell = - i z$), and we will comment on this unexpected relation in section~\ref{DSregime}. This is an indication that the oscillatory non-perturbative wavefunction has a more natural interpretation in such a cosmological setup.

On the other hand, if we are to interpret the wavefunction as the finite temperature partition function of a dual system, we run into the difficulty of having an oscillatory partition function as a function of the dual temperature $\beta = \ell$ for small temperatures $T \sim 1/\ell$. This seems to be related to the well known problem of the difficulty of assigning a probabilistic interpretation to the WdW wavefunction. Related to this, as we will see in the next section~\ref{DOSofdual}, the dual density of states $\rho_{dual}(E)$ is manifestly positive definite, but the spectral weight has support both on $E>0$ and $E<0$ and is an even function of $E$. The non-perturbative effects are precisely the ones that make these ``two worlds" communicate. We can remedy this by fiat, demanding the spectral weight to have support only at positive energies $E>0$. In this case the resulting positive frequency wavefunction $\Psi^+_{WdW}(\ell)$ does admit a probabilistic interpretation. We describe the consequences of this choice for the partition function in subsection~\ref{OnesidedLL}. The infinite temperature limit $\ell \rightarrow 0$ in \eqref{oneloopMQM} is singular\footnote{The non normalisability of the Hartle-Hawking wavefunction in a similar context was observed and discussed in~\cite{Cotler:2019dcj}.}, (needs a UV-cutoff $\Lambda$ corresponding to putting the inverted oscillator in a box). Nevertheless such a limit still makes sense physically, from a third quantised point of view. The reason is that the geometries then smoothly close and reproduce a closed string theory partition function of compact manifolds~\cite{Klebanov:1991qa}.

\begin{figure}[tb!]
\begin{center}
\includegraphics[width=0.7\textwidth]{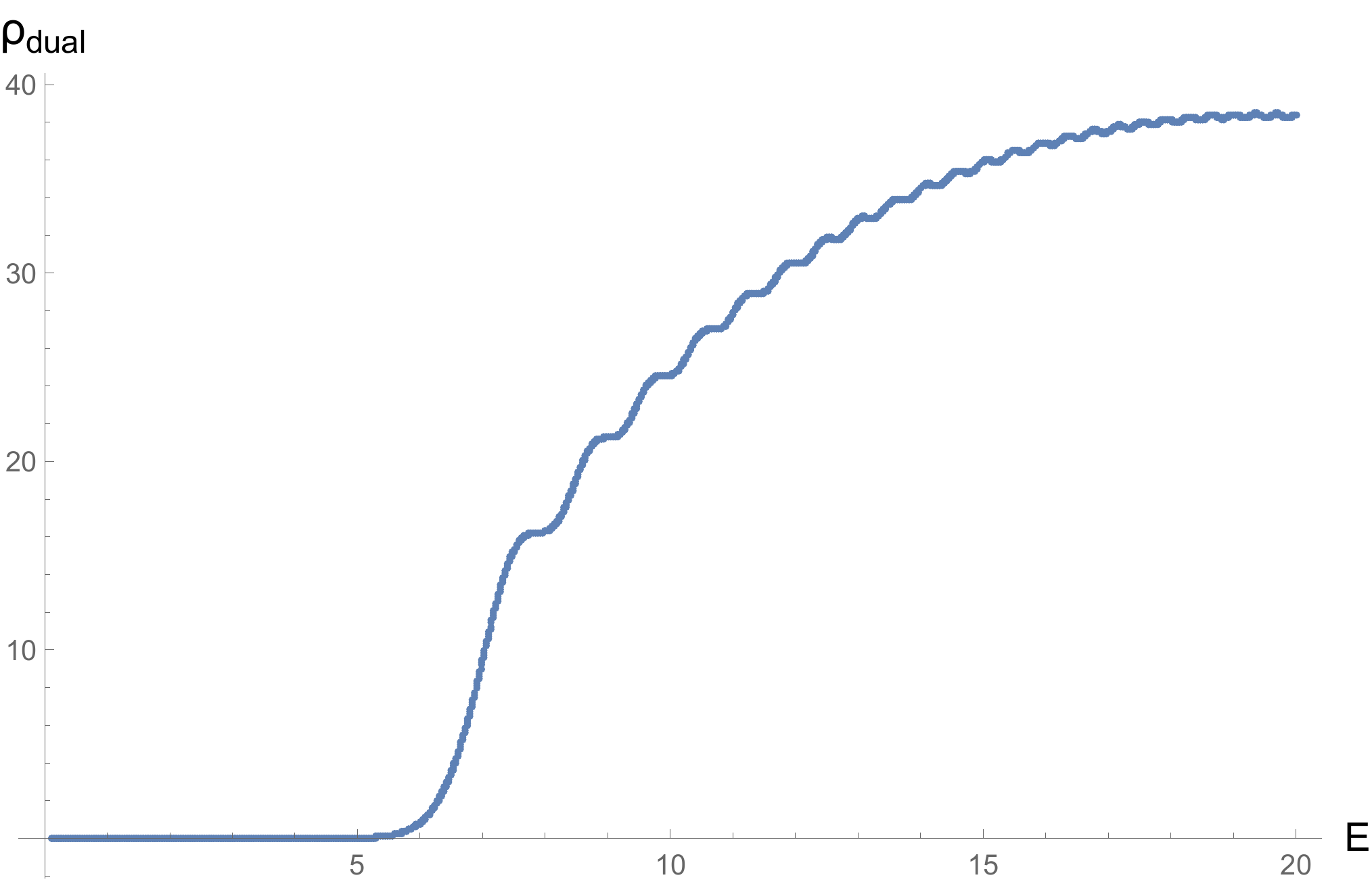}%
\caption{The density of states for $\mu=10$ as a function of the energy $E$. It exhibits an exponential growth that is then transitioning to a Dyson semicircle law with superimposed oscillations.}
\label{fig:DOS}
\end{center}
\end{figure}

\subsection{The density of states of the holographic dual}\label{DOSofdual}

We shall now compute the density of states of the dual quantum mechanical theory, assuming that the renormalised wavefunction of loop length $\ell$ corresponds to the dual thermal partition function, according to our interpretation. This also dovetails with the fact that the loop length is related to the zero-mode of the conformal mode of the metric $\ell = e^{\phi_0}$ and hence its dual variable is the trace of the boundary theory stress-tensor (energy in the case of a purely quantum mechanical system).  Since the length $\ell= \beta$ of the loop is related to the inverse temperature of the dual field theory, we can define the density of states through the inverse Laplace transform\footnote{Another equivalent way of defining the dual density of states is provided in appendix~\ref{DualResolvent}.}
\be
\rho_{dual}(E) = \int_{c - i \infty}^{c + i \infty} \frac{d \ell}{2 \pi i} e^{\ell E} \Psi_{WdW}(\ell) \, .
\ee
It is more clarifying instead of directly performing the inverse Laplace transform (or fourier transform in terms of $z$) in the final expression for the wavefunction \eqref{oneloopMQM}, to go back to the original definition of the Loop operator \eqref{Loopoperatordfn} and Laplace transform it. This then results into the following expression for the density of states in terms of parabolic cylinder wavefunctions (see also appendix~\ref{DualResolvent})
\be\label{dualdos}
\rho_{dual}(E) = \int d x \langle \mu | \hat{\psi}^\dagger(x, E) \hat{\psi}(x, E) | \mu  \rangle = \sum_{s = \pm} \int_{-\infty}^\infty d \omega \,  \Theta(\omega - \mu) \psi^s (\omega, E) \psi^s (\omega, E) \, .
\ee
We therefore observe the remarkable fact that the energy of the dual field theory $E$ corresponds to the fermionic field theory coordinate $\lambda$, since it is a conjugate variable to the loop length $\ell$. In addition the natural interpretation of the parameter $\mu$ from this point of view, is that of an IR mass gap to the energy spectrum (since the inverted oscillator potential provides an effective cutoff $[2 \sqrt{\mu}, \infty)$ to the allowed range of the eigenvalues). In more detail, by expanding the parabolic cylinder functions in the region $E \gg \sqrt{\mu}$, the eigenvalue distribution is found to follow a smooth Wigner-Dyson envelope on top of which  many rapid oscillations of small amplitude are superimposed
\be
\rho_{dual}(E)|_{E \gg \sqrt{\mu}} \, \simeq \,  \frac{1}{2 \pi} \sqrt{E^2 - 4 \mu^2} + \textit{Osc.} \, .
\ee
This behaviour is similar to the large-$N$ limit of a random Hermitian matrix Hamiltonian, but the small rapid oscillations is an indication that we have also incorporated some additional non-universal effects\footnote{Similar effects were also observed in~\cite{Blommaert:2019wfy}, when fixing some of the matrix eigenvalues to take definite values. Here they come from the IR-cutoff $\mu$ that depletes the spectrum.}. Moreover, the oscillations become more pronounced in the limit $\mu \rightarrow \infty$ and diminish as $\mu \rightarrow 0$ when the sum over topologies breaks down. In addition, there is a tail of eigenvalues that can penetrate the forbidden region outside the cut $[2 \mu^\half , \infty)$. In this region $E^2 \ll \mu$, there is an exponentially growing density of states
\be
\rho_{dual}(E)|_{E \ll \sqrt{\mu}} \, \simeq \, \frac{1}{2 \pi} \frac{e^{- \pi \mu + 2 \sqrt{\mu} E}}{(\pi \mu)} \, .
\ee
This is a Hagedorn growth of states, but only for a small window of energies. In other words it is important that there is a transition from an exponential to an algebraic growth, that makes the density of states Laplace transformable and the WdW wavefunction well defined. In fig.~\ref{fig:DOS} the complete behaviour of the density of states is depicted.

The exact integral \eqref{dualdos} can be further manipulated to give an integral expression ($c$ is an infinitesimal regulating parameter)
\bea
\rho_{dual}(E) = - i  \int_{- \infty}^\infty \frac{d p}{\sqrt{2 \pi}}  \frac{ e^{- i\mu p}}{p - i c}    \int_{-\infty}^\infty d \omega \,  e^{i \omega p} \sum_s \psi^s (\omega, E) \psi^s (\omega, E) \nn \\
= - i  \int_{- \infty}^\infty \frac{d p}{\sqrt{2 \pi}}  \frac{ e^{- i\mu p}}{p - i c} \frac{1}{\sqrt{4 \pi i \sinh p}} e^{ \frac{i}{2} E^2 \tanh (p/2) } \nn \\
= \Re \left( i  \int_{0}^\infty \frac{d \xi}{2 \pi}  \frac{ e^{ i\mu \xi }}{\xi} \frac{1}{\sqrt{- 2 i \sinh \xi}} e^{ - \frac{i}{2} E^2 \tanh (\xi/2) } \right)
\eea
This then matches the fourier transform of \eqref{oneloopMQM} as expected. The complete spectral weight $\rho_{dual}(E)$ is manifestly an even function of $E$ (as well as positive definite). It would be very interesting to try to give a Hilbert space interpretation for this density of states but we will not attempt that here. The only comment we can make is that since we also have the presence of negative energy states, the dual boundary theory needs to have a fermionic nature so that one can define a Fermi/Dirac sea and a consistent ground state. 

\subsubsection{Comparison with minimal models and JT gravity}

Let us now compare this result with the density of states found in the Schwarzian limit of the SYK model dual to JT gravity, as well as with the result for the minimal models. The first point to make is that the growth of states near the edge of the support of the spectrum is in fact faster than that of JT gravity, since 
\be
\rho_{J.T.}^{Sch.}(E, \gamma) = \frac{\gamma}{2 \pi^2} \sinh \left( 2 \pi \sqrt{2 \gamma E} \right) \, ,
\ee
where now $\gamma$ provides an energy scale and in the exponent one finds merely a square root growth with the energy.

In fact we can also make a further comparison, using our perturbative expansion in $1/\mu$ of the exact density of states \eqref{dualdos}, with the ones related to minimal strings. This is because the inverse Laplace transform of the genus zero result \eqref{Genuszerowavefunction} and in fact of every term in the perturbative expansion of the exact partition function \eqref{oneloopMQM} can be performed via the use of the identity ($\sqrt{\mu} >0$)
\be
\int_{c - i \infty}^{c + i \infty} \, \frac{d \ell}{2 \pi i} \, e^{\ell E} \, \frac{1}{\ell} K_\nu( 2 \sqrt{\mu} \ell) \, = \, \frac{1}{\nu} \sinh \left(\nu \cosh^{-1}(E/2\sqrt{\mu}) \right) \qquad E > 2 \sqrt{\mu} \, .
\ee
This gives a spectral curve for each genus that is very similar to the ones related to the minimal strings, discussed in~\cite{Saad:2018bqo}. Nevertheless none of them can capture the non-perturbative $\sim e^{- \mu}$ effects that give rise to the oscillations both in the exact partition function and density of states plotted in fig.~\ref{fig:DOS}. These are also effects that make the two worlds $E \lessgtr 0$ communicate through eigenvalue tunneling processes in the matrix quantum mechanics model. Similar non-perturbative effects were also discussed in the context of JT gravity and SYK~\cite{Saad:2019lba}. In that case it has been argued that they are \emph{doubly non-perturbative} in $\exp ( c \, e^{N_{SYK}})$, while in the present example this could arise only if $\mu$ was a parameter having a more microscopic description (as happens in the model of~\cite{Betzios:2016yaq}\footnote{In that case $\mu \sim R^2_{BH}/G_N$ is related to a \emph{four-dimensional} black hole entropy.}). We will come back to a discussion of a possible interpretation of such a double layered asymptotic expansion from the point of view of the bulk theory in the conclusions section~\ref{Conclusions}.

\subsubsection{The one sided Laplace transform}\label{OnesidedLL}

As we mentioned previously, we can define the dual density of states to have support only for $E>0$. This is equivalent to demanding that no eigenvalues can penetrate to the other side of the potential. In such a case the dual partition function is given by the Laplace transform of \eqref{dualdos}, with support at $E>0$ so that
\be 
Z_{dual}^{(+)} (\ell) =  - \Re \left(  \frac{i}{2}  \int_{0}^\infty d \xi  \frac{ e^{ i\mu \xi }}{\xi} \frac{ e^{ -\frac{i}{2} \ell^2 \coth (\xi /2) }}{ \sinh \xi /2} \,  \textit{Erfc} \left[\frac{\ell}{  \sqrt{2 i \tanh \xi /2}} \right]  \right) \, .
\ee
This is a positive real function exhibiting a decaying behaviour with no large amplitude oscillations\footnote{We expect the presence of oscillations with very small amplitude but it is hard to probe them numerically.}. Hence by restricting the spectral weight to positive energies $\rho^{(+)}_{dual}(E)$, the WdW wavefunction does acquire a well defined probabilistic interpretation. On the other hand this modification and non-perturbative definition of the model, does not seem to make sense for the analytic continuation $z = i \ell$ into the cosmological regime of section~\ref{DSregime}. The reason is that after this analytic continuation the wavefunction $\Psi^{(+)} (z = i \ell)$ remains non oscillatory and decays to zero for large $\ell$. We believe that this is an indication that microscopic models of $AdS_2$ and $dS_2$ geometries, should be inherently different at a non-perturbative level.

\section{The case of two asymptotic regions}\label{twoasregions}

We will now analyse observables in the presence of two asymptotic boundaries. One should include both disconnected and connected geometries. We focus mainly in the density two-point function and in the spectral form factor (SFF) related to its fourier transform.

\subsection{Density two-point function}\label{DOStwopoint}

We first analyse the density two-point function $\langle \mu| \hat{\rho}_{dual}(E_1) \hat{\rho}_{dual}(E_2) | \mu \rangle$.  A quite thorough discussion of similar correlation functions from the point of view of quantum gravity can be found in~\cite{Cotler:2016fpe}. This correlation function is defined in appendix~\ref{Densitycorrelationfunctions}. The disconnected part is given by the product of \eqref{dualdos} with itself.

\begin{figure}[!tb]
\begin{center}
\includegraphics[width=0.49\textwidth]{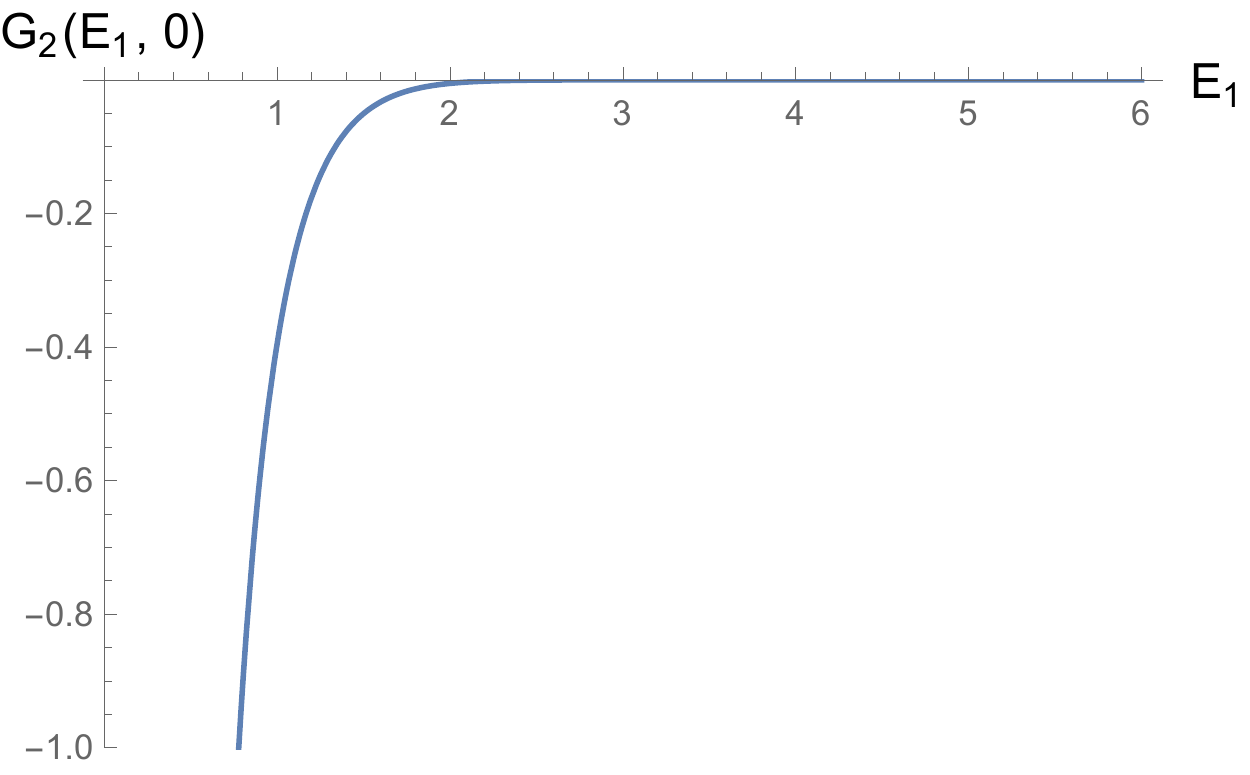}%
\hspace{2mm}\includegraphics[width=0.49\textwidth]{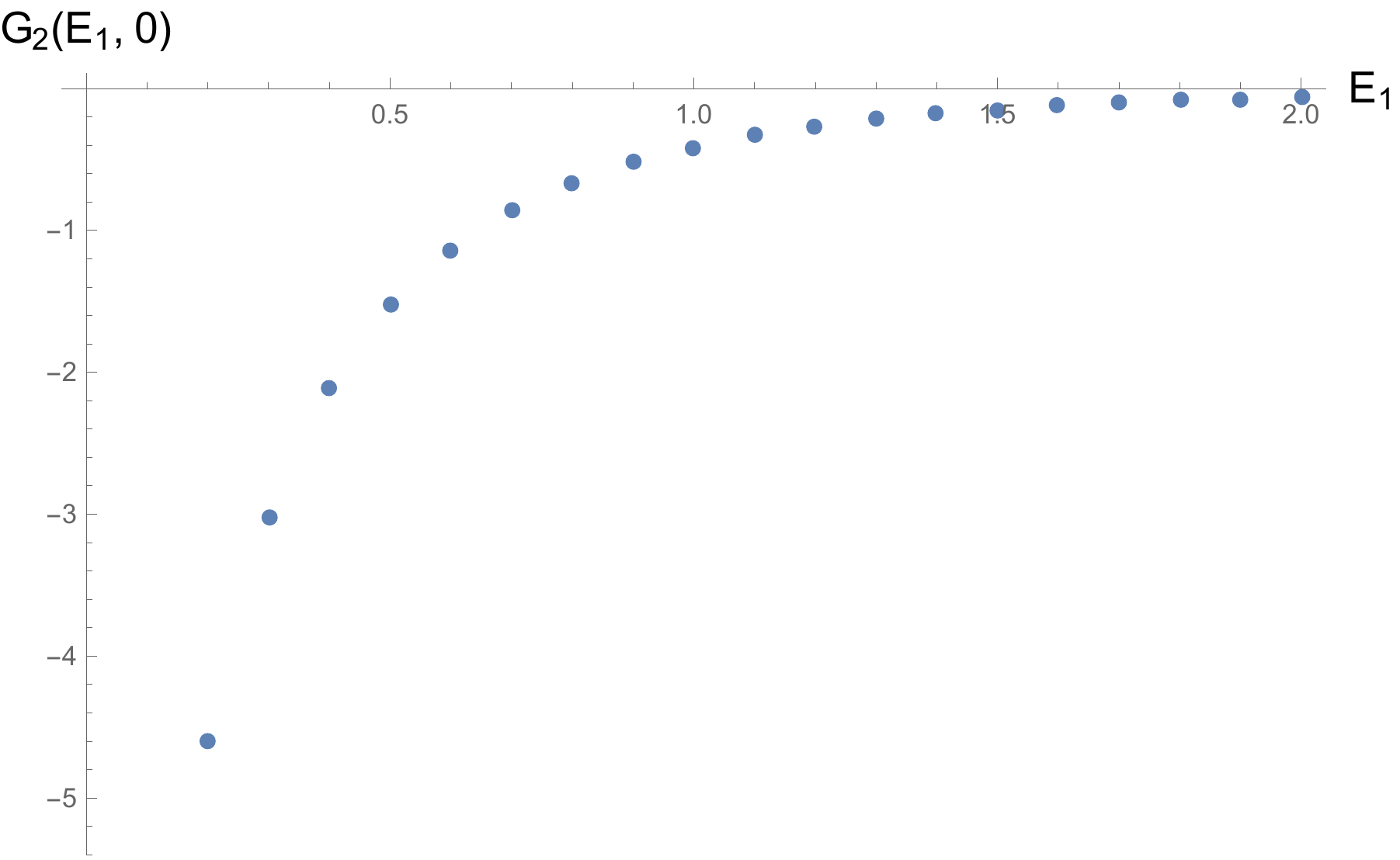}
\end{center}
\caption{Left: The genus zero connected eigenvalue correlation function for $\mu=1, E_2=0$. Right: The behaviour of the non-perturbative correlation function is quite similar at short spacings.}
\label{fig:doscorrelatorg0}
\end{figure}

\begin{figure}[!tb]
\begin{center}
\includegraphics[width=0.49\textwidth]{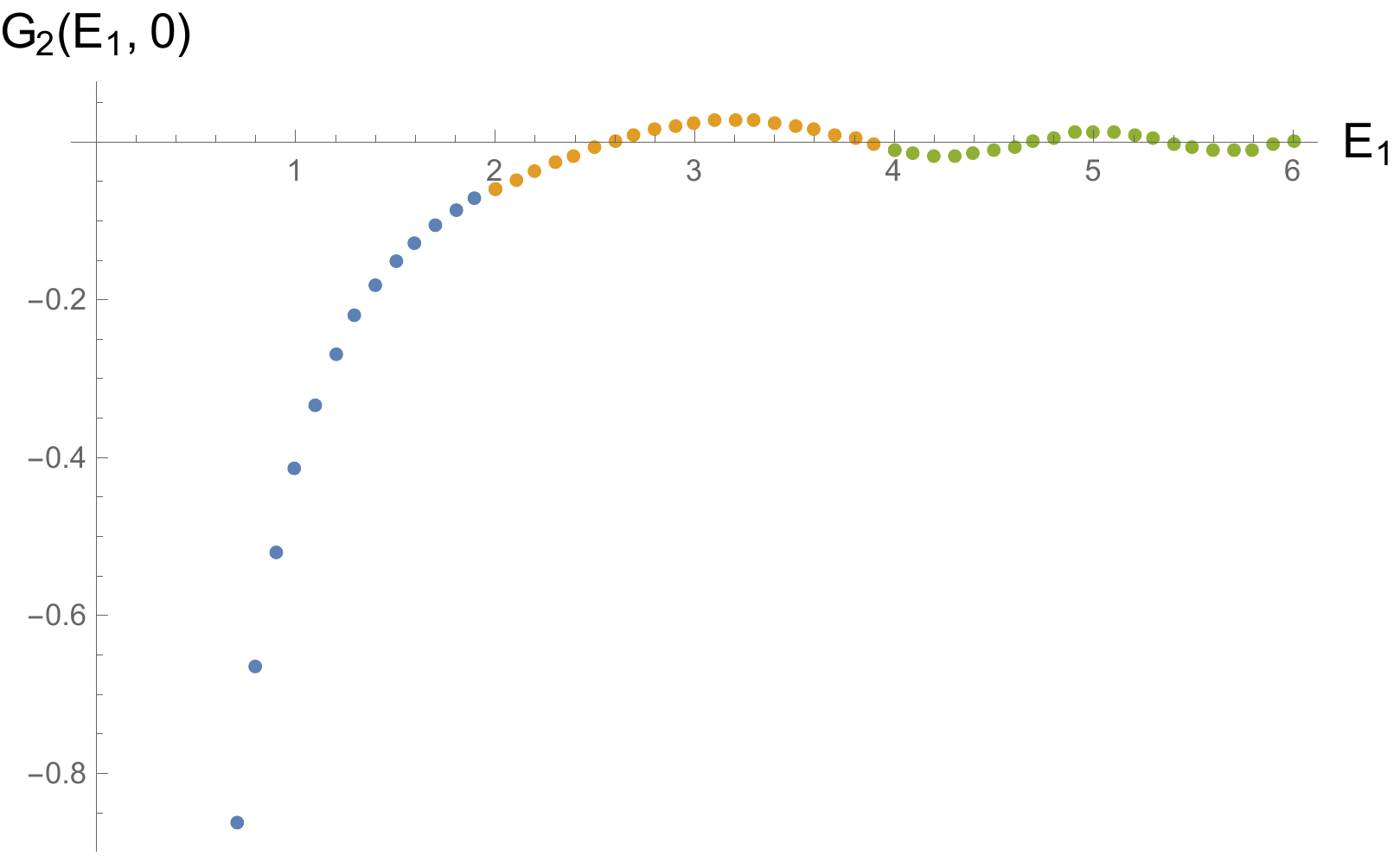}%
\hspace{2mm}\includegraphics[width=0.49\textwidth]{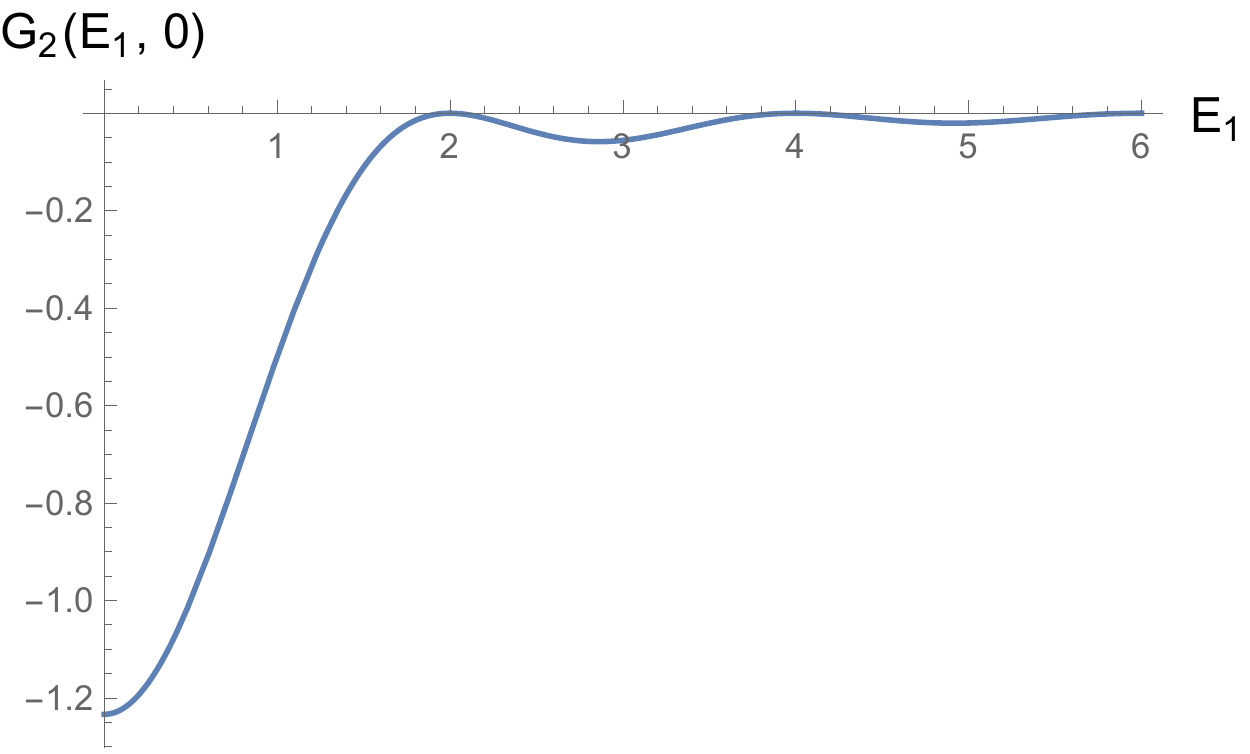}
\end{center}
\caption{The non-perturbative correlator for $\mu =1$ vs. the sine kernel. While they are qualitatively similar, there do exist differences between them.}
\label{fig:doscorrelator}
\end{figure}

The genus zero result can be computed analytically and is missing any oscillatory behaviour. A plot is given in fig.~\ref{fig:doscorrelatorg0}. The non-perturbative result is plotted in figs.~\ref{fig:doscorrelatorg0} for the short spacing behaviour $\delta E = E_1 - E_2$ of energy eigenvalues, and~\ref{fig:doscorrelator} for larger spacings. It has the behaviour of the \emph{sine-kernel} indicative of short range eigenvalue repulsion and chaotic random matrix statistics for the eigenvalues. For an easy comparison we have also plotted the sine kernel on the right hand side of fig.~\ref{fig:doscorrelator}. We observe a qualitative similarity, but there do exist differences that will become more pronounced in the SFF leading to a slightly erratic oscillatory behaviour.

These results combined with the ones of section~\ref{DOSofdual}, indicate that if we would like to endow the boundary dual with a Hilbert space and a Hamiltonian, its spectrum is expected to be quite complicated and resemble those found in quantum chaotic systems.

\subsection{Spectral form factor due to disconnected geometries}

Another interesting quantity we can compute is the spectral form factor (SFF), first due to disconnected bulk geometries. This corresponds to the expression
\be
SFF_{disc}(\beta, t) \, = \, |Z_{dual}(\beta +i t)|^2  \, = \, \Psi_{WdW}(\ell = \beta + i t)  \Psi_{WdW}(\ell = \beta - i t) \, .
\ee
In the genus zero case we can use the analytic expression \eqref{Genuszerowavefunction} to compute it. This results in a power law $\sim 1/t^3$ decaying behaviour at late times, with a finite value at $t=0$, due to the non-zero temperature $\beta$.

\begin{figure}[!tb]
\begin{center}
\includegraphics[width=0.49\textwidth]{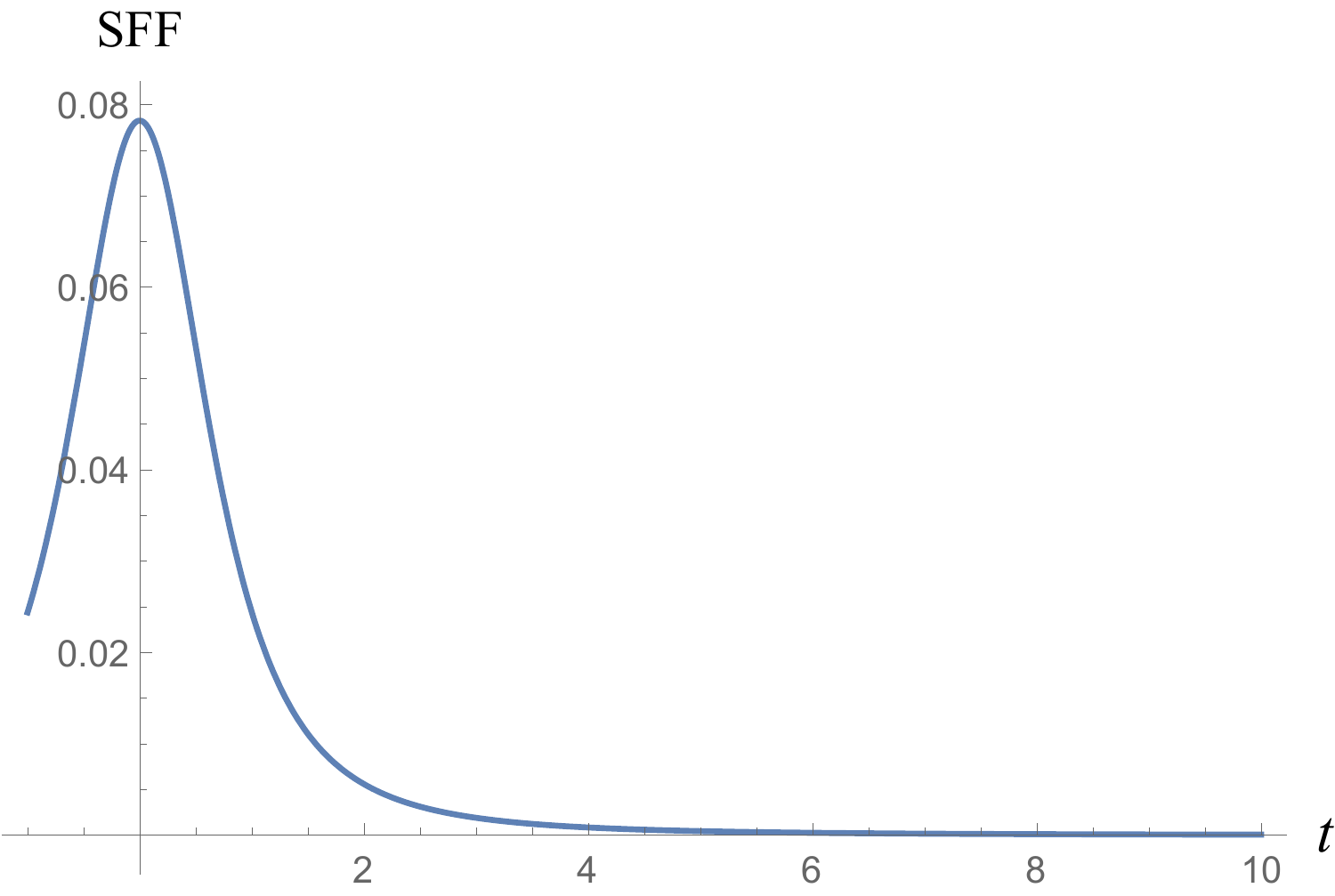}%
\hspace{2mm}\includegraphics[width=0.49\textwidth]{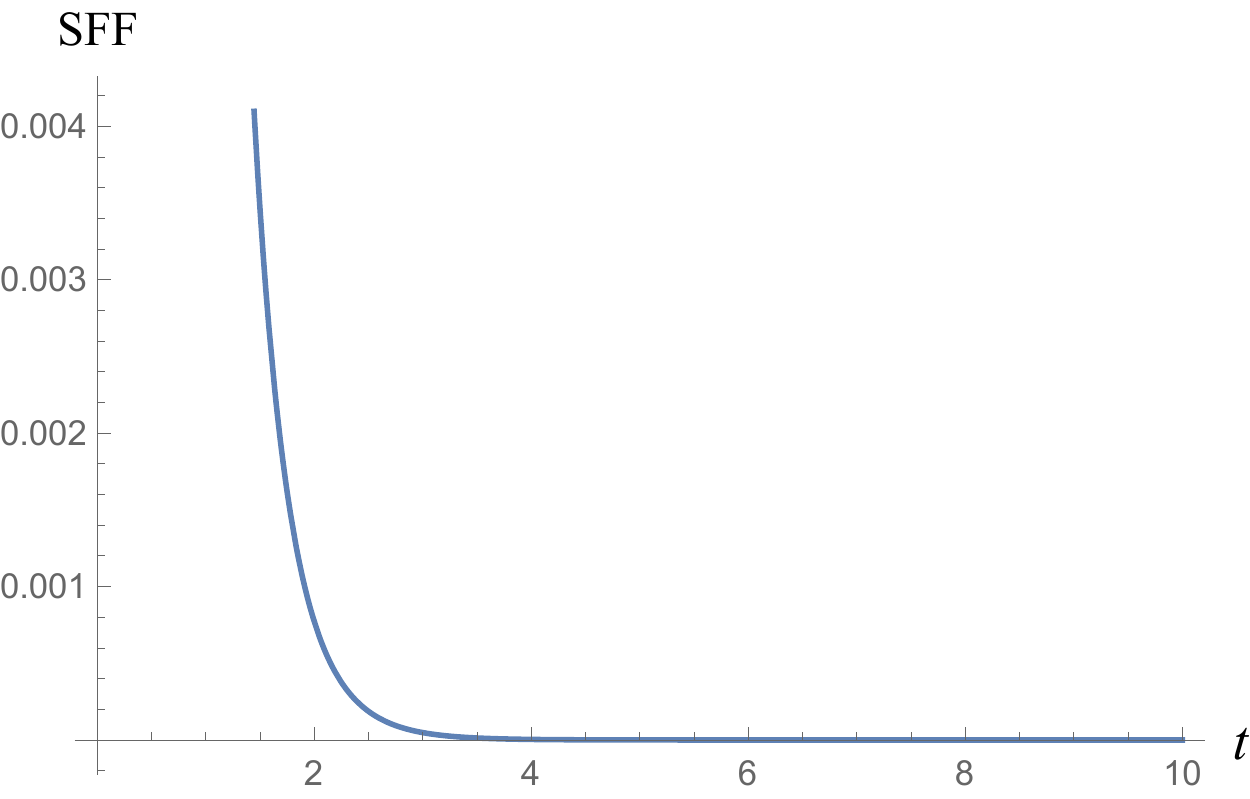}
\end{center}
\caption{Left: The SFF from the topology of two disconnected disks for $\mu=1, \, \beta=1$ as a function of the time $t$. Right: The non-perturbative disconnected SFF exhibiting again a decay at late times.}
\label{fig:SFFdisconnected}
\end{figure}

The non perturbative result can be computed at three different limits using a steepest descent analysis. The first is for $\mu  \gg t, \beta$ that is equivalent to the genus zero result. The early time limit is for $\beta \gg t, \mu$ that gives a result that is again very similar to the genus zero answer. The last is the late time result for $t \gg \beta, \mu$ which is plotted on the right hand side of fig.~\ref{fig:SFFdisconnected}. The decay in this case, has the scaling behaviour $\sim 1/ t \log^4 t$ as $t \rightarrow \infty$.

\subsection{Euclidean wormholes and the loop correlator}

We now pass to the case of the connected loop-loop correlator $\langle \hat {\mathcal{W}}(\ell_1,q)\hat {\mathcal{W}}(\ell_1,q) \rangle \, = \, M_2(\ell_1, q , \ell_2, -q)$. From the bulk quantum gravity point of view according to  \cite{Saad:2018bqo,Saad:2019lba}, this observable corresponds to a correlator of partition functions that does not factorise. For the SYK model this happens due to the intrinsic disorder averaging procedure when computing this observable. In non-disordered theories with complicated spectra it was argued~\cite{Penington:2019kki} that this could arise from Berry's diagonal approximation \cite{Berry} that effectively correlates the two partition functions, even though the exact result does factorise\footnote{The geometric avatar of Berry's interpretation, is that connected wormhole saddles capture only the diagonal approximation to the full path integral.}. On the other hand in~\cite{Betzios:2019rds} it was proposed that such multi-boundary geometries could in fact correspond to a single partition function of a system of coupled QFT's. A previous work by \cite{Maldacena:2004rf} also considered such geometries and discussed various possibilities for their possible interpretation. Since there is no ultimate resolution given to this question yet, it is of great importance to analyse such correlators in the present simple example where we can compute them (in principle) at the full non-perturbative level.

In particular for two loops we find the following expression for the derivative of the correlator $ \partial M_2(z_1, q , z_2, -q) /  \partial \mu$~\cite{Moore:1991sf}
\bea\label{loopcorrelatorzeta}
\Im \int_0^\infty \frac{d \xi}{\sinh (\xi/2)} e^{i \mu \xi + \half i (z_1^2+z_2^2) \coth (\xi/2)} \int_0^\infty d s e^{- |q|s} \left(e^{i z_1 z_2 \frac{\cosh(s-\xi/2)}{\sinh (\xi/2)}} - e^{i z_1 z_2 \frac{\cosh(s+\xi/2)}{\sinh (\xi/2)}}  \right) \nn \\
\eea
The prescription for analytic continuation is $z_i \rightarrow i \ell_i$ together with $\Im \rightarrow  i/2$. One can also express the second integral over $s$ as an infinite sum of Bessel functions giving\footnote{This expression has a smooth limit as $q \rightarrow n \in \mathbb{Z}$.}
\be
I(\xi, z_1,z_2;q) = 2 \pi e^{- i \pi |q|/2} \frac{\sinh(|q|\xi/2)}{\sin \pi |q|} J_{|q|}(x) + \sum_{n=1}^\infty \frac{4 i^n n}{n^2 - q^2} J_{n}(x) \sinh( n \xi/2) \, ,
\ee
with $x = z_1 z_2/\sinh(\xi/2)$. This expression can be used to extract the genus zero-result. In addition the integral \eqref{loopcorrelatorzeta} does have a nice behaviour for large $\xi$, and all the corresponding integrands vanish for large $\xi$ exponentially. A similar property holds for large-$s$ for each term of the $s$-integral at the corresponding quadrant of the complex-$s$ plane. 
One can also pass to the position basis $q \leftrightarrow \Delta x$ via the replacement 
\be
e^{- |q| s} \leftrightarrow \frac{s}{\pi (s^2 + (\Delta x)^2)}
\ee
These two bases reflect the two different choices for the matter field $X$, either Dirichlet $x=$fixed, or Neumann $q=$ fixed at the boundary and hence to the two basic types of correlation functions.

At genus zero the expression for the correlator simplifies drastically. In particular it can be written in the following equivalent forms
\bea\label{g0loopcorrelator}
M(\ell_1, q , \ell_2, -q) &=& \int_{-\infty}^\infty d p \, \frac{1}{q^2 + p^2}  \, \frac{p}{\sinh  (\pi p)} \, \Psi^{(macro)}_p(\ell_1) \, \Psi^{(macro)}_p(\ell_2) \, \nn \\
&=& \frac{\pi q}{\sin \pi q} I_q(2 \sqrt{\mu}\ell_1) K_q(2 \sqrt{\mu}\ell_2) \, + \,  \sum_{r=1}^\infty \frac{2 (-1)^r r^2}{r^2 - q^2} I_r(2 \sqrt{\mu}\ell_1) K_r(2 \sqrt{\mu}\ell_2) \nn \\
&=& 4 q \sum_{r=0}^\infty \frac{(-1)^r}{r!} \Gamma(- q - r) \left(\frac{\mu \ell_1 \ell_2}{\sqrt{\mu(\ell_1^2 + \ell_2^2)}} \right)^{q+ 2 r} K_{q+ 2 r}\left(2 \sqrt{2 \mu (\ell_1^2 + \ell_2^2)}\right) \nn \\
\eea
The expressions above elucidate different aspects of this correlation function. The first expression has an interpretation in terms of a propagation of states between macroscopic boundary wavefunctions $\Psi^{macro}_p(\ell)$ \eqref{MacroscopicWdW}. The result can also be expanded in an infinite sum of microscopic states as the second line indicates. The final expression is a superposition of single wavefunctions corresponding to singular geometries (microscopic states). It also shows that there might be an interpretation for which the complete result corresponds to a single partition function of a coupled system. It would be interesting to see whether the exact result \eqref{loopcorrelatorzeta} can be manipulated and written in a similar form, for example using the exact wavefunctions \eqref{ExactwavefunctionsWdW} or  \eqref{Exactwavefunctionoperatorinsertion}. 
We now turn to the study of the spectral form factor arising from such connected geometries.

\begin{figure}[!tb]
\begin{center}
\includegraphics[width=0.49\textwidth]{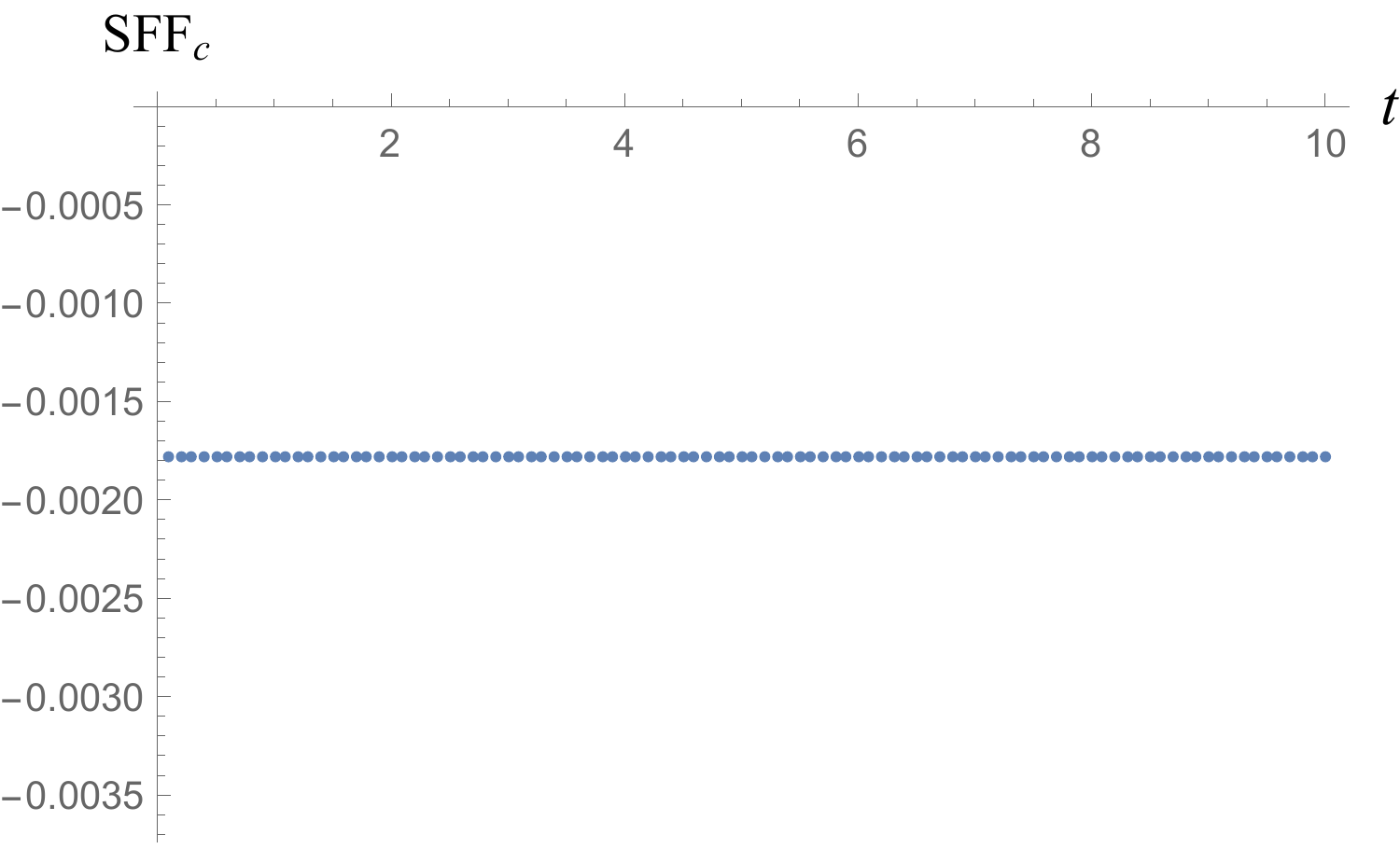}%
\hspace{2mm}\includegraphics[width=0.49\textwidth]{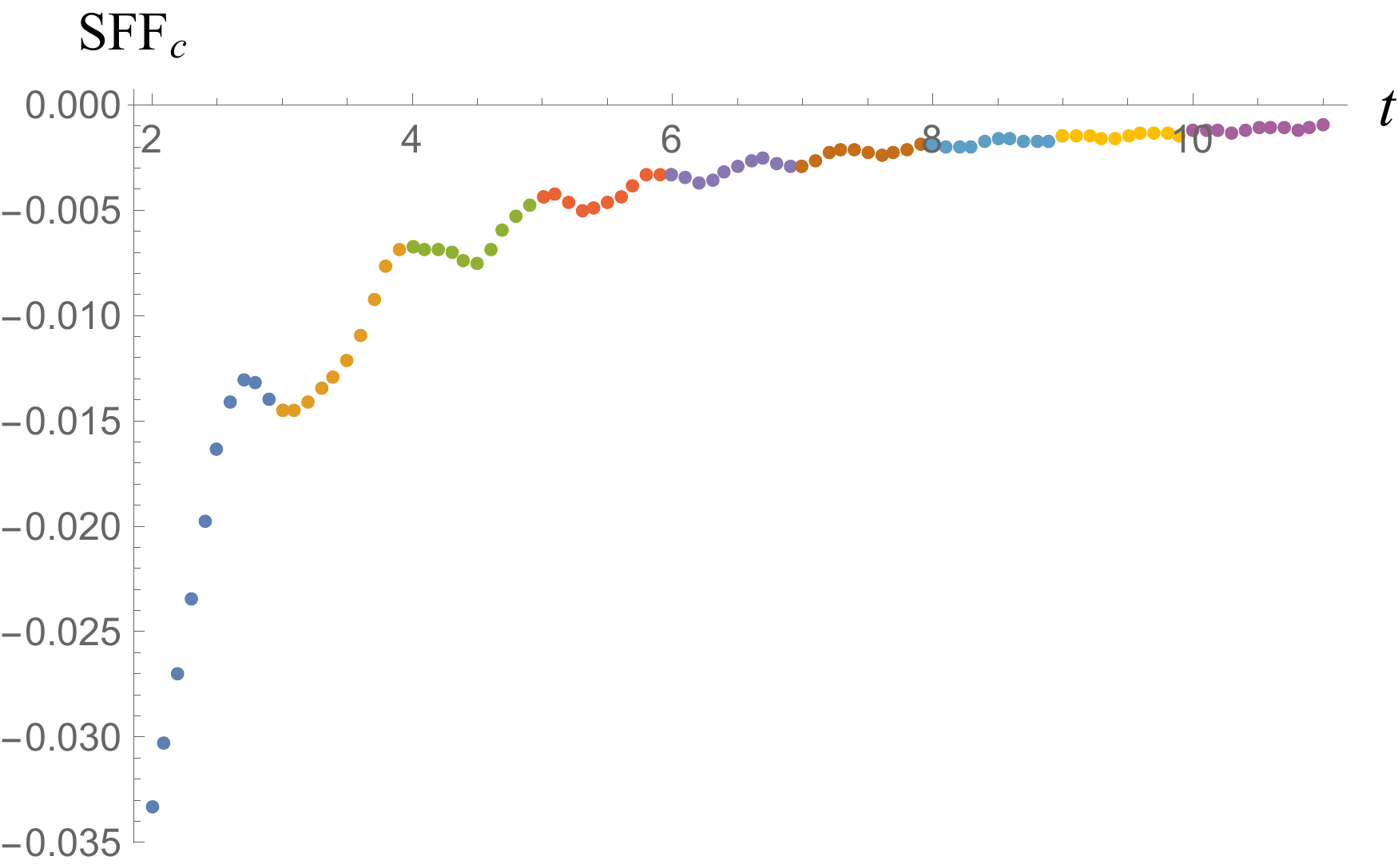}
\end{center}
\caption{Left: The connected SFF for $\mu=1, \, \beta=1$ as a function of the time $t$ from the connected wormhole geometry of cylindrical topology. Right: The non-perturbative connected SFF exhibiting a ramp plateau behaviour with persistent oscillations. At late times (around $t \sim 10$) it approaches approximately the constant value shown in the left figure. Beyond that point the numerical algorithm converges very slowly if we wish to keep the relative error under control.}
\label{fig:SFFconnected}
\end{figure}

\begin{figure}[!tb]
\begin{center}
\includegraphics[width=0.49\textwidth]{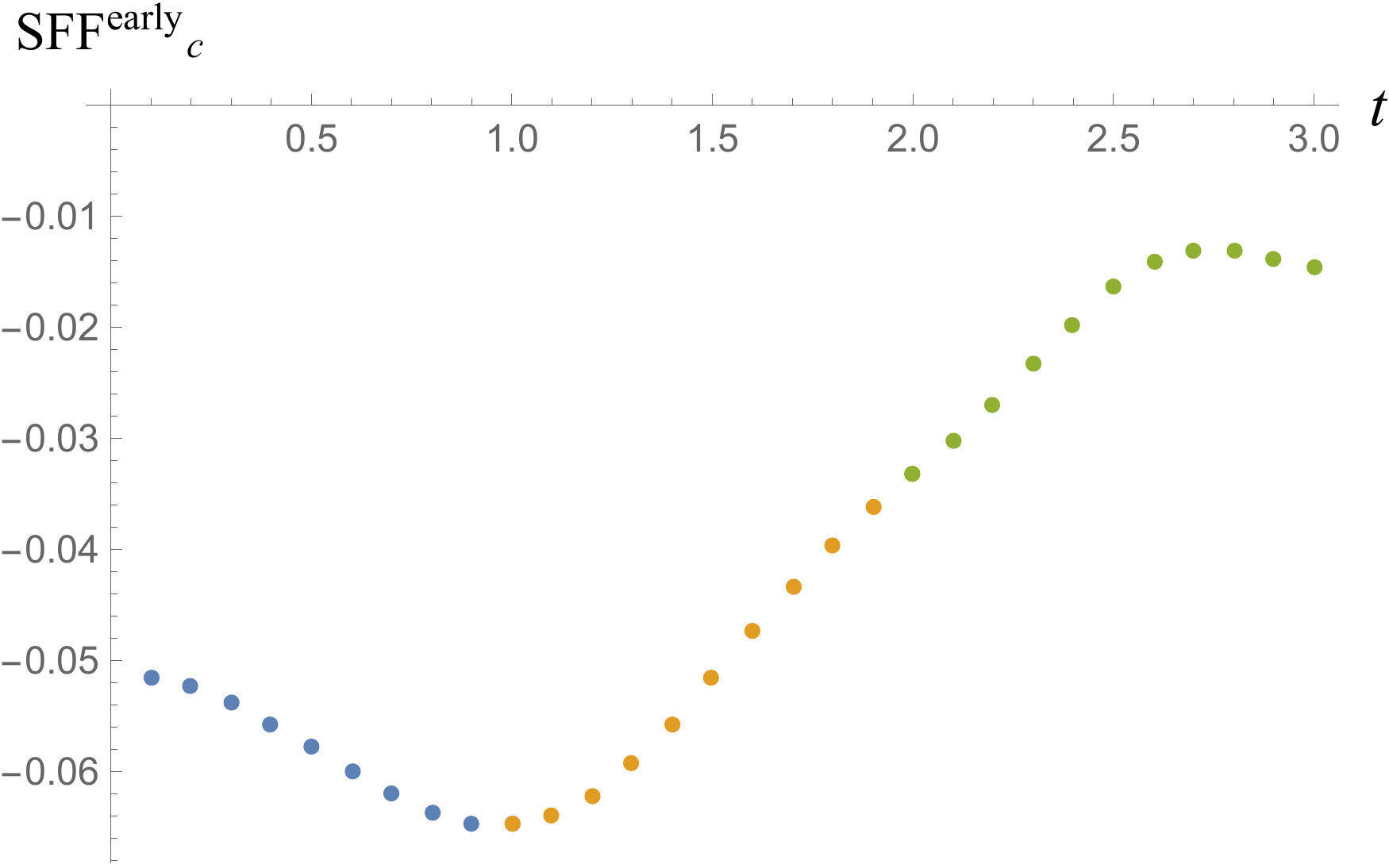}%
\hspace{2mm}\includegraphics[width=0.49\textwidth]{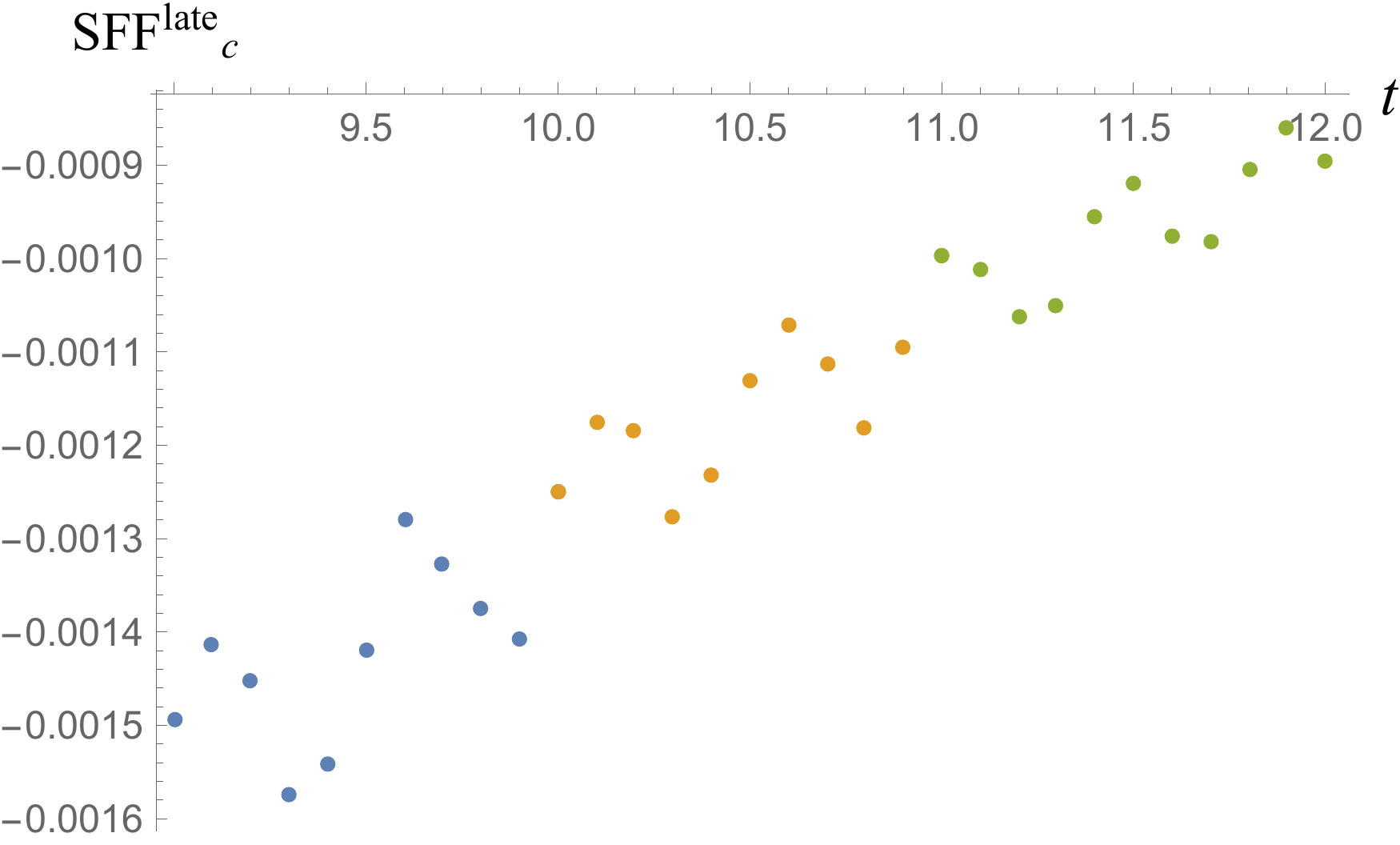}
\end{center}
\caption{Left: The early time behaviour of the exact connected SFF for $\mu=1, \, \beta=1$. Right: The late time behaviour for which the fluctuations become $O(|SFF_c|)$.}
\label{fig:SFFconnectedearlylate}
\end{figure}

\subsection{Spectral form factor due to connected geometries}\label{SFFconnected}

In this subsection we analyse the part of the SFF corresponding to a sum over connected bulk geometries. According to the discussion in appendix~\ref{Densitycorrelationfunctions}, for the unrefined SFF it is enough to study the limit $q \rightarrow 0$ of the more general expression for the momentum dependent loop correlator, eqn. \eqref{loopcorrelatorzeta}. Before doing so, we first define the parameters of the spectral form factor through $z_{1,2} = i (\beta \pm i t) $
\be
z_1^2 + z_2^2 = - 2(\beta^2 - t^2) \, , \quad z_1 z_2 = - (\beta^2 + t^2) \, . 
\ee
We can then distinguish the three basic timescales: $t \gg \beta$, $t \approx \beta$ and $\beta \gg t$. We will denote them as \emph{early}, \emph{median} and \emph{late} time-scale respectively. The spectral form factor can then be expressed as a double integral using \eqref{loopcorrelatorzeta} as
\be\label{SFFmain}
\Im \int_0^\infty d s \, \int_0^\infty \frac{d \xi}{\xi \, \sinh (\xi/2)} e^{i \mu \xi - i (\beta^2 - t^2) \coth (\xi/2) - i (\beta^2 + t^2) \coth (\xi/2) \cosh s}  \sin  \left((\beta^2 + t^2) \sinh s   \right) \, .
\ee
The genus zero result is shown in fig.~\ref{fig:SFFconnected} and is found to be a time independent function. The graph can also be obtained by directly integrating \eqref{g0loopcorrelator} for $q=0$. One can notice that the $g=0$ SFF captures the plateau behaviour, unlike the case in \cite{Saad:2019lba} where the $g=0$ SFF captures the ramp behaviour, for $t>>\beta$. The plateau behaviour arises due to the repulsion among neighboring energy eigenvalues. On the other hand, the ramp behaviour arises due to the repulsion among eigenvalues that are far apart\footnote{This long range repulsion is know as the phenomenon of long-range rigidity \cite{Mehta,Dyson,Guhr:1997ve}. } \cite{Cotler:2016fpe}. The difference between our case and the one in \cite{Saad:2019lba} is not surprising, since in our case the genus zero part is obtained in the limit $\mu\rightarrow \infty$ where effects involving eigenvalues that are far apart are suppressed.

\begin{figure}[!tb]
\begin{center}
\includegraphics[width=0.6\textwidth]{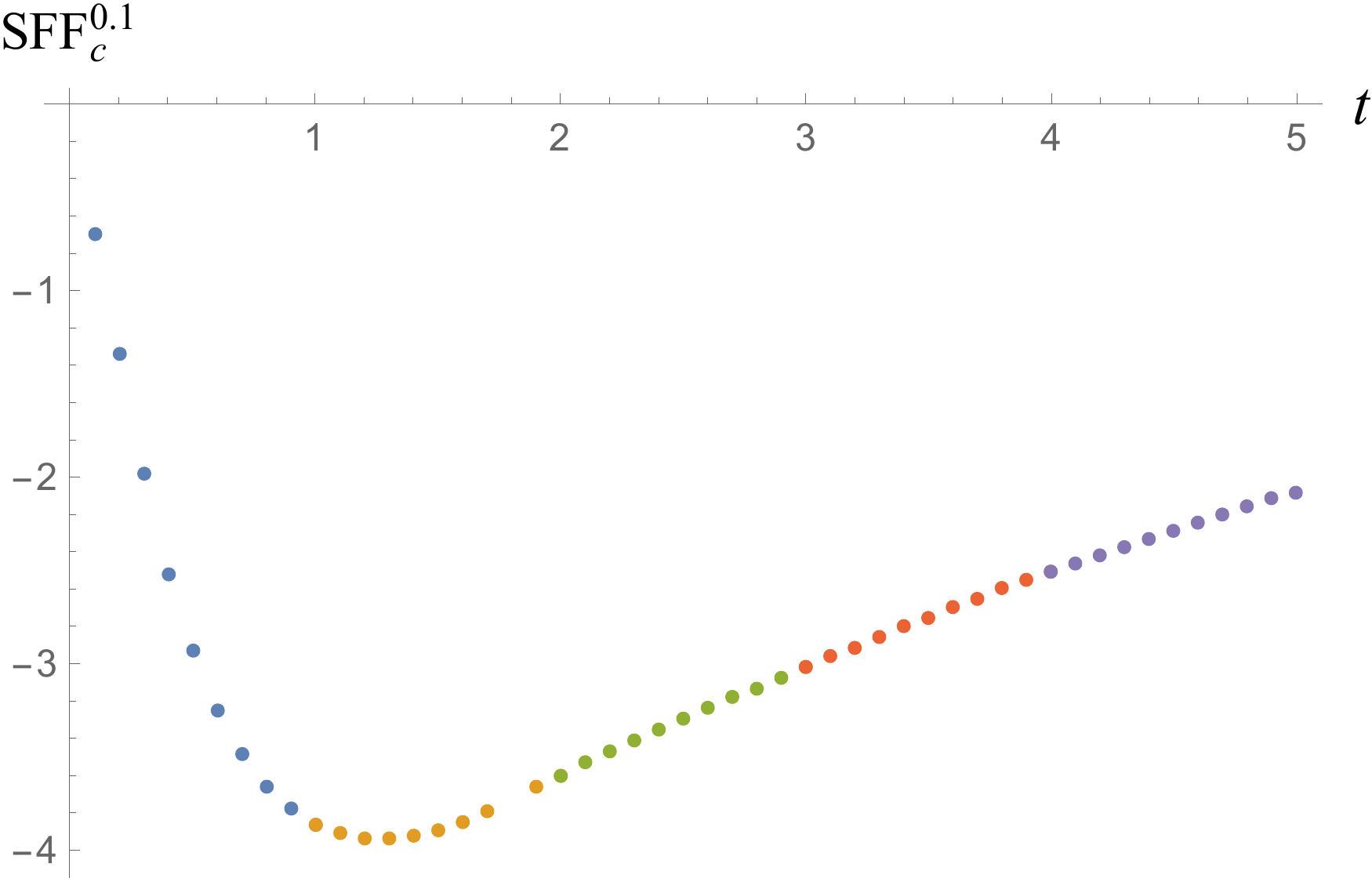}
\end{center}
\caption{Left: The non-perturbative correlator for $\mu=1, \, \beta=0.1, \, q=0.1$ as a function of the time $t$. It exhibits an initial dip transitioning into a smooth ramp behaviour.}
\label{fig:SFFconnectedq}
\end{figure}

\begin{figure}[!tb]
\begin{center}
\includegraphics[width=0.49\textwidth]{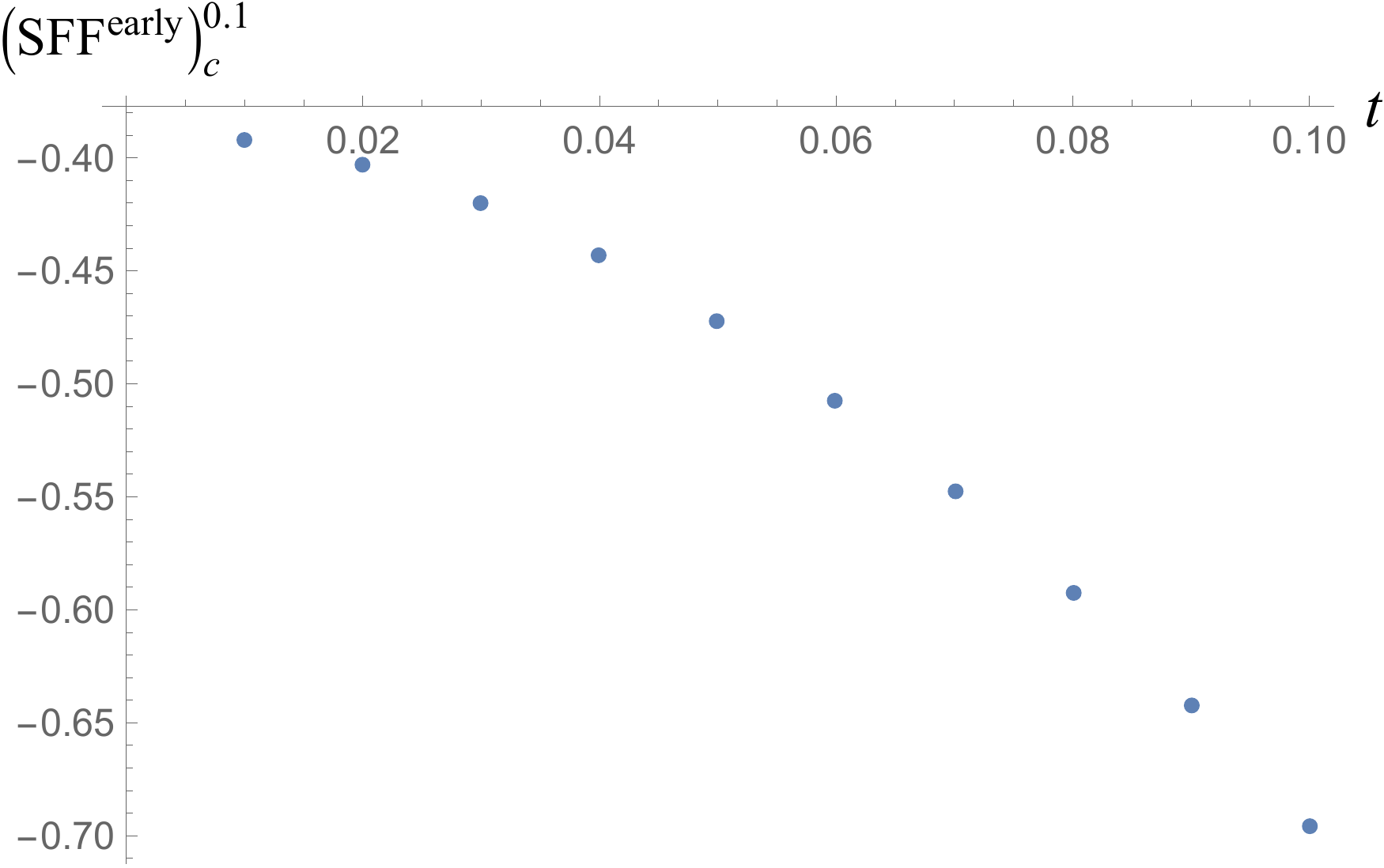}%
\hspace{2mm}\includegraphics[width=0.49\textwidth]{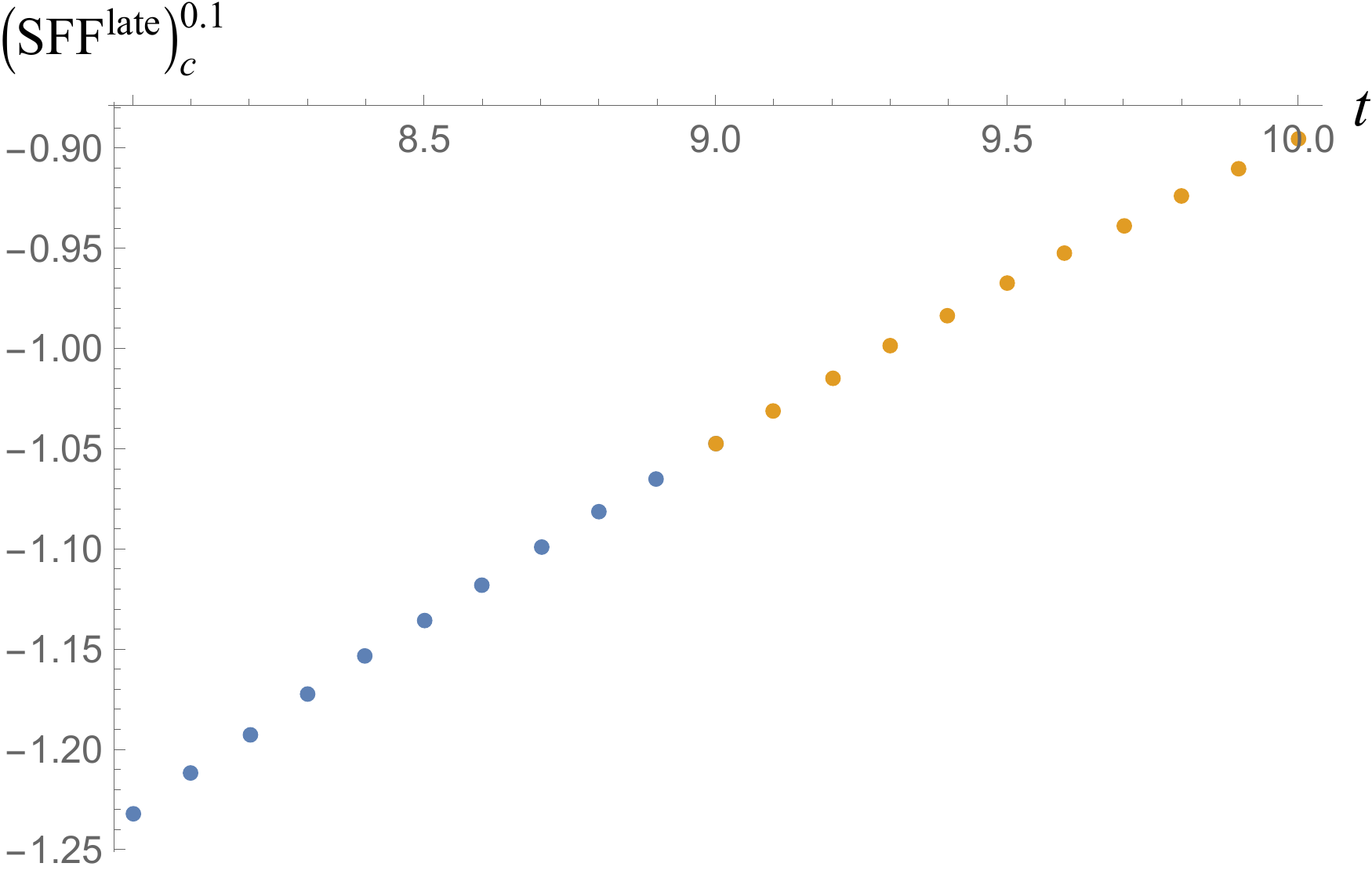}
\end{center}
\caption{Left: The early time behaviour of the exact correlator for $\mu=1, \, \beta=0.1, \, q=0.1$. Right: Zooming in the late time behaviour. The behaviour is smooth, in contrast with the $q=0 \,$ SFF. We expect the correlator to saturate in a plateau but we cannot access this very late time regime $t \gg  10$ with our numerics.}
\label{fig:SFFconnectedearlylateq}
\end{figure}

The double integral describing the exact result contains a highly oscillatory integrand, which needs further manipulation so that it can be computed numerically with good accuracy using a Levin-rule routine. We have kept the relative numerical errors between $10^{-3}$ and $10^{-1}$ relative to the values shown in the plots. In order to do so, it is useful to perform the coordinate transformation $u = \coth \xi/2$, that effectively ``stretches out" the oscillatory behaviour near $\xi = 0$. 

For the SFF, the numerical result is contrasted with the genus zero analytic computation in fig.~\ref{fig:SFFconnected}. It exhibits a ramp - plateau like behaviour with erratic oscillations near the onset of the plateau, that become more regular at late times as seen in fig.~\ref{fig:SFFconnectedearlylate}. At relatively late times $ t \geq 10$ the oscillations are of the same order as the function itself: $\, \Delta SFF_c(t) / SFF_c(t) \rightarrow O(1)$. These oscillations can be trusted since the relative error is always bound $\delta_{err} SFF_c(t) / SFF_c(t)  < 10^{-1}$ at late times. This is an indication that the boundary dual could be a theory with no disorder averaging.

We have also studied a refined SSF, or better said the $q \neq 0$ correlator. Unexpectedly, its behaviour is qualitatively different and much smoother from the $q=0$ case, even for small values such as $q=0.1$. It exhibits an initial dip at early times that transitions into an increasing ramp behaviour. The result is shown in fig.~\ref{fig:SFFconnectedq} and in fig.~\ref{fig:SFFconnectedearlylateq}. We expect a plateau saturation at late times but it is numerically hard to access this regime. We conclude that it would be interesting to further improve the accuracy of the numerics and access the very late time regime.

\section{Comments on the cosmological wavefunctions}\label{DSregime}

In this section we analyse the possibility of giving a $dS_2$ or more general cosmological interpretation for the WdW wavefunctions, after discussing the various possibilities for analytically continuing the $AdS_2$ results\footnote{Bang-Crunch cosmologies on the target space (that is now the superspace) were described in~\cite{Betzios:2016lne}.}. A similar analysis in the context of JT-gravity can be found in~\cite{Maldacena:2019cbz,Cotler:2019nbi,Cotler:2019dcj}.

\begin{figure}[tb!]
\begin{center}
\includegraphics[width=0.9\textwidth]{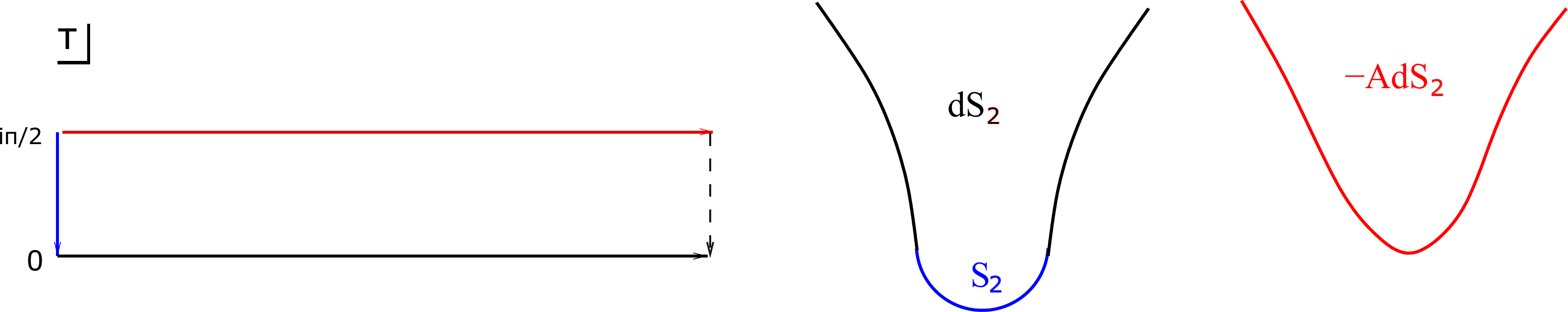}%
\caption{The two different contours one can take in the complex metric space parametrised by $\tau$. The blue line describes a Euclidean $S_2$, the black Lorentzian $dS_2$ and the red negative Euclidean $AdS_2$. The dashed line is a complex geometry connecting the last two types of geometries. (Adapted from~\cite{Maldacena:2019cbz}).}
\label{fig:contours}
\end{center}
\end{figure}

The analytic continuation we consider, is going back to the parameter $z =  i \ell$ in section \ref{MQMandfermions}, and using the fourier transformed operators to compute the partition function and correlators.
In \cite{Maldacena:2019cbz}, the authors explained why this analytic continuation describes ``negative trumpet" geometries by analysing how the geometries change in the complex field space. Let us first define the $dS_2$ global metric
\be
ds^2 = - d \tau^2 + \cosh^2 \tau d \phi^2 \, ,
\ee
and consider then the case of complex $\tau$, so that we can describe Euclidean geometries as well. The usual Hartle-Hawking \cite{Hartle:1983ai} contour for $dS_2$ involves gluing a half-sphere to $dS_2$. This is indicated by the blue and black lines in fig.~\ref{fig:contours}. On the other hand the geometries obtained by the continuation $z =  i \ell$ are ``negative trumpet" geometries that again smoothly cap-off much similarly to what happens in the no-boundary geometries of Hartle and Hawking. These are indicated via the red line in fig.~\ref{fig:contours}. Even though these are not asymptotically $dS_2$ geometries, nevertheless they can be used to define an appropriate no-boundary wavefunction $\Psi_{WdW}(z = i \ell) =  \langle \Tr e^{i z \hat{H}} \rangle$, where $\hat{H}$ is now the generator of space translations at the boundary. In addition according to fig.~\ref{fig:contours}, one can reach the same point in field space (describing a large $dS_2$ universe), either through the usual Hartle-Hawking prescription, or according to a different contour that passes through the ``negative trumpet" geometries which then continues along the imaginary axis so that it connects them to the $dS_2$ geometry\footnote{This contour was also described in~\cite{Hertog:2011ky} for higher dimensional examples.}. One can also imagine the presence of obstructions, in the sense that the two paths in field space might not commute and therefore give different results for the wavefunction. In fact this is precisely what happens in the present example, for the genus-zero wavefunctions. The mathematical counterpart for this, are the properties and transitions between the various Bessel functions as we analytically continue their parameters.

In order to clarify this further, we should also mention a slightly different approach of analysing bulk $dS_2$ geometries proposed in \cite{DaCunha:2003fm} and \cite{Martinec:2003ka}. In the latter case the author performed an analytic continuation of the Liouville theory: $b \rightarrow i b , \, \phi \rightarrow i \phi$, resulting in the supercritical case for which $c \geq 25$. The appropriate minisuperspace wavefunctions describing the $dS_2$ geometries (the conformal boundary is still at $\phi \rightarrow \infty$), are given by the analytic continuation of those in \eqref{MacroscopicWdW} and result into the $dS_2$ Hankel wavefunctions $\Psi^{(macro)}_{dS}(z) \sim H^{(1)}_{i q }(z)$. These wavefunctions are disk one-point functions that describe an asymptotically large $dS_2$ universe that starts at a Big-Bang singularity (whose properties are determined via the vertex operator inserted at the disk - the label $q$). They also correspond to geometries having a hyperbolic class monodromy. The case with an insertion of the cosmological operator, corresponds to the Hartle-Hawking wavefunction.

\begin{figure}[tb!]
\begin{center}
\includegraphics[width=0.8\textwidth]{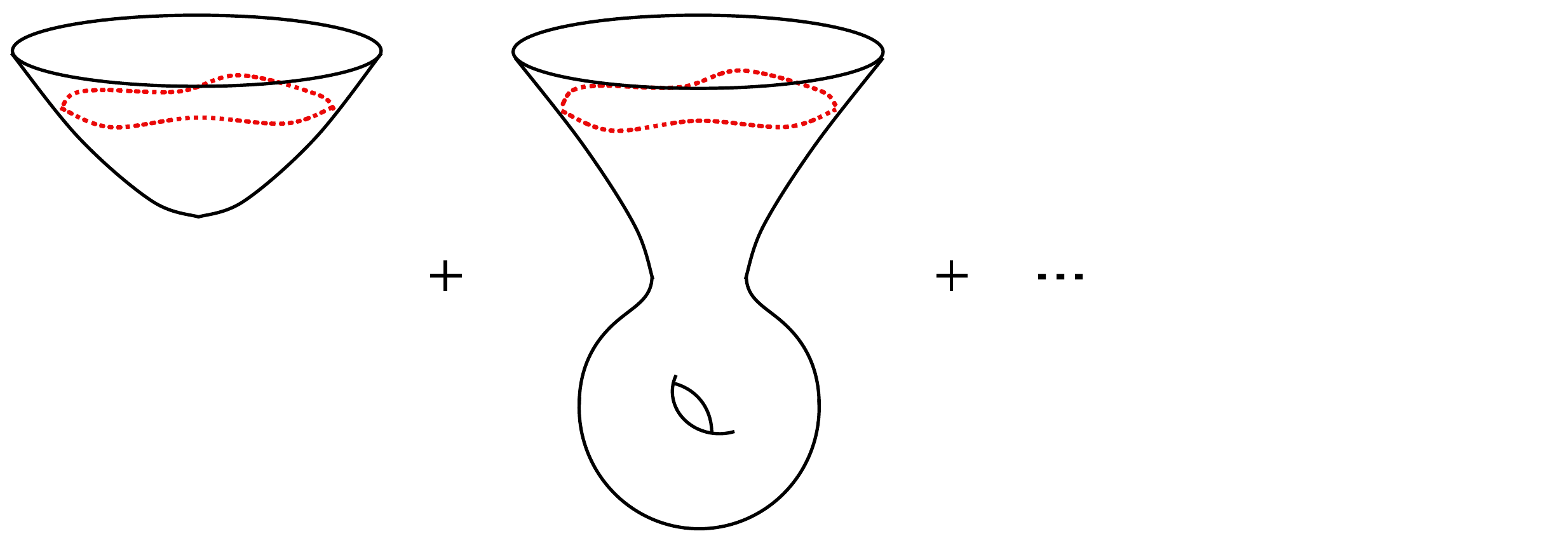}%
\caption{The summation of all the possible smooth Euclidean geometries that asymptote to a ``trumpet" geometry.}
\label{fig:trumpets}
\end{center}
\end{figure}
 
\begin{figure}[!tb]
\begin{center}
\includegraphics[width=0.49\textwidth]{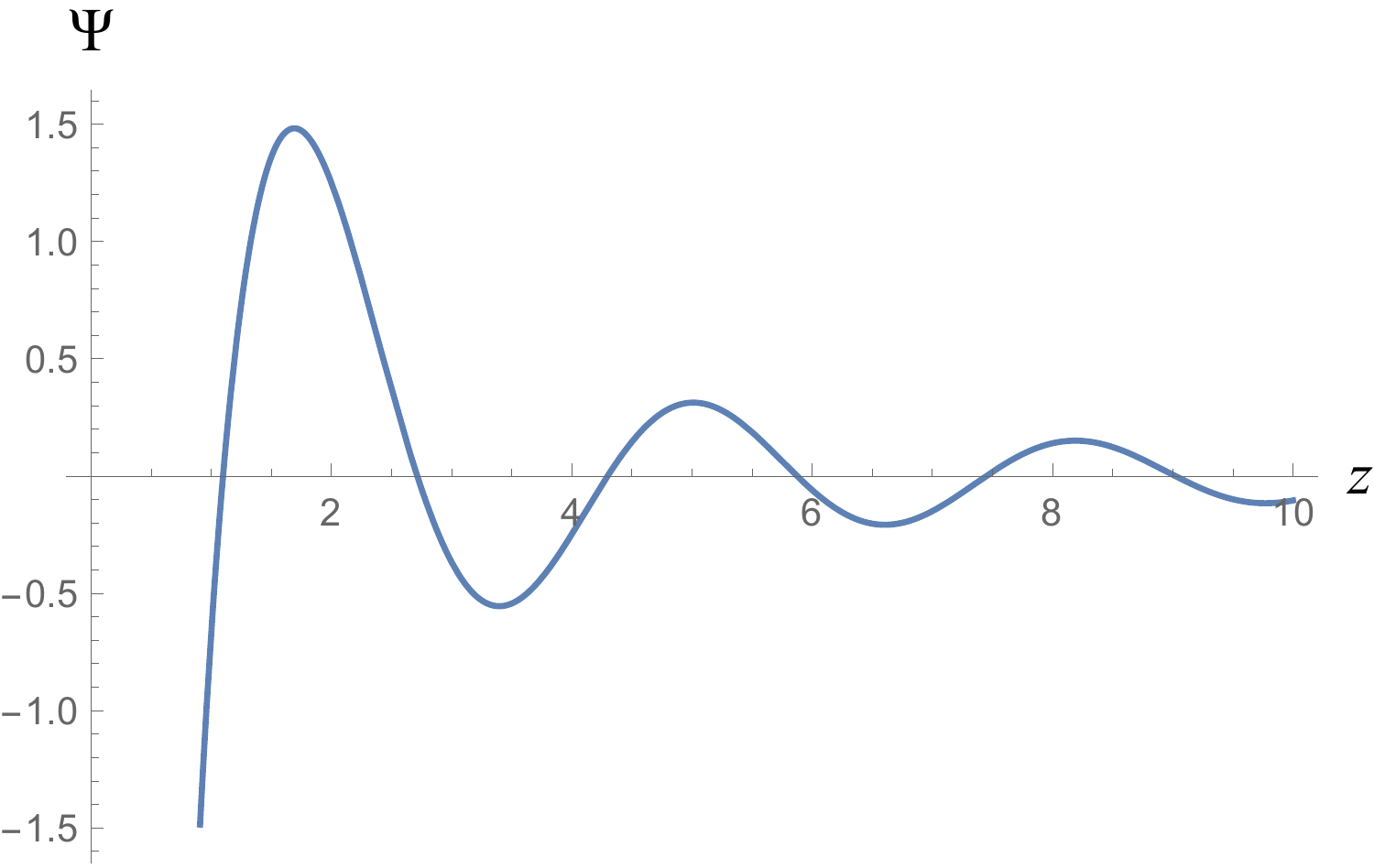}%
\hspace{2mm}\includegraphics[width=0.49\textwidth]{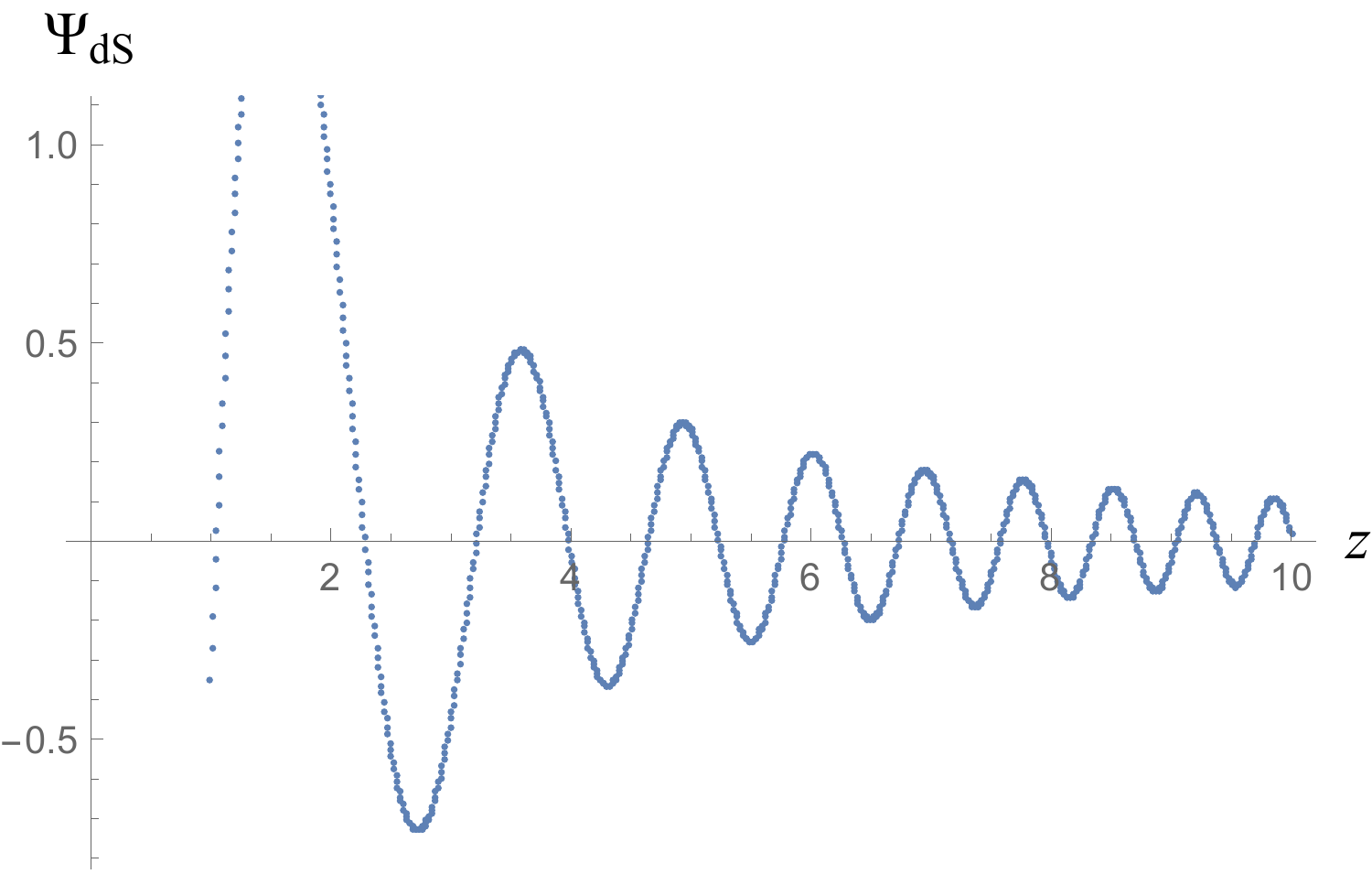}
\end{center}
\caption{Left: The genus-zero cosmological ($-AdS_2$) wavefunction for $\mu=1$ as a function of the boundary parameter $z$. Right: The non-perturbative wavefunction exhibiting a faster oscillatory behaviour.}
\label{fig:dSwavefunction}
\end{figure}

In the present $c=1$-model at the level of the genus zero wavefunctions, the approaches of \cite{Maldacena:2019cbz} and \cite{Martinec:2003ka} remarkably seem to coincide, since they give the same type of Hankel functions as the appropriate cosmological WdW wavefunctions. This can be seen using the formula
\be
K_a(\ell) = K_a(-i z) = \frac{\pi}{2} i^{a+1} H_a^{(1)} (z) \, , \quad  - \pi < \arg( - i z ) < \frac{\pi}{2} \, ,
\ee
on eqn.\eqref{MacroscopicWdW}, that holds in particular for $z \in \mathbb{R}$. If we apply this formula to the genus zero Euclidean result corresponding to the cosmological operator eqn. \eqref{Genuszerowavefunction}, we find
\be
\Psi_{cosm.}(z) \, = \, - i \frac{\pi \sqrt{\mu}}{z} H_1^{(1)} (2 \sqrt{\m} z) \, .
\ee
The plot of its real part is shown in the left panel of fig.~\ref{fig:dSwavefunction}. At the non-perturbative level, we can analytically continue the wavefunction of eqn.\eqref{oneloopMQM}. This result is resumming the geometries plotted in fig.~\ref{fig:trumpets}. A plot of this non-perturbative wavefunction is given on the right panel of fig.~\ref{fig:dSwavefunction}. The behaviour is again qualitatively similar to the genus zero Hankel functions but with more rapid oscillations. This is in contrast with the Euclidean $AdS_2$ case, leading to the conclusion that the most reasonable non-perturbative choice describing the dual of the $AdS_2$ geometries is the one having only one side of the potential filled as described in subsection~\ref{OnesidedLL}, whilst the two-sided fermi-sea is better suited for describing \emph{cosmological} types of geometries. This is also in line with the fact that in $\Psi_{WdW}(z = i \ell) =  \langle \Tr e^{i z \hat{H}} \rangle$, $\hat{H}$ is performing space translations and there is no constraint on the positivity of this trace.

So far we discussed a particular analytic continuation. Nevertheless as we mentioned above, the story is more complicated and there exist various other physical choices in choosing an appropriate wavefunction. In particular at genus zero, all the rest of the Bessel functions appear and play the role of various other choices for states depending on the boundary conditions imposed~\cite{DaCunha:2003fm}. These possibilities are summarised here:

\begin{itemize}

\item $\Psi_{n.b.} \sim J_\nu(z), \, \nu \in \mathbb{R}$, is the no-boundary wavefunction corresponding to the Hartle-Hawking contour of fig.~\ref{fig:contours}. It is real, goes to zero at small $z$ and oscillates at large $z$ (see appendix~\ref{WdWcompendium} for more details on this no-boundary wavefunction).

\item $\Psi_{t.} \sim H_\nu^{(2)}(z)$ corresponds to the tunneling proposal. In particular it is complex and increasing at small $z$ (see appendix~\ref{WdWcompendium} for more details on the tunneling wavefunction).

\item Finally there is the option of demanding only expanding universes at small scales with the resulting wavefunction $\Psi_{exp.} \sim J_{i \nu}(z), \, \nu \in \mathbb{R}$. This has a more fitting minisuperspace interpretation as giving the ``birth of a superspace quantum"~\cite{DaCunha:2003fm}, that can be used in a ``scattering process in superspace".

\end{itemize}

A natural question regards the possibility of realising all these wavefunctions as various limits of a complete non-perturbative description. This is what seems to be happening, since the more general Whittaker WdW wavefunctions appearing in \eqref{ExactwavefunctionsWdW}, \eqref{Exactwavefunctionoperatorinsertion}, can be related to various types of Bessel functions as we send $\mu \rightarrow \infty,$ along different regions of the complex $\mu, \, z, \, q$ planes. We then expect the presence of interesting Stokes phenomena and transitions between the various asymptotic genus zero wavefunctions. Before that, one should also understand the unexpected relation between the sub-critical and supercritical regime. We plan to revisit this problem in the future.

\section{Conclusions}\label{Conclusions}

We will now make some further comments on our findings and discuss some interesting future directions. We have discussed MQM from the point of view of selecting a non perturbative completion of the $c=1$ Liouville bulk quantum gravity path integral (which is not unique). The bulk quantum gravity path integral seems not to factorise even at the full non-perturbative level and hence Euclidean wormholes and the physics of multiple boundaries are inevitable consequences of this ``third quantised" point of view. Unitarity can still be kept intact as exemplified by the use of MQM, but only in such an extended Hilbert space.

The Schwarzian alone is known not to be a consistent quantum mechanical model, since the path integral on the circle cannot be interpreted as $\Tr e^{-\beta \hat{H}}$ for any $\hat H$~\cite{Stanford:2017thb,Harlow:2018tqv}. On the other hand similar boundary duals of Liouville theory coupled to matter, have a possibility of being consistent quantum mechanical theories of their own. In the present $c=1$ case, this is a problem that relies on a Hilbert space interpretation of the exact function for the density of states given by eqn. \eqref{dualdos} or equivalently of the dual resolvent of appendix~\ref{DualResolvent} (the usual resolvent can be interpreted in terms of the inverted oscillator Hamiltonian). Such a system could have both a continuous and a discrete spectrum, the small persistent oscillations being indicative of the discrete part. Since the density of states resembles that of large-N random matrices, any dual system is expected to have approximate chaotic properties. This is also evident from an analysis of the exact connected SFF of eqn. \eqref{SFFmain}, displaying a ramp plateau behaviour with the presence of erratic oscillations.

An important property of the spectral form factor is that the ramp and plateau are not self-averaging. It has been argued that for a fixed Hamiltonian system, even though we are summing over many energy levels, the result should be a function with $O(1)$ fluctuations. The smooth ramp and plateau are results of a time or disorder average, but the exact function is expected to be erratic \cite{Saad:2018bqo}. In our example we have remnants of such an erratic behaviour even in the double scaling limit. While they are not as pronounced as the ones expected in higher dimensional examples they are still present and affect both the DOS and the SFF pointing to a non-disorder averaged dual theory. Similar effects were observed and analysed in~\cite{Blommaert:2019wfy}, when fixing some of the matrix eigenvalues to take definite values. Here an additional effect is the IR-cutoff $\mu$ that depletes the spectrum. 

It is also interesting to notice that while there exists a complete description of the bulk path integral in superspace in the form of MQM, a single boundary dual theory alone does not seem to be able to capture all its intricacies. For example, while the interpretation of Euclidean wormholes is rather straightforward from a superspace perspective, it is not obvious how and if this information is encoded in a \emph{single boundary} dual theory. Perhaps the most reasonable attitude on this, is that one needs to first choose the types of boundary conditions in superspace that are allowed and depending on this choice there could be a single boundary dual description. The reason is that from a third quantised point of view the dual partition function and density of states correspond to operators, and one can form various expectation values out of them. The complete Hilbert space is hence naturally larger than naively expected. Nevertheless this is not an argument against holography. Holography simply seems to hold slightly differently than expected, involving matrix degrees of freedom and an enlarged third quantised Hilbert space. By imposing certain restrictions on that complete Hilbert space we can then use a more usual holographic correspondence.

A preliminary analysis of the cosmological regime described after the analytic continuation $z = i \ell$ in section~\ref{DSregime}, reveals that while there is a single WdW wavefunction that takes the form of an integral of a Whittaker function, there exist various asymptotic expansions/limits of the parameters of the wavefunction leading to interesting Stokes phenomena. This seems to be the reason behind the appearance of several types of semi-classical wavefunctions in superspace and boundary conditions, such as the Hartle Hawking or the tunneling proposal etc., while all of them could arise from a unique progenitor in this simple two dimensional example. This is clearly a point that deserves further study.

Let us also mention that we could also have considered other types of ``target space" minisuperspace geometries, that would correspond to other types of matter content from the two dimensional bulk point of view. For example one could try to analyse WZW-models in the presence of boundaries and again interpret the worldsheet as a bulk spacetime. It would be interesting to see what are the differences with the present example.

\paragraph{Geometries inside geometries and double layered expansions -}

Another question that we briefly alluded to in the introduction, is the possibility of having a double layered expansion, or what one could call ``doubly non-perturbative effects". One natural quantum gravitational setting in which such an expansion could arise is the idea of having ``worldsheets inside worldsheets" or ``geometries inside geometries". For example, it would be interesting to understand whether the two dimensional Liouville theory worldsheets (quantum gravity target spaces from our WdW point of view) can emerge themselves from an underlying string theory in a fashion similar to the proposal in~\cite{Green:1987cw}\footnote{This can be realised in theories with $\mathcal{N}=2$ SUSY as shown in~\cite{Kutasov:1996fp}.}. This would be based on the sequence of mappings $\sigma, \bar{\sigma} \rightarrow z, \bar{z} \rightarrow X, \phi$ and result in a third quantised theory of universes inside of which strings propagate. That said, the resulting theory should a priori have a \emph{two parameter} expansion: the genus expansion of the strings $g_s$ as well as the topology expansion of the resulting target space, which in our interpretation was the $1/\mu $ expansion. Then for a fixed topology of the target space, one would have to sum over all the string worldsheet genera. Once resummed, non-perturbative effects could then take the form $\exp( - \mu e^{- 1/g_{s}})$, since one can introduce both boundary branes for the space ($SD$-branes) as well as worldsheet boundaries ($D$-branes).

It is much harder to concoct a dual model that can realise such a setup. If there is any matrix/tensor model realisation of this idea, it should involve two free parameters in the appropriate continuum scaling limit. While the tensor models naturally exhibit similar rich scaling limits~\cite{Gurau:2016cjo}, they are much harder to study. Something analogous is also realised in rectangular $N \times M$ matrix models~\cite{Myers:1992dq,DiFrancesco:2002mvz}, where one can tune the parameters $M,N \rightarrow \infty$ together with a coupling constant $g$ independently, to reach new critical points. There is an obvious problem in this idea related to the fact that when the shape of the matrices becomes very narrow the surfaces degenerate into branched polymers. To overcome this obstacle, we can use several large-N limits in conjuction with the idea of breaking the original Hermitean matrix $M_{i j}$ into blocks, for example
\be
M_{N \times N} = 
\begin{pmatrix}
X_{n \times n} & \Phi_{n \times (N- n)} \\
\Phi^\dagger_{(N - n) \times n} & Y_{(N -n) \times (N- n)} 
\end{pmatrix} \, .
\ee
In this splitting there is a breaking of the $U(N)$ symmetry into $U(n) \times U(N-n)$, when the off-diagonal elements are zero, that results in two distinct geometrical objects (surfaces) if we take the double scaling limit in both of them. On the other hand when $N \approx n$ the off-diagonal elements become very narrow (leading to a surface interacting with a particle through branched polymers in the scaling limit). This then means that we can try to break the original matrix into n-blocks $N = N_1 + N_2 + ... N_n$ and introduce a chemical potential $\mu_N$ for $N$ and $\mu_n$ for $n$. While $\mu_N$ governs a genus expansion in superspace as before, the parameter $\mu_n$ playing the role of a genus expansion for any fixed value of $\mu_N$. More hierarchical matrix embeddings could lead into analogous hierarchical surface embeddings. Such embeddings also always satisfy a general version of the stringy exclusion principle~\cite{McGreevy:2000cw}: submatrices are always smaller than the matrix they are embedded into.

\paragraph{Higher dimensions -}

There is a crucial difference of the present models with higher dimensional examples. All the theories in two dimensions could be related to some limit of string theories with some particular form of matter content. For example it is not clear whether we can consistently define higher dimensional models of geometries propagating in some form of superspace. This is a first crucial point to understand if one wishes to extrapolate the present results into higher dimensions, and makes even more pertinent the analysis of higher dimensional examples even at a truncated mini-superspace level.

The most probable conclusion consistent with the present results is that the bulk quantum gravity path integral is in fact \emph{richer} than that of a single quantum field theory (in the present case boundary quantum mechanics). Imposing certain restrictions and boundary conditions at the superspace level, one can indeed reduce the path integral in a subsector that is dual to a single QFT as in the usual implementations of holography. In the most general case though the non-perturbative bulk path integral can compute more complicated objects than that of a single QFT partition function, that might or not have a single QFT interpretation. This is not in contradiction though, either with the holographic interpretation at fixed genus given in~\cite{Betzios:2019rds}, or with the possibility that there could exist a complicated matrix/tensor model, such as MQM in the present case, that is able to capture all the non-perturbative quantum gravity information. Such a model would not be the boundary dual of the bulk quantum gravity theory in the usual AdS/CFT sense, but a model describing directly the dynamics of geometries in superspace. In fact variants of the BFSS matrix model~\cite{Mtheory} could very well have the potential to serve as a dual of this kind.

\section*{Acknowledgements}\label{ACKNOWL}
\addcontentsline{toc}{section}{Acknowledgements}

We wish to thank Costas Bachas and Elias Kiritsis for the discussion that gave us the stimulus to initiate this project. We also thank Dionysios Anninos for reading a preliminary draft and his very useful comments and suggestions. We happily acknowledge the hospitality provided by APC and ENS Paris during the initial stages of this work.

\noindent This work is supported in part by the Advanced ERC grant SM-GRAV, No 669288.

\newpage
\appendix
\renewcommand{\theequation}{\thesection.\arabic{equation}}
\addcontentsline{toc}{section}{Appendix\label{app}}
\section*{Appendices}

\section{Matrix models for minimal models}\label{Minimalmodels}

\subsection{Conformal maps and Integrable hierarchies}\label{conformalmaps}

There exists a well studied relation of contour dynamics in two dimensions with the dispersionless limit of the Toda Hierarchy~\cite{Conformalmaps}. We now briefly review the results of these works and then discuss their relation to the two dimensional quantum gravity path integral.

An equation for a curve on the complex plane $F_{\mathcal{C}}(z,\bar{z}) = 0$ can be resolved locally with the help of the Schwarz function as $\bar{z} = S(z)$. We assume this curve to bound a simply connected domain $D^+$ and we label the exterior domain with $D^-$. The problem of multiple domains is described in~\cite{Krichever:2003ys} (in terms of the Schottky double). The Schwarz function obeys a unitarity condition $\bar{S}(S(z)) = z$ and can be decomposed into two functions $S(z) = S^+(z) + S^-(z)$ that are holomorphic in the interior/exterior of the domain.  Let us also define a conformal map $z(w)$ that maps the exterior of the unit disk to the exterior domain $D^-$.

We then define a function $\Omega(z)$ via $S(z) = \partial_z \Omega(z)$. It plays the role of the generating function of the canonical transformation from the unit disk to the region bounded by the curve. Its differential defines a multi-time Hamiltonian system through
\be\label{Hamiltoniansystem}
d \Omega = S(z) d z + \log w dt + \sum_{k=1}^\infty \left( H_k d t_k - \bar{H}_k d \bar{t}_k \right) \, .
\ee 
In this formula
\be
H_k = \frac{\partial \Omega}{\partial t_k} \, , \quad H_k = - \frac{\partial \Omega}{\partial \bar{t}_k} \, ,
\ee
and $t_k$ are ``time variables" corresponding to the moments of the region and $t$ is the zero time dual to the area, for more details see~\cite{Conformalmaps}. This indicates a dual way of describing the curve via the moments $\mathcal{C}(t, t_k, \bar{t}_k)$ and a dual ``prepotential" $F(t, t_k , \bar{t}_k)$. The relevant set of equations that governs this system corresponds to the dispersionless limit of the Toda hierarchy $\hbar \rightarrow 0$.

In this dispersionless limit, the $\tau$ function is then defined as $\tau = \exp(F/\hbar^2)$ and the Baker-Akhiezer wavefunction as $\Psi = \exp(\Omega/\hbar) $. A particular representation of a $\tau$-function of the Toda hierarchy is in terms of a two-matrix model
\be\label{complexMatrix}
\tau_N = \mathcal{Z}(N) = \int d M \, d \bar{M} e^{-N \left( \Tr ( M \bar{M}) + \sum_{k>0} ( t_k \Tr M^k + \bar{t}_k  \Tr \bar{M}^k) \right)} \, .
\ee
The interface dynamics is then described by the dispersionless limit of this matrix model i.e. $\hbar = 1/N \rightarrow 0$, which is the large-N limit. Its free energy is thus the prepotential $F(t_k, \bar{t}_k)$. An equivalent definition of the $\tau$-function is encoded in the Schwarzian derivative of the conformal map $w(z)$ through the following relation
\be
\frac{w'''(z)}{w'(z)} - \frac{3}{2} \left( \frac{w''(z)}{w'(z)} \right)^2 = 6 z^{-2} \sum_{k, n \geq 1} z^{-k -n} \frac{\partial^2 \log \tau }{\partial t_k \partial t_n} \, .
\ee

\subsection{The $(p,q)$ minimal models}

The matrix model \eqref{complexMatrix} describes also the dynamics of the $(p,q)$-minimal models coupled to gravity~\cite{Seiberg:2003nm,Maldacena:2004sn,Kazakov:2004du}. The $(p,q)$ minimal models arise when the square of the Liouville parameter becomes a rational $b^2 = p/q $. In this series of works it was understood that all the possible $(p,q)$ models can be described in terms of a $\mathcal{M}_{(p,q)}$ Riemann surface at the perturbative level. Let us first define the dual cosmological constants via
\be
\tilde{\mu} = \mu^{1/b^2} \, , \quad \frac{\tilde{\mu}_B}{\sqrt{\tilde{\mu}}} = \cosh \frac{\pi \sigma}{b} \, ,
\ee
which describe a symmetry of the physical observables under $b \rightarrow 1/b$. If we then use the parameters
\be
x = \frac{\mu_B}{\sqrt{\mu}} \, , \quad y = \frac{\partial_{\mu_B} Z_{FZZT}}{\sqrt{\tilde{\mu}}} \, , \quad \tilde{x} = \frac{\tilde \mu_B}{\sqrt{\tilde \mu}} \, , \quad \tilde y = \frac{\partial_{\tilde \mu_B} \tilde{Z}_{FZZT}}{\sqrt{\mu}} \, ,
\ee
with $Z_{FZZT}$ the disk partition function of the FZZT-brane, the duality means that $\tilde{x} = y$ and $\tilde{y} = x$.  The set of equations that determine the partition functions is
\bea\label{partitionfunctions}
Z_{FZZT} = \int^{x(\mu_B)} y d x \, , \qquad Z^{ZZ}_{(m,n)} = \oint_{B_{m, n}} y d x \, \nn \\
\tilde{Z}_{FZZT} = \int^{y(\tilde \mu_B)} x d y \, , \qquad \tilde{Z}^{ZZ}_{(m,n)} = \oint_{\tilde{B}_{m, n}} x d y
\eea
the second integral corresponds to an integral through the pinched cycles of $\mathcal{M}_{(p,q)}$. This is where the ZZ-branes reside. An equivalent way of rewriting all these equations is through
\be
F_{p,q}(x,y) \equiv T_p(x) - T_q(y) = 0 \, , \quad \tilde{F}_{p,q}(x,y) \equiv T_q(\tilde x) - T_p(\tilde y) = 0 \, ,
\ee
This means that the function $F$ describes a curve corresponding to $\mathcal{M}_{(p,q)}$. This surface has $(p-1)(q-1)/2$ singularities when
\be
F_{p,q} = \partial_x  F_{p,q} = \partial_y  F_{p,q} = 0 \, ,
\ee
that correspond to pinched cycles of the surface where the ZZ-branes reside. The presence of a background of such branes has the effect of opening up the pinched cycles. At the non-perturbative level in the string coupling, Stokes phenomena change the picture drastically and the surface $\mathcal{M}_{(p,q)}$ is replaced by the simple complex plane $\mathbb{C}$~\cite{Maldacena:2004sn}. For example the exact FZZT partition function is described by an Airy function $Ai(x+1)$ which is an entire function of $x$. In order to understand precicely how this happens, the matrix model comes at rescue, since one can compute the appropriate loop operator expectation values. The loop operator is defined through
\be 
W(x) = \frac{1}{N}  \Tr \log (x - M) \, .
\ee
In particular at genus zero $\langle W(x) \rangle = Z_{FZZT}$ corresponds to the disk amplitude and $y = \partial_x Z_{FZZT} $ to the resolvent of the matrix model. The full non-perturbative FZZT brane corresponds to an exponential of the loop operator
\be
\Psi(x) = \det( x - M) = e^{N W(x)}
\ee
It is known that the expectation value of the determinant operator, in the double scaling limit that corresponds to forming continuous surfaces, corresponds to a Baker-Akhiezer function that is an entire function of $x$ and hence the complete non-perturbative moduli space of the FZZT branes is the complex plane $\mathbb{C}$.

\subsection{Deformations}

We can also turn on an arbitrary number of closed string couplings $t_k, \bar{t}_k$ to deform the closed string background as seen from the matrix model \eqref{complexMatrix}. The notation here is due to the fact that the deformed quantities such as the partition functions are generically related to $\tau$-functions of the KP (a single set of ``times") or Toda hierarchies with $t_n$'s playing the role of ``times".

The new differential to be integrated is then defined via a deformation of $y d x$:
\be\label{deformeddifferential}
d \Phi  \, = \, y d x \, + \, \sum_{k \geq 1}  H_k(x) \, dt_k  - \bar{H}_k d \bar{t}_k	 \, ,
\ee
where $H_m(x)$ are mutually commuting ``Hamiltonians" dual to the time variables $t_k$ and so forth for the bar quantities. In particular one can now replace the previous differential $y d x$ by $d \Phi$ in all the formulae \eqref{partitionfunctions}. Notice the equivalence to the Hamiltonian system~\ref{Hamiltoniansystem} that allows to transition between the physics of the minimal models and that of interface dynamics.

We therefore conclude that the dispersionless limit captures a universal sector of the interface dynamics as well as of  $(p,q)$-minimal models coupled to gravity. We can further consider a small sector of the ``Goldstone hydrodynamic modes" that describe small ripples of the boundary curve (interface) geometry. Nevertheless there is a huge class of integrable-deformations governed by the time parameters $t_k$, $\bar t_k$, that capture finite deformations of the geometry.

\section{Properties of the WdW equation}\label{WdWcompendium}

In this appendix we will describe some basic properties of the Wheeler DeWitt equation as well as the most common boundary conditions employed in the literature. Some reviews can be found in~\cite{BabyUniverses}. A complete higher dimensional superspace WdW equation is not really well defined and suffers from various problems such as infinite configuration space, operator ordering ambiguities, action unbounded from below etc. We will hence restrict to a minisuperspace formulation of the problem which reduces the degrees of freedom to a finite number, fortunately the two dimensional example of Liouville theory is very well described just by the minisuperspace approximation due to the small number of physical degrees of freedom in two dimensions. 

The general minisuperspace action takes the form 
\be
S[q^a(r), N(r) ] = \int d r N \left(\frac{1}{2 N^2} G_{a b}(q) \dot q^a \dot q^b - U(q)\right) \, ,
\ee
where $q^a(r)$ are the finite number of variables, $N(r)$ is a non-dynamical Lagrange multiplier and $G_{a b}(q)$ is a reduced form of the metric in superspace. The momentum constraints are satisfied trivially in the minisuperspace ansatze, so what is left is the Hamiltonian constraint coming from the Lagrange multiplier $N(r)$
\be
H = \half  G_{a b} \pi^a \pi^b + U(q) = 0
\ee
Upon quantising one should replace the momenta with
\be
\pi_a = - i  \frac{\partial}{\partial q^a} \, ,
\ee
but generically there is a non-trivial operator ordering problem because of the metric $G_{ab}(q)$ on minisuperspace. Assuming a freedom of field redefinitions, the appropriately ordered operator takes the form
\be\label{minisuperspaceWdW}
\hat H = - \half \nabla^2_{G}  + U(q) \, , \quad \hat{H} \Psi_{WdW}(q^a) = 0 \, .
\ee
In this expression the covariant Laplacian is computed using the minisuperspace metric. This minisuperspace metric has generically indefinite signature and hence one can find both exponential and oscillatory solutions.

Let us also note that there is a path integral representation of the wavefunction. First there is a reparametrisation symmetry due to $N(r)$. It can be shown that only the zero mode plays a role and the path integral takes the form (in the Lorentzian case)
\be
\Psi(q^a_{r_{max}}) = \int d N \int \mathcal{D} p_a \mathcal{D} q^a e^{i S_L (p_a, q^a , N) } \, = \, \int d N \Psi(q^a_{r_{max}}, N)
\ee
where the boundary conditions are $q^a(r_{max}) = q^a_{r_{max}} $ and the other variables are free. There are also extra boundary conditions to be specified at any other boundary of the $r$ variable. The wavefunction $\Psi(q^a_{r_{max}}, N)$ satisfies the time dependent Schroendinger equation with $N$ playing the role of the time variable. In particular
\be
\hat{H} \Psi(q^a_{r_{max}}) = i \Psi(q^a_{r_{max}}, N) \vert_{N_1}^{N_2} \, ,
\ee
so that we need to take either an infinite contour in the $N$-space or a closed contour. Some usual physical requirements on $\Psi(q^a_{r_{max}})$ employed in the literature are that it is peaked near the classical configurations, and the interference between two configurations is small (decoherence).

To pass over to the classical limit it is convenient to search for a semi-classical WKB ansatz for the wavefunction of the Hamilton-Jacobi form $\Psi = e^{- I_R(q)/\hbar + i S(q)/\hbar}$, with $I_R(q), \, S(q)$ real. Plugging this inside \eqref{minisuperspaceWdW} one then finds the set of equations
\bea\label{HJequations}
- \half  (\nabla_{G} I_R)^2 + \half  (\nabla_{G} S)^2  + U(q) = 0 \, , \nn \\
\nabla I_R \cdot \nabla S = 0 \, .
\eea
The second equation is equivalent to the usual condition of steepest descent analysis: the steepest contours of the real part are orthogonal to those of the imaginary part. As for the first equation, in the regions where $U(q) > 0$ we find
\be
\Psi_\pm^{c.f}(q) \sim e^{\pm \frac{1}{\hbar} \int^q \sqrt{U(q')} d q'} \, ,
\ee
while for $U(q) < 0$ we have oscillating solutions
\be
\Psi_\pm^{c.a}(q) \sim e^{\pm \frac{i}{\hbar} \int^q \sqrt{-U(q')} d q' \, \mp i \pi/4} \, ,
\ee
The subscript $c.a$ stands for classically allowed region of the minisuperspace and $c.f$ for classically forbidden or tunneling. The classically forbidden case exhibits a tunneling wavefunction, for which quantum effects are important. On the other hand the classically allowed oscillatory wavefunctions are strongly peaked on a \emph{set of classical solutions} of the Hamilton-Jacobi equation \eqref{HJequations}. In addition if $|\nabla_G I_R| \ll |\nabla_G S|$ the amplitude is changing very slowly and the classical approximation becomes extremely accurate.

Let us now discuss some caveats that one should be careful about. A first complication comes from the fact that the metric in minisuperspace is not positive definite (due to the conformal mode of the metric) and generically the boundary value problem is of the hyperbolic type. Another issue related to this is the choice of contour in mini-superspace which is quite subtle. The most common ones are:
\\
\\
The \emph{no-boundary proposal} that passes from the Euclidean regime and posits a smooth Euclidean completion of the boundary geometry, it is CPT invariant and consists out of two WKB modes, so that it corresponds to a real wavefunction.
\\
\\
The \emph{tunneling} wavefunction that posits an outward probability flux near the singular boundary of superspace that gives only one WKB mode\footnote{The boundary of superspace consists of both regular geometries (such as the poles of a sphere) and  singular geometries.}. The probability current cannot be defined globally but for the WKB approximation it takes the form $J \sim - \nabla S$.
\\
\\
Let us now give a few more details about these two most common options.

The \emph{tunneling wavefunction} can be described is the classically forbidden/allowed regions via
\bea\label{tunnelingwavefunction}
\Psi_T^{c.f}(q) &=& \Psi_+^{c.f}(q) - \frac{i}{2} \Psi_-^{c.f}(q) \, , \nn \\
\Psi_T^{c.a}(q) &=&  \Psi_-^{c.a}(q)  \, ,
\eea
which in the semi-classical approximation becomes
\be
\Psi_{T}(q) \sim \cosh( I_R(q)/\hbar) e^{i S(q)/\hbar} \,.
\ee

The \emph{no-boundary proposal} of Hartle-Hawking, consists of the following solutions in the classically forbidden/allowed regions
\bea\label{HHwavefunction}
\Psi_{HH}^{c.f}(q) &=&  \Psi_-^{c.f}(q) \, , \nn \\
\Psi_{HH}^{c.a}(q) &=&  \Psi_+^{c.a}(q) + \Psi_-^{c.a}(q)  \, .
\eea
or in terms of the semi-classical variables
\be
\Psi_{HH}(q) \sim e^{- I_R(q)/\hbar} \cos(S(q)/\hbar) \,.
\ee
This is a manifestly real wavefunction.
\\
\\
So far we had been careful not to use Lorentzian vs. Euclidean geometries, what is important is that the tunneling behaviour incorporates \emph{both classically forbidden modes}, whilst the no-boundary has \emph{both classically allowed oscillatory modes}. Moving in the field space it is also possible to cross Stokes-lines, so extra care needs to be taken in the choice and interpretation of the contour. In addition a complex saddle can have different interpretations: if we fix two points in superspace we can choose various contours to connect them. This is used in the main text in section~\ref{DSregime}.

\section{Parabolic cylinder functions}\label{Parabolic}

These are the eigenfunctions of the inverted harmonic oscillator time independent Schroendinger equation. The specific differential equation is
\be
\label{inverted}
\left( \frac{d^2}{d \lambda^2} + \frac{\lambda^2}{4} \right) \psi(\omega, \lambda) = \omega \psi(\omega, \lambda) \,.
\ee

\subsection{Even - odd basis}

A useful basis of solutions for real $\lambda$ are the delta function normalised even/odd parabolic cylinder functions~\cite{Moore:1991sf} which we will denote by $\psi^\pm (\omega, \l)$
\bea
\psi^+(\omega, \l) = \left(\frac{1}{4 \pi \sqrt{(1+e^{2 \pi \omega})}}\right)^\half 2^{1/4} \bigg| \frac{\Gamma(1/4+ i \omega/2)}{\Gamma(3/4+ i \omega/2)} \bigg|^{1/2} e^{- i \l^2/4} {_1F_1} (1/4-i \omega /2 , 1/2 ; i \l^2/2) \nn \\
=\frac{e^{- i \pi/8}}{2 \pi} e^{- \omega \pi /4} |\Gamma(1/4 + i \omega/2)| \frac{1}{\sqrt{|\l|}} M_{i \omega/2 , - 1/4}(i\l^2/2) \nn \\
\psi^-(\omega, \l) = \left(\frac{1}{4 \pi \sqrt{(1+e^{2 \pi \omega})}}\right)^\half  2^{3/4}  \bigg| \frac{\Gamma(3/4+ i \omega/2)}{\Gamma(1/4+ i \omega/2)} \bigg|^{1/2} \lambda e^{- i \l^2/4} {_1F_1} (3/4-i \omega /2 , 3/2 ; i \l^2/2) \nn \\
=\frac{e^{-3 i \pi/8}}{ \pi} e^{- \omega \pi /4} |\Gamma(3/4 + i \omega/2)| \frac{\l}{|\l|^{3/2}} M_{i \omega/2 ,  1/4}(i\l^2/2)\,. \nn \\
\eea
Their normalisation is
\be\label{norm1}
\int_{-\infty}^\infty d \lambda \sum_{s=\pm} \psi^s (\omega_1 , \lambda) \psi^s (\omega_2 , \lambda) = \delta(\omega_1 - \omega_2),
\ee
and
\be\label{norm2}
\int_{-\infty}^\infty d\omega \sum_{s=\pm} \psi^s (\omega , \lambda_1) \psi^s (\omega , \l_2) = \delta(\l_1 - \l_2)\,.
\ee

\subsection{Complex basis}\label{Complexbasis}

Another possible set of solutions to use, is a complex set which we label by $\Phi(\omega, \l)$.

These solutions can be expressed in terms of parabolic cylinder functions as follows
\bea\label{Paraboliccomplex}
\Phi(\omega, \l) &=& e^{\pi \omega/4 + i \phi_2 /2} e^{i \pi/8} U(i \omega, \, e^{- i \pi/4} x)  \, , \nn \\
\phi_2(\omega) &=& Arg \, \Gamma(1/2 + i \omega) \, , \quad U(i \omega, \, e^{- i \pi/4} \lambda) = D_{-i \omega - 1/2}( e^{- i \pi/4} \lambda)\, ,
\eea
with $\phi_2(\omega = 0) = 0$.
\\
\\
Another useful representation is in terms of Whittaker functions
\be
\Phi(\omega, \l) = \frac{2^{- i \omega/2}  e^{\pi \omega/4 + i \phi_2 /2} e^{i 5 \pi/8}}{x^{1/2}} W_{- i \omega/2 , \, - 1/4}(- i \lambda^2/2) \, .
\ee

The complex solutions obey the following relations
\bea
\mathcal{W} \lbrace \Phi(\omega, \l), \, \Phi^*(\omega, \l) \rbrace &=& - i \, , \qquad \Phi^*(\omega, \l) = e^{- i( \phi_2 + \pi/4)} \Phi(-\omega, i \l) \nn \\
\sqrt{\Gamma(1/2+ i \omega)}\Phi^*(\omega, \l)  &=& e^{- i \pi/4} \sqrt{\Gamma(1/2 - i \omega)} \Phi(-\omega, i \l)
\eea
so they form an appropriate orthonormal set as well (up to the factor of $- i$).


\subsection{Resolvent and density of states}\label{Resolvent} 

In this subsection we define the usual resolvent and density of states corresponding to the inverted oscillator equation \eqref{inverted} that defines the Hamiltonian $\hat{H}$. The resolvent operator is defined as
\be
\hat{R}(\zeta) = \frac{1}{\hat{H} - \zeta - i c}
\ee
The fixed energy amplitude between two position states is then
\bea
\langle \lambda_1 | \hat{R}(\zeta) | \lambda_2   \rangle \, = \, R(\zeta ; \, \lambda_1, \lambda_2) &=& \int_{-\infty}^\infty d \omega \frac{1}{\omega - \zeta} \sum_s \psi^s(\omega , \lambda_1) \psi^s(\omega , \lambda_2) \nn \\
&=& - i \int_0^{- \infty \sgn (\Im \zeta)} d s \,  e^{- i  s \zeta} \, \langle \lambda_1 |  e^{- 2 i s \hat{H}}| \lambda_2 \rangle \, .
\eea
The expression for the propagator is given by the Mehler formula for parabolic cylinder functions~\cite{Moore:1991sf}:
\be
\langle \lambda_1 | e^{-2 i T \hat{H}}| \lambda_2 \rangle= \int_{-\infty}^\infty d \omega e^{i \omega T} \sum_{s=\pm} \psi^s(\omega , \lambda_1) \psi^s(\omega , \lambda_2) = \frac{1}{\sqrt{4 \pi i \sinh T}} \exp \frac{i}{4} \left[\frac{\lambda_1^2 + \lambda_2^2}{\tanh T} - \frac{2 \lambda_1 \lambda_2}{\sinh T} \right],
\ee
which is computing the real-time $T$ inverted H.O. propagator. This holds for $-\pi< \Im T < 0$ or $\Im T=0$ with $\Re T \neq 0$. To prove it one can use the general expression (7.694) in~\cite{Jeffrey:2007}. Notice that the same expression is also related to the Euclidean propagator for the normal oscillator upon double analytic continuation. 

Another useful formula is (Sokhotski-Plemelj)
\be
\frac{1}{x-y \pm i c} = \mathcal{P} \frac{1}{x-y} \, \mp \, i \pi \delta(x-y)
\ee
The density of states is then given by ($c$ is an infinitesimal number)
\be
\rho(\omega) = \frac{1}{\pi} \Im \, \Tr \left( \frac{1}{\hat{H} - \omega - i c} \right) = \frac{1}{\pi} \Im \int_{-\infty}^\infty d \lambda \, R(\omega + i c; \, \lambda, \lambda) \, .
\ee

\subsection{Dual resolvent and density of states}\label{DualResolvent}

The expressions of the previous section were useful for performing computations using the usual interpretation of the theory as a string theory in a linear dilaton background. In our interpretation it will be instead useful to define a dual version of the resolvent that is related with the notion of time and energy for the boundary theory ``living" on the macroscopic loop. This ``duality" is in a sense exchanging the two indices of the Parabolic cylinder functions, since the new notion of energy is conjugate to the loop length and hence is directly identified with the coordinate label of the parabolic cylinder function. If we wish to explicitly compute the integrals below it turns out that instead of using the even/odd parabolic cylinder function basis, it is more convenient to use the complex basis defined by \eqref{Paraboliccomplex}.

We first define the dual resolvent operator as
\be
\hat{R}_{dual}(\zeta) = \frac{1}{\hat{H}_{dual} - \zeta - i c} \, ,
\ee
where $\hat{H}_{dual}$ is the position operator $\hat \lambda$. Once should also understand the role of the chemical potential $\mu$ from this point of view, though. What we find is that it provides an IR mass-gap for the eigenvalues, at the semi-classical level\footnote{Quantum mechanically the eigenvalues can tunnel to the other side, and there are still eigenvalues in the excluded region as we found in section~\ref{DOSofdual}}. We can define again a fixed ``energy" amplitude now as a transition amplitude between the previous oscillator energy eigenstates
\bea
\langle \omega_1 | \hat{R}(\zeta) | \omega_2   \rangle \, = \, \tilde{R}_{dual}(\zeta ; \, \omega_1, \omega_2) &=& \int_{-\infty}^\infty d \lambda \frac{1}{\lambda - \zeta} \sum_s \psi^s (\omega_1, \lambda) \psi^s(\omega_2, \lambda) \nn \\
&=& - i \int_0^{- \infty \sgn (\Im \zeta)} d z \, e^{- i  z \zeta} \,  \langle \omega_1 |  e^{-  i z \hat{H}_{dual}}| \omega_2 \rangle \, . \nn \\
\eea
with the new dual version of the Mehler formula being
\be
\langle \omega_1 | e^{- i z \hat{H}_{dual}}| \omega_2 \rangle= \int_{-\infty}^\infty d \l \, e^{i \l z} \sum_{s=\pm} \psi^s(\omega_1 , \l) \psi^s(\omega_2 , \l) \, .
\ee
This is most easily computed in the complex basis \eqref{Paraboliccomplex} where it takes the form
\be\label{dualMehler}
\langle \omega_1 | e^{- i z \hat{H}_{dual}}| \omega_2 \rangle= \int_{-\infty}^\infty d \l \, e^{i \l z}  \Phi(\omega_1 , \l) \Phi^*(\omega_2 , \l) \, .
\ee
By expressing the complex wavefunctions in terms of the Whittaker function $W_{\m , \n}(z)$, one can then use (6.643) of~\cite{Jeffrey:2007} to express them in terms of Bessel functions. By changing the order of the integrals one finds the following expression for \eqref{dualMehler}
\be
\frac{\pi^2}{2^{i(\omega_1 - \omega_2)}} \frac{e^{\frac{i}{2}(\phi_2(\omega_1) - \phi_2(\omega_2))}}{\Gamma(\frac{1}{4} + i \frac{\omega_1}{2})\Gamma(\frac{3}{4} + i \frac{\omega_1}{2}) \Gamma(\frac{1}{4} - i \frac{\omega_2}{2})\Gamma(\frac{3}{4} - i \frac{\omega_2}{2})} \, \int_0^\infty  \frac{d y \, e^{i z^2/2 - z y}}{y^{\half + i \omega_2} (i z - y)^{\half - i \omega_1}} \,.
\ee
The first thing to notice is that in the limit $z \rightarrow 0$ we recover $\delta(\omega_1 - \omega_2)$ due to the orthonormality of the complex wavefunctions. Unfortunately we are not aware of a more compact way of expressing the dual
Mehler kernel \eqref{dualMehler}.

The new dual density of states ($E \equiv \lambda$) is
\bea
\rho_{dual}(E) &=& \frac{1}{\pi} \Im \, \Tr \left( \frac{1}{\hat{H}_{dual} - E - i c} \right) = \frac{1}{\pi} \Im \int_{-\infty}^\infty d \omega \, \tilde{R}_{dual}(E + i c; \, \omega, \omega) \, \nn \\
 &=& \int_{-\infty}^\infty d \omega \, \Theta(\omega - \mu) \, \sum_{s = \pm} \psi^s(\omega, E) \psi^s(\omega, E) \, .
\eea
This is exactly equivalent to the formula \eqref{dualdos} of the main text. We also observe the importance of the mass gap $\mu$ and the definition of the fermionic vacuum state $| \mu \rangle$ for obtaining a non-trivial result.

Our final expression, is the definition of the density of states as an \emph{operator}. This is obtained from the general time dependent fermionic bi-linear operator \eqref{fermiondensityoperator} upon Euclidean time averaging i.e.
\be
\hat{\rho}_{dual}(E) = \int_{- \infty}^\infty d x \,  \hat{\psi}^\dagger(x, E) \hat{\psi}(x, E) \, .
\ee
This then means that we can compute any correlator of the dos operator upon sending $q_i \rightarrow 0$ in eqn. \eqref{densitycorrelators} of appendix~\ref{fermioniccorrelators}.

\subsection{Density correlation functions}\label{Densitycorrelationfunctions}

Using the previous formulae it is also easy to compute the density correlation functions or resolvent correlation functions. It is useful to use \eqref{norm1} and \eqref{norm2} to resolve the identity operator accordingly. For example the two point function of the dual resolvent is
\be
\langle \mu | \hat{R}_{dual}(\zeta_1) \hat{R}_{dual}(\zeta_2)  | \mu \rangle \, .
\ee
As another example the two point function of the dual density of states is given by
\be
\langle \mu | \hat{\rho}_{dual}(E_1) \hat{\rho}_{dual}(E_2)  | \mu \rangle =  \,
\ee
$$
\int_{-\infty}^\infty d \omega_1 \, \Theta(\omega_1 - \mu) \,  \int_{- \infty}^\infty d \omega_2 \,  \Theta(\omega_2 - \mu)  \, \sum_{s,s' = \pm}  \psi^s (\omega_1 , E_1) \psi^s (\omega_2 , E_1)  \psi^{s'} (\omega_2 , E_2) \psi^{s'} (\omega_1 , E_2)    \, .
$$
Nevertheless in this case it is most convenient to use the general formula eqn.\eqref{densitycorrelators} sending $q_i \rightarrow 0$ and use the inverted oscillator propagator to get
\be
\langle \mu | \hat{\rho}_{dual}(E_1) \hat{\rho}_{dual}(E_2)  | \mu \rangle =  \,
\ee
$$
=  \Re \, \frac{1}{2 \pi }  \int_{0}^\infty d s_1 \int_{0}^\infty d s_2  \frac{e^{i \mu (s_1 + s_2)}}{s_1 + s_2} \frac{1}{\sqrt{ \sinh s_1 \sinh s_2} } e^{ - \frac{i}{4} \left(  \frac{\sinh(s_1 + s_2)}{\sinh s_1 \, \sinh s_2}  (E_1^2 + E_2^2) -  E_1 E_2 \left[ \frac{2 }{\sinh s_1} + \frac{2 }{\sinh s_2} \right] \right)}
$$
The connected piece to the spectral form factor is given by the fourier transform of this expression
\be
SFF(t, \beta) = \int_{- \infty}^\infty d E_1 \int_{- \infty}^\infty d E_2 \, \langle \mu | \hat{\rho}_{dual}(E_1) \hat{\rho}_{dual}(E_2)  | \mu \rangle \, e^{- (\beta + i t) E_1 - (\beta - i t) E_2} \, ,
\ee
that is found to match the expression in the main text~\ref{SFFmain}. 



\section{Correlation functions from the fermionic field theory}\label{fermioniccorrelators}

In this appendix we review the computation of various correlation functions through the use of the fermionic field theory described in section \ref{MQMandfermions}. It was first described in detail in~\cite{Moore:1991sf}, which we closely follow almost verbatim.

The connected correlation function of any number of density operators \eqref{fermiondensityoperator}, is found by
\be
G_E (q_1, \lambda_1; q_2, \lambda_2;....;q_n, \lambda_n ) = \prod_{i=1}^n \int d x_i e^{i q_i x_i} \langle \mu | \hat \rho(x_1, \lambda_1) ... \hat \rho(x_n, \lambda_n) |\mu\rangle_c  \, .
\ee
We chose to work in momentum space with momenta $q_i$, since this is where this correlation function takes the most simple form. The idea is to first compute this expression and then perform the rest of the integrals. From Wick's theorem it is reduced in combinations of the two-point Euclidean time ordered correlation function
\bea
G_E(q_j , \lambda_j) &=& \frac{1}{n}\int d x_i e^{i q_i x_i} \sum_{\sigma \in S_n} \prod \langle \mu |T_E \hat \psi^\dagger(x_{\sigma(i)}, \lambda_{\sigma(i)}) \hat \psi(x_{\sigma(i+1)}, \lambda_{\sigma(i+1)}) |\mu\rangle \nn \\
&=& \frac{1}{n}\delta(\sum q_i)\int d q  \sum_{\sigma \in S_n} \prod_{k=1}^n R(\tilde{Q}^\sigma_k , \lambda_{\sigma(k)}, \lambda_{\sigma(k+1)})\, ,
\eea
with $\tilde{Q}^\sigma_k = q+ q_{\sigma(1)+ ...+q_{\sigma(k)}}$ and $R(\zeta, \lambda_i, \lambda_j)$ the fixed energy amplitude related to the harmonic oscillator resolvent, see~\ref{Resolvent}. The two-point correlation function is related to the resolvent and the H.O. propagator through
\bea
\langle \mu |T_E \hat \psi^\dagger(x_1, \lambda_1) \hat \psi(x_2, \lambda_2) |\mu\rangle &=& \int d \e  \, e^{- (\e-\mu)\Delta x} \left[\theta(\Delta x)\theta(\e - \mu) - \theta(-\Delta x) \theta(\mu - \e)\right] \times \nn \\
&\times & \psi^{\dagger s}(\e, \lambda_1) \psi^s(\e, \lambda_2) \nn \\
&=& i \int_{- \infty}^\infty \frac{d p}{2 \pi} e^{-i p \Delta x} \int_0^{\sgn (p) \infty} ds e^{- s p + i \mu s} \langle \lambda_1 | e^{i 2 s \hat H} | \lambda_2 \rangle \nn \\
&=& i \int_{- \infty}^\infty \frac{d p}{2 \pi} e^{-i p \Delta x} R(\mu + i p, \lambda_1 , \lambda_2)
\eea
which uses the explicit form of $ R(\mu + i p, \lambda_1 , \lambda_2)=  \langle \lambda_1 |\frac{1}{\hat{H}-\mu -i p}  | \lambda_2 \rangle $, the fixed energy amplitude for $p>0$. 

Using this, the generic formula for the n-point density correlator then takes the form
\bea\label{densitycorrelators}
\frac{\partial G_E(q_i, \lambda_i)}{\partial \mu} = i^{n+1} \delta(\sum_i q_i) \sum_{\sigma \in S_n} \int_{-\infty}^\infty d \xi e^{i \mu \xi} \int_0^{\sgn_1 \infty} d s_1 ... \int_0^{\sgn_{n-1} \infty} d s_{n-1} \nn \\
\times e^{- \sum_{k=1}^{n-1} s_k Q_k^\sigma} \langle \lambda_{\sigma(1)} | e^{2 i s_1 \hat H}|  \lambda_{\sigma(2)} \rangle ..... \langle \lambda_{\sigma(n)} | e^{2 i (\xi - \sum_k^{n-1} s_k) \hat H}|  \lambda_{\sigma(1)} \rangle
\eea
with $\xi = \sum_{i=1}^n s_i$ and $Q^\sigma_k = q_{\sigma(1)}+ ...+q_{\sigma(k)} $.

From this expression it is possible to describe the general correlation function, expressed through the formula ($q_i, z_i$ are the correlator parameters)
\bea\label{correlatorfinal}
\frac{\partial \langle \mathcal{\hat O}(q_1, z_1)...  \mathcal{\hat O}(q_n, z_n) \rangle}{\partial \mu} = i^{n+1} \delta(\sum_i q_i) \sum_{\sigma \in S_n} \int_{-\infty}^\infty d \xi e^{i \mu \xi} \int_0^{\sgn_1 \infty} d s_1 ... \int_0^{\sgn_{n-1} \infty} d s_{n-1} \nn \\
\times e^{- \sum_{k=1}^{n-1} s_k Q_k^\sigma} \Tr \left( f(z_{\sigma(1)}, \hat \lambda) e^{2 i s_1 \hat H} f(z_{\sigma(2)}, \hat \lambda)  e^{2 i s_2 \hat H} ..... f(z_{\sigma(n)}, \hat \lambda) e^{2 i (\xi - \sum_k^{n-1} s_k) \hat H} \right) \nn
\eea
In particular to obtain the correlation function of loop-operators, we just need to perform gaussian integrals, since $f(z, \hat{\lambda}) = e^{i z \hat{\lambda}}$. These gaussian integrals are most easily expressed in operator form and computed to give the result
\bea
&\Tr (e^{z_n \hat \lambda} e^{- x_1 \hat H} e^{z_1 \hat \lambda} e^{- x_2 \hat H}....e^{z_{n-1} \hat \lambda} e^{- x_n \hat H}) \nn \\
&= \frac{1}{2 \sinh \omega T/2} \exp \left[\frac{ \coth \omega T/2}{4 \omega} \sum_{1}^n z_i^2\right] \exp \left[ \sum_{i<j} \frac{\cosh (\omega(x_i+....x_{j+1}-T/2))}{2 \omega \sinh \omega T/2} z_i z_j \right]
\eea
with $\hat H = \half \hat p^2 + \half \omega \hat \lambda^2$ so that in the end one analytically continues $\omega $ to get the result for the inverted oscillator. One can also analytically continue $x$ to get the real time result.

One then finds a compact expression for the derivative of the n-point loop correlator in terms of an n-fold integral
\bea\label{Anyloopcorrelator}
\frac{\partial M(z_i , q_i)}{\partial \mu} = \half i^{n+1} \delta(\sum q_i) \sum_{\sigma \in S_n} \int_{-\infty}^\infty d \xi \frac{e^{i \mu \xi}}{|\sinh \xi/2|} \int_0^{\sgn_1 \infty} d s_1 ... \int_0^{\sgn_{n-1} \infty} d s_{n-1}  \nn \\
\times \exp \lbrace - \sum_{k=1}^{n-1} s_k Q_k^\sigma +\frac{i}{2}\coth(\xi/2)\sum z_i^2 + i \sum_{1\leq i<j \leq n} \frac{\cosh(s_i+..s_{j-1}-\xi/2)}{\sinh(\xi/2)} z_{\sigma(i)} z_{\sigma(j)} \rbrace \nn \\
\eea
So far in this expression we have used both sides of the inverted oscillator potential, so that $\xi \in (-\infty, \infty)$. If we wish to describe the bosonic theory, we can focus on the one side of the potential $\xi \in [0, \infty)$. In the integral \eqref{Anyloopcorrelator}, we can analytically continue $z_i = i \ell_i$ in the positive $\xi$ region and $z_i = - i \ell_i$ in the negative $\xi$ region to obtain convergent answers.

This is the expression we use in the main text in the specific cases of a single or two loops.

\section{Correlation functions for a compact boson}\label{finiteTcorrelators}

The analysis so far was performed in the case of a non-compact boson $X(z, \bar{z})$. If we take this to be compact with a radius $R$, it is still possible to compute non-perturbatively all the correlation functions as before~\cite{Klebanov:1991ai}. In order to do so one needs to use the formalism of free fermions at finite temperature. The thermal vaccum satisfies (notice that $\beta = 2 \pi R$ is not related with $\ell$, the temperature of the holographic boundary theory, but is related with a ``temperature in superspace")
\bea
\langle \mu | b^\dagger_s(\omega) b_{s'}(\omega)| \mu \rangle_R = \delta_{s s'}  \frac{1}{e^{\beta(\mu - \omega)} +1} \, , 
\eea
which means that in all the formulas of appendix~\ref{fermioniccorrelators} one just needs to replace the strict occupation $\theta(\omega-\mu)$ with the fermi distribution $f(\omega) =  \frac{1}{e^{\beta(\mu - \omega)} +1} $.
It also results in having discrete frequencies $\omega_n = (n+\half)/R$, which certifies the anti-periodicity of the correlator around the compact Euclidean time direction 
\be
\langle \mu |T_E \hat \psi^\dagger(\lambda_1, x_1) \hat \psi(\lambda_2, x_2) |\mu\rangle_R 
= \frac{i}{2 \pi R} \sum_{\omega_n}  e^{-i \omega_n \Delta x} \int_0^{\sgn (\omega_n) \infty} ds e^{- s \omega_n + i \mu s} \langle \lambda_1 | e^{i 2 s \hat H} | \lambda_2 \rangle \, ,
\ee
In the end one finds the following formula
\bea
\frac{\partial G_E(q_i, \lambda_i)}{\partial \mu} = i^{n+1} R \delta_{(\sum_i q_i)} \sum_{\sigma \in S_n} \int_{-\infty}^\infty d \xi e^{i \mu \xi} \frac{\xi/2R}{\sinh(\xi/2R)} \int_0^{\sgn_1 \infty} d s_1 ... \int_0^{\sgn_{n-1} \infty} d s_{n-1} \nn \\
\times e^{- \sum_{k=1}^{n-1} s_k Q_k^\sigma} \langle \lambda_{\sigma(1)} | e^{2 i s_1 \hat H}|  \lambda_{\sigma(2)} \rangle ..... \langle \lambda_{\sigma(n)} | e^{2 i (\xi - \sum_k^{n-1} s_k) \hat H}|  \lambda_{\sigma(1)} \rangle \nn \\
\eea
with $\xi = \sum_{i=1}^n s_i$ and $Q^\sigma_k = q_{\sigma(1)}+ ...+q_{\sigma(k)} $. This simply means that one can just replace $\delta(\sum q_i) \rightarrow R \delta_{\sum q_i}$, adding an extra factor $\frac{\xi/2R}{\sinh(\xi/2R)}$ in the formulae of the previous appendix~\ref{fermioniccorrelators}. For example one finds the simple interesting relation between the infinite/finite radius observables
\bea
\langle \mathcal{O}_1(q_1)...\mathcal{O}_n(q_n) \rangle = \delta(\sum q_i) M(q_i ; \mu)\, , \nn \\
\langle \mathcal{O}_1(q_1)...\mathcal{O}_n(q_n) \rangle_R =R \delta_{(\sum q_i)} M_R(q_i ; \mu)\, , \nn \\
M_R(q_i ; \mu) =\frac{\frac{1}{2R} \partial_\mu}{\sin(\frac{1}{2R} \partial_\mu)}   M(q_i ; \mu) \, .
\eea
The genus expansion is achieved using
\be
\frac{\frac{1}{2R} \partial_\mu}{\sin(\frac{1}{2R} \partial_\mu)} = 1+ \sum_{k=1}^\infty \frac{(1 - 2^{1-2k}) |B_{2k}|}{(R)^{2 k} (2k)!} \left( \frac{\partial}{\partial \mu} \right)^{2k}
\ee

\section{Steepest descent}\label{SteepestDescent}

We will briefly review here the method of steepest descent that can be used to obtain asymptotic expansions of integrals as one parameter goes to infinity. 

Let us define
\be\label{Steepest1}
I(t) = \int_{\mathcal{C}} h(s) e^{t \rho(s)} d s \, ,
\ee
an integral supported in a contour $\mathcal{C}$ of the complex plane. We are interested in the $t \rightarrow \infty$ asymptotics of this integral. Depending on the analyticity properties of the functions $h(s)$, $\rho(s)$ we can deform the contour to another contour $\mathcal{C}'$. If $\rho(s) = \phi(s) + i \psi(s)$ with $\phi(s), \psi(s)$ real functions, it is useful to deform the contour such that we fix $\Im \, \rho(s) = const.$. This is because the $\psi = const.$ contours are parallel to the steepest contours in $\phi$. Since the full asymptotic expansion of the integral is governed by the neighbourhood of $s$ on which $\phi(s)$ acquires its maximum value, it is then useful to use precicely these steepest contours to approximate the integral. For example in case $\phi(s)$ attains its maximum at the endpoint/s of the contour, one can use integration by parts on \eqref{Steepest1} and approximate the integral via the boundary term/s
\be
I(t) \sim e^{i t \psi} \frac{h(s)}{t \phi'(s)} e^{ t \phi(s)} \bigg|_{s=a}^{s=b} \, , \quad t \rightarrow \infty \, .
\ee
If the maximum is in a middle point $c$, one can use a quadratic saddle point expansion around $s^*$ and so forth. The generic case for $\phi'(c) = \phi''(c) = ... = \phi^{(p-1)}(c)=0$ gives
\be
I(t) \sim h(c) e^{i t \psi} e^{t \phi(c)} \left[ \frac{p!}{-t \phi^{(p)}(c)} \right]^{1/p} \frac{\Gamma(1/p)}{p} \, , \quad t \rightarrow \infty \, ,
\ee
This idea can fail in case $\rho'(s^*) = 0$. In such a case distinct steepest contours can intersect at the point $s^*$ and one needs to make a careful choice depending on the precise way one takes the $t \rightarrow \infty$ limit on the complex $t$ plane. This can lead to the Stokes phenomenon. This phenomenon is simply the fact that the analytic continuation of an asymptotic expansion of a function does not agree with the asymptotic expansion of the exact function's analytic continuation. 

A special case that will be of interest here is the case of a movable maximum. Such a case arises for example when at the saddle point $\phi'(s)=0$ the function $h(s)$ goes to zero exponentially fast. In such cases one is instructed to find the maxima of the total exponent $\log ( h(s) ) + t \, \phi(s)$. These maxima are then dependent on $t$, $s^* = s^{*}(t)$. In such a case one needs to rescale $s$ appropriately, so that the maxima no-longer depend on $t$. Another special case is when at the maximum $\phi(s^*)$ blows up. This case can be treated in the same fashion with the movable maxima.

\subsection{Stationary phase approximation}\label{StationaryPhase}

In the main text we encountered various integrals with highly oscillatory behaviour as we send $t \rightarrow \infty$. Here for concreteness we collect some useful results of an asymptotic analysis of such integrals. This covers a subcase of integrals for which we can apply the more general steepest descent method.

We first define
\be\label{SPintegral}
I(t) = \int_a^b d s f(s) e^{i t g(s)}  \, ,
\ee
the integral whose asymptotic properties as $t \rightarrow \infty$ we wish to analyse. Asumming that there is no stationary point $g'(s) = 0$ in the region $s \in [a, b]$, then we can use the general Riemann-Lebesgue lemma and its corollaries
\begin{itemize}

\item If $f(s)$ is integrable at the region of support and if $g(s)$ is continuously differentiable and non-constant in any sub-interval then $I(t) \rightarrow 0$ as $t \rightarrow \infty$.

\item In particular integration by parts gives the leading asymptotic behaviour of \eqref{SPintegral} provided that $f(s)/g'(s)$ is smooth for $s \in [a,b]$, that is
\be
I(t) \sim \frac{f(s)}{i t g'(s)} e^{i t g(s)} \bigg|_{s=a}^{s=b} \, , \quad t \rightarrow \infty \, .
\ee

\item If there is a stationary point of $g(s)$ denoted by $c$, then if $g'(c) = g''(c) = ... = g^{(p-1)}(c)=0$ and $g^{(p)}(c) \neq 0$ the leading asymptotic behaviour is given by
\be
I(t) \sim f(c) e^{i t g(c) \pm \frac{i \pi}{2 p}} \left[ \frac{p!}{t |g^{(p)}(c)|} \right]^{1/p} \frac{\Gamma(1/p)}{p} \, , \quad t \rightarrow \infty \, ,
\ee
where the sign of the phase is the same as the sign of $g^{(p)}(c)$.

\end{itemize}

\newpage
\addcontentsline{toc}{section}{References}
 
\end{document}